\PassOptionsToPackage{table}{xcolor}
\documentclass[10pt,journal,compsoc]{IEEEtran}
\usepackage{graphicx}
\usepackage{amsmath}
\usepackage{amssymb}
\usepackage{booktabs}
\usepackage{color}
\usepackage{multirow}

\usepackage{forest}

\usepackage{adjustbox} 

\usepackage{supertabular}
\usepackage{longtable}
\usepackage{booktabs}
\usepackage{multirow}
\usepackage{array}
\usepackage{siunitx}

\usepackage[table]{xcolor}
\usepackage{colortbl}
\definecolor{lightblue}{RGB}{242,247,255}   
\definecolor{lightgreen}{RGB}{242,255,242}  
\definecolor{lightpurple}{RGB}{250,242,255} 
\usepackage{color}
\definecolor{mygray}{gray}{0.5}  

\newcommand{\pubinfo}[1]{{\color{mygray}\scriptsize{#1}}}
\usepackage{cite}
\usepackage{hyperref}
\hypersetup{hypertex=true,
            colorlinks=true,
            linkcolor=blue,
            anchorcolor=blue,
            citecolor=blue}

\ifCLASSINFOpdf
\else
\fi

\begin{document}
%


\title{Composed Multi-modal Retrieval: A Survey of Approaches and Applications}

%
%
%
%

\author{Kun Zhang$^{*}$,
        Jingyu Li$^{*}$,
        Zhe Li$^{*}$,
        ~Jingjing Zhang$^{*}$,
        Fan Li,
        Yandong Liu, 
        Rui Yan,
        Zihang Jiang, \\
        Nan Chen,
        Lei Zhang,
        Yongdong Zhang,~\IEEEmembership{Fellow,~IEEE,} 
        Zhendong Mao$^{\dagger}$, and
        S.Kevin Zhou$^{\dagger}$~\IEEEmembership{Fellow,~IEEE}
        
\IEEEcompsocitemizethanks{\IEEEcompsocthanksitem All authors are from the University of Science and Technology of China.\protect\\
$*$: Contributed equally.
$\dagger$: Corresponding authors. \\
E-mails: \{kkzhang@, jingyuli@iai., lizhe777@mail., zjj1029@mail., zdmao@,  skevinzhou@\}ustc.edu.cn}
}

%
%

\markboth{Journal of \LaTeX\ Class Files,~Vol.~14, No.~8, August~2015}%
{Shell \MakeLowercase{\textit{et al.}}: Bare Demo of IEEEtran.cls for Computer Society Journals}
%



\IEEEtitleabstractindextext{%
\begin{abstract}
The burgeoning volume of multi-modal data necessitates advanced retrieval paradigms beyond unimodal and cross-modal approaches. Composed Multi-modal Retrieval (CMR) emerges as a pivotal next-generation technology, enabling users to query images or videos by integrating a reference visual input with textual modifications, thereby achieving unprecedented flexibility and precision. This paper provides a comprehensive survey of CMR, covering its fundamental challenges, technical advancements, and applications. CMR is categorized into supervised, zero-shot, and semi-supervised learning paradigms. We discuss key research directions, including data construction, model architecture, and loss optimization in supervised CMR, as well as transformation frameworks and linear integration in zero-shot CMR, and semi-supervised CMR that leverages generated pseudo-triplets while addressing data noise/uncertainty. Additionally, we extensively survey the diverse application landscape of CMR, highlighting its transformative potential in e-commerce, social media, search engines, public security, etc. Seven high impact application scenarios are explored in detail with benchmark data sets and performance analysis. Finally, we further provide new potential research directions with the hope of inspiring exploration in other yet-to-be-explored fields. A curated list of works is available at: \href{https://github.com/kkzhang95/Awesome-Composed-Multi-modal-Retrieval}{Awesome Composed Multi-modal Retrieval}.
\end{abstract}

}




\maketitle

\IEEEdisplaynontitleabstractindextext

%
\IEEEpeerreviewmaketitle

\IEEEraisesectionheading{\section{Introduction}\label{introduction}}
\IEEEPARstart{N}{owadays}, the unprecedented prosperity of social media, short video platforms, and e-commerce (such as Facebook, Weibo,  YouTube, Amazon, and Alibaba) has greatly promoted the rapid growth of multi-modal data, encompassing various modalities such as texts, images, and videos. Faced with this data flood composed of heterogeneous information, content-based retrieval, as a key technology to search and utilize these vast resources~\cite{liu2007survey}, not only plays a core role in the field of e-commerce, such as achieving accurate matching of goods, but also greatly enriches the user experience of the social media, allowing users to quickly and accurately find the content of interest. 
Due to its widespread applications in everyday life that almost impact everyone, content-based retrieval has attracted great attention from both academia and industry \cite{ dubey2021decade, hameed2021content}.

\begin{figure*}
  \centering
\includegraphics[width=1.0\textwidth]{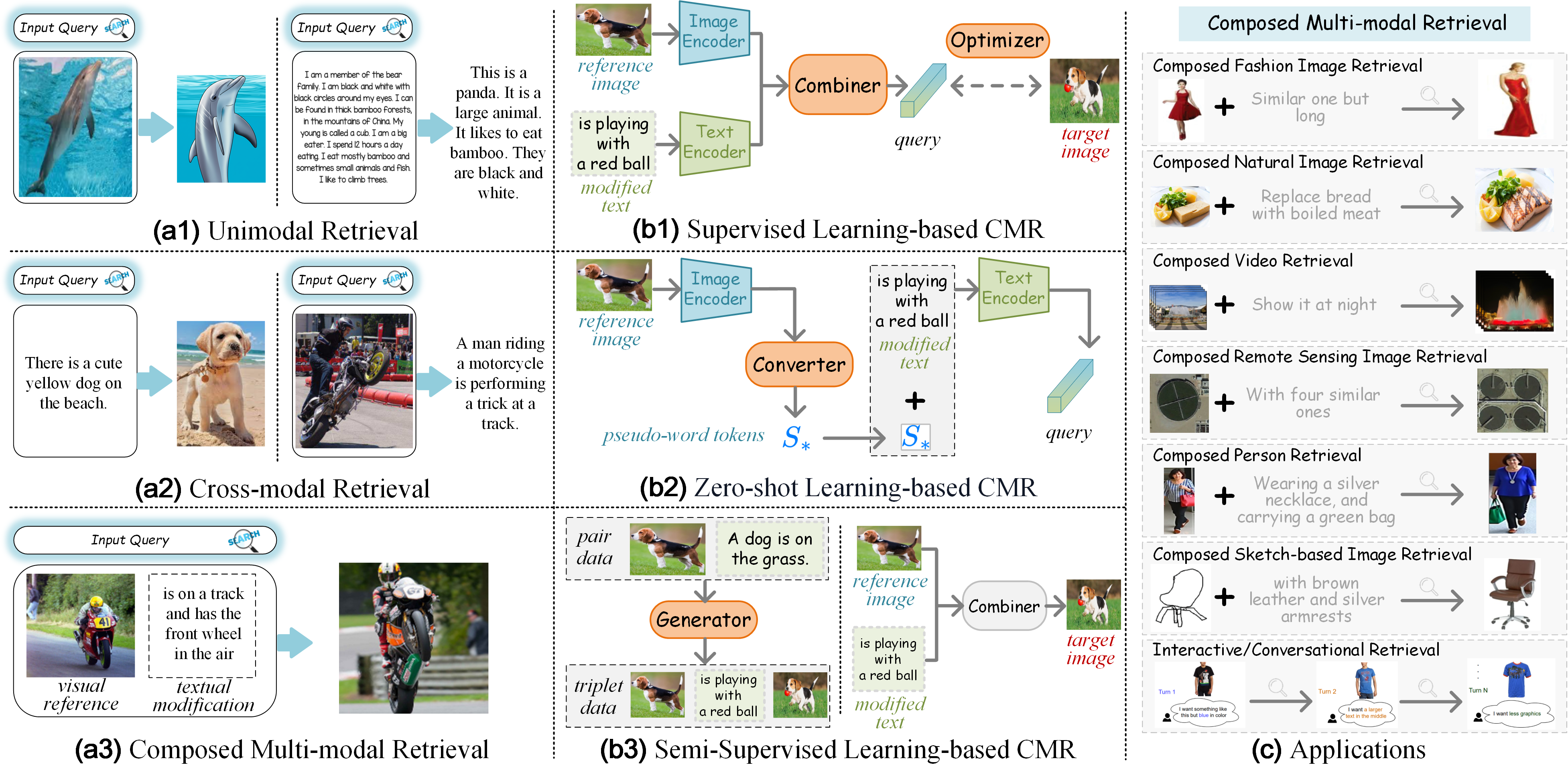}
  \caption{(a) The evolution of content-based retrieval technology. (b) In the current research on composed multimodal retrieval (CMR), three main paradigms have been developed.  (c) The CMR applications, broadly categorized based on application scenarios and image domains. 
  }
  \label{intro}
\end{figure*}

Generally, the evolution of content-based retrieval technology has witnessed the transformation from Unimodal Retrieval (UR) to Cross-modal Retrieval (CR), and then to Composed Multi-modal Retrieval (CMR). Compared with early-stage unimodal retrieval, which was limited to querying information within the same modality~\cite{datta2008image, gordo2016deep, zhao2024dense}, as shown in Fig.~\ref{intro}(a1), cross-modal retrieval has achieved remarkable accuracy and widespread application in the present era. This enables the search for semantically relevant content in one modality based on the instance query from another modality~\cite{wang2016comprehensive, zhen2019deep, lee2018stacked,chen2020fine, ma2022x}, e.g., using text search on images in Fig.~\ref{intro}(a2), allowing users to make full use of these heterogeneous data. In recent years, composed multi-modal retrieval has emerged as a thriving content-based retrieval technology. Within this technical framework~\cite{ saito2023pic2word}, as depicted in Fig.~\ref{intro}(a3), the system aims to discover images/videos that not only bear resemblance to the given reference image/video but also allow for specific modifications based on the provided textual feedback from the user. In this sense, CMR pioneers an advanced level of interactive and conditional retrieval mechanisms, leveraging deep integration of visual and linguistic information. This integration greatly enhances the flexibility and precision of user-expressed search intents, injecting new vitality into domains such as internet search and e-commerce. Consequently, CMR exhibits vast potential and far-reaching impact as the next-generation content-based retrieval engine in real-world application scenarios.

A core of CMR is that it requires a synergistic understanding and composition of both input vision and language information as the multi-modal query. The earliest closely related studies of CMR are in the field of attribute-based fashion image retrieval~\cite{islam2024survey, liu2016deepfashion, yan2022attribute, wan2023dual, shin2019semi, ak2018efficient, hou2021learning}, where the key difference is that the textual feedback in attribute-based fashion image retrieval is limited to the predefined attribute value (e.g., `mini', `white', `red'), while CMR is the natural language with multiple words (e.g., `showing this animal of the input image facing the camera under sunlight'), which is more flexible yet challenging. 
The pioneering CMR works are proposed in \cite{vo2019composing}, where the input query is specified in the form of an image plus some natural language that describes desired modifications to the input image, leading to a series of subsequent approaches. Current research in CMR is primarily focused on three paradigms: (1) supervised learning-based CMR (SL-CMR), which focuses on how to design a better combination mechanism of vision and language through supervised training of annotated data; (2) zero-shot learning-based CMR (ZSL-CMR), which focuses on how to simulate and build a visual-linguistic multi-modal information combination framework without annotated data; and (3) semi-supervised learning-based CMR (SSL-CMR), which focuses on how to enhance the learning of visual-linguistic combination via generated pseudo-labeling data.



For the SL-CMR pipeline, a notable characteristic is the requirement of annotated triplet data $(I_{r}, T_{m}, I_{t})$, which denotes the reference query image, the modified text, and the ground-truth target image, respectively. 
As illustrated in Fig.~\ref{intro}(b1), for the given inputs $I_{r}$ and $T_{m}$, SL-CMR involves mining the content that should be modified in the reference image $I_{r}$ according to the text $T_{m}$, so as to learn a multi-modal compositional embedding to find the interested target image $I_{t}$. 
Thus, the challenges faced by SL-CMR mainly lies in addressing two issues: ``Where to see", which refers to attending to the content in the reference image that needs change, and ``How to change", which aims to modify the reference image based on the textual information while preserving the remaining information. 
In recent years, research on SL-CMR has primarily focused on three aspects: 
(1) data construction~\cite{vo2019composing, liu2021image, suhr2019corpus, baldrati2023zero, forbes2019neural, isola2015discovering, wu2021fashion, han2017automatic, guo2018dialog, ventura2024covr, hummel2024egocvr, zhang2024localizing}, focusing on labeling triples with accurate semantic difference descriptions;
(2) model architecture~\cite{vo2019composing, dodds2022training, baldrati2022effective, baldrati2022conditioned, baldrati2023composed, lin2023clip, lee2021cosmo, anwaar2021compositional, xu2024align, hosseinzadeh2020composed, han2022fashionvil, zhang2021heterogeneous, pang2022heterogeneous, zhang2022composed, jandial2022sac, zhao2024neucore, wang2024negative, chen2020image, dodds2020modality, jandial2020trace, yang2023composed, xu2021hierarchical, li2023multi, huang2023language, liu2021image, wen2021comprehensive, goenka2022fashionvlp, wen2023self, shin2021rtic, zhang2022comprehensive, jiang2024cala, chen2024spirit, zhang2021geometry, bai2023sentence, neculai2022probabilistic, xu2024set}, focusing on designing a better vision-language combiner via cross-modal feature alignment and fusion strategies, as well as the design of other novel frameworks that can be plugged; 
(3) loss optimization~\cite{vo2019composing, zhu2023amc, yang2023decompose, liu2021image, hosseinzadeh2020composed, yang2021cross, chen2020learning, chen2020image, zhang2021geometry, faghri2017vse, zhang2024multimodal, wen2023target, zhou2024vista, zhang2022composed}, focusing on the design of more reasonable feature combination constraints. 
Although supervised training relying on these carefully labeled data often offers high performance, SL-CMR inherently faces two shortcomings: 1) annotating such triplets is both difficult and labor-intensive, and 2) the supervised approaches trained on the collected limited and specific triplets are also hard for generalization.



Recently, ZSL-CMR has been proposed to address the above limitations, where the model is trained solely on easily obtainable large-scale image-caption pairs or unlabeled images. As depicted in Fig.~\ref{intro}(b2), its training process usually revolves around learning the modality converters that simulate the combination of visual and linguistic information in the test. The training and testing phases typically involve different network structures. Thus, a main challenge of ZSL-CMR lies in designing transformation frameworks that achieve accurate vision-language combination in the absence of supervision signals, aiming to maximize the zero-shot generalization ability. To address this challenge, the academic community has developed strategies across three key aspects: (1) image-side transformation~\cite{karthik2023vision, sun2023training, yang2024ldre, baldrati2023zero, saito2023pic2word, agnolucci2024isearle, tang2024context, lin2024fine, du2024image2sentence, suo2024knowledge}: this approach focuses on learning the implicit or explicit visual-to-linguistic transformation using images as input. During testing, it converts the reference image into a query that can be integrated with the relative textual information; (2) text-side transformation~\cite{gu2024language, byun2024reducing, thawakar2024composed, liimproving,li2024motadual}: in this approach, text is used as input to simulate image features, constructing a training framework that relies solely on language. During testing, the model directly takes image inputs; (3)  linear interpolation~\cite{jia2021scaling, jang2024spherical, chen2023pretrain,jangtext}: this approach explores the simple yet effective linear weighted combination strategies of visual and textual features. 





Although zero-shot approaches do not rely on labeled data, their performance is often lower than supervised training, which brings obstacles to the application of the model. To alleviate this problem, as shown in Fig.~\ref{intro}(b3), some CMR works have proposed the semi-supervised learning paradigm based on generated pseudo-labeling data. In this setting, relying on the relatively easy-to-obtain image-text data, existing SSL-CMR works mainly generate triplet data from two aspects: (1) 
generating text~\cite{liu2023zero, hou2024pseudo, zhang2024multimodal, liu2024bi, jiang2024hycir, jang2024visual, levy2024data}, such as describing the difference caption between the two input images; and (2) generating images~\cite{zhang2022tell,gu2023compodiff,zhou2024vista}, such as editing the input reference image according to the conditional text to create the target image. Besides, alleviating the noise in generated data is also a focus of this paradigm~\cite{liu2023zero, jiang2024hycir, gu2023compodiff, lin2023clip, hou2024pseudo, levy2024data, wen2023target, zhang2024collaborative, chen2023ranking, chen2022composed, baldrati2023zero, dodds2022training}. 
In this way, the model can not only capture the combination of vision and language more accurately during learning, but also avoid the limitations of cumbersome annotation. Although the generated data may introduce uncertainty, this paradigm combines the advantages of supervision and zero-shot learning, which is a promising direction.

Research in CMR has vast application potential. 
As illustrated in Fig.~\ref{intro}(c), it can be broadly categorized based on application scenarios and image domains, including fashion and E-commerce  images~\cite{wu2021fashion, han2017automatic, guo2018dialog, dodds2022training, yu2014fine, yu2017semantic,  rostamzadeh2018fashion, yang2020fashion, vasileva2018learning}, natural images~\cite{vo2019composing, liu2021image, baldrati2023zero, levy2024data, liu2023zero, gu2023compodiff, vaze2023genecis}, videos~\cite{ventura2024covr, ventura2024covr2, thawakar2024composed, karthik2023vision, hummel2024egocvr}, remote sensing images~\cite{psomas2024composed, wang2024scene}, person images~\cite{liu2025automaticsyntheticdatafinegrained}, sketch images~\cite{koley2024you}, and interactive conversation~\cite{guo2018dialog, tan2019drill, wu2021deconfounded, yuan2021conversational, levy2024chatting, pal2023fashionntm, wei2023conversational, chen2023fashion, feng2023vqa4cir, barbany2024leveraging}. 
The specific application can be personalized product shopping, media search, event discovery, environmental monitoring, law enforcement, customer service bots, and so on. 
In summary, CMR represents a paradigm shift in search systems by integrating visual and textual modalities. These systems enable fine-grained, context-aware, and user-centric searches across diverse domains, offering significant improvements in both retrieval accuracy and user satisfaction.

Compared to related surveys~\cite{Wan2025, song2025comprehensive} that appeared around the same period of early 2025 on arXiv, the novelty and contributions of this survey are: (1) We comprehensively summarize methods and techniques of visual-language retrieval in recent years, especially cross-modal semantic alignment and combination learning, covering more than 250 works, providing a more in-depth summary for the community. In particular, for the first time, we comprehensively review 6 types of combiners, introduce more than 7 insighful losses, 3 zero-shot combinations, including a new linear interpolation perspective, noise and uncertainty considerations in semi-supervision; (2) We comprehensively emphasize 7 applications and potential scenarios of CMR, and provide more than 26 commonly used datasets and detailed performance statistics for each application, which is the most covered so far; (3) We further provide new potential research directions and guidance for CMR, hoping to inspire exploration in other unexplored scenarios.



\section{Overview} \label{over}

In this section, we provide a concise overview of the taxonomy of current Composed Multi-modal Retrieval (CMR) methods. To better track the latest advancements in CMR, this survey categorizes existing methods from both technical and application perspectives. (1) \textbf{From a Technical Perspective}: we summarize different paradigms of multimodal compositional learning, including sophisticated module designs and the state-of-the-art innovative solutions. (2) \textbf{From an Application Perspective}: we examine various application scenarios where CMR is employed, such as fashion e-commerce, video platforms, geospatial data, and others. It highlights how these methods are applied in real-world contexts, showcasing their versatility and practical utility. By providing a comprehensive summary and reference for researchers in both the technical and application aspects, this survey aims to facilitate deeper thinking in this field and promote further development.

\subsection{CMR Methodology}

From the technical perspective, current CMR methods can be grouped into three main paradigms: supervised learning (SL), zero-shot learning (ZSL), and semi-supervised learning (SSL). Each paradigm addresses the understanding and integration of multiple modalities for effective retrieval, with distinct strategies and trade-offs.

SL-CMR relies on annotated triplet data, i.e.,  (reference image $I_{r}$, modification text $T_{m}$, target image $I_{t}$), to learn how to combine input modalities so that the composed content is semantically aligned with the target image. This process can be summarized as:
\begin{equation} \label{E1}
\begin{aligned} &X_{composed}=Combiner(I_{r}, T_{m}),\\
& \text{s.t.} \quad Optimizer(X_{composed}, I_{t}),
\end{aligned}
\end{equation}
where $Combiner(\cdot)$ denotes the combination of visual and textual inputs, and $Optimizer(\cdot)$ enforces semantic alignment between the composed content and the visual target.
This paradigm benefits from explicit supervision, enabling precise modality combination and high performance. However, it requires large-scale, labor-intensive annotations, which limit its generalization to new domains.

ZSL-CMR overcomes the need for annotated triplets by using readily available image-text pairs. Here, they mainly focus on  converting one modality into another, e.g., mapping a reference image to a textual representation:
\begin{equation} \label{E2}
T_{r}^{*}=Converter(I_{r}),
\end{equation}
and then combines this with the modification text $T_{m}$ to form a composite query. This shifts the task to cross-modal text-to-image retrieval. While ZSL-CMR offers easy data acquisition and better generalization, it lacks direct supervision for modality combination, leading to lower performance on specific tasks compared to supervised learning.

SSL-CMR combines elements of both supervised and zero-shot approaches. It uses modality generation (e.g., image-to-text or text-to-image) to automatically create triplet data from easily accessible image-text pairs $(I, T)$:
\begin{equation} \label{E3}
\hat{I}_{r}, \hat{T}_{m}, \hat{I}_{t} = Generator(I,T).
\end{equation}
Although this paradigm yields high performance, it faces challenges in generating large volumes of high-quality triplets, especially those with subtle semantic differences. Existing work highlights that generated content often contains errors and noise, which remains an ongoing issue.

\begin{figure*}[thp]
  \centering
\includegraphics[width=0.8\textwidth]{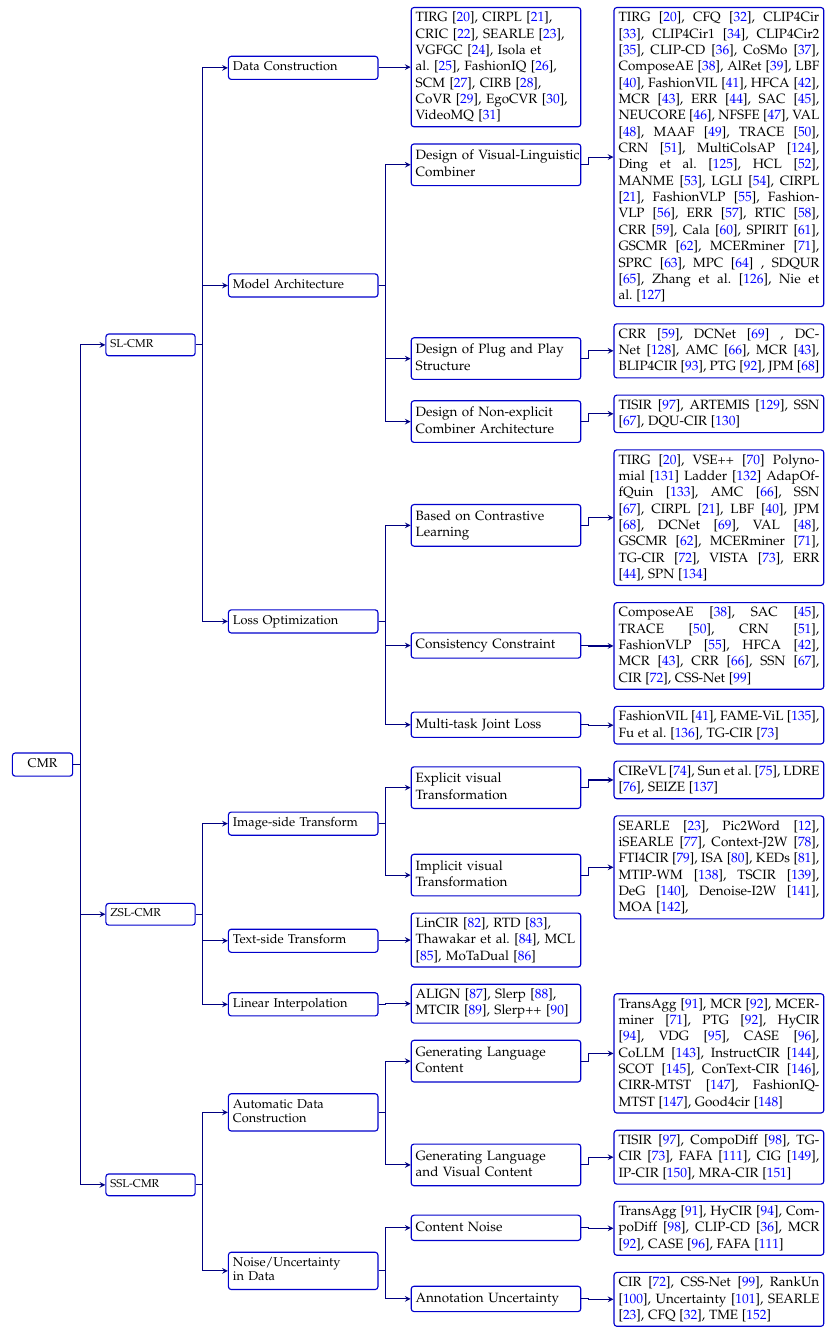}
  \caption{Composed Multi-modal Retrieval Methods Taxonomy.}
  \label{fig:taxonomy}
\end{figure*}

\subsection{CMR Applications}
CMR can be broadly categorized based on application scenarios and image domain differences. 
Specifically, (1) In fashion and e-commerce fields, CMR enables users to search for products by combining reference images with descriptive text, improving personalization and shopping accuracy. (2) In natural image domains, CMR helps users refine searches with contextual cues (e.g., "same scene in autumn"), supporting creative and personal content discovery. (3) In composed video retrieval, composed queries allow users to locate dynamic content with greater precision by specifying conditions like time or activity.
(4) In remote sensing image domains, CMR deals with satellite imagery, where text adds geographic or temporal specificity for tasks like urban planning or disaster monitoring. 
(5) Composed person retrieval combines visual input from a specific person with attribute-based descriptions to support identity matching in surveillance or public safety. 
(6) Sketch-based CMR leverages user-drawn inputs and textual hints, particularly useful in design and creative industries, or law enforcement when photos are unavailable. 
(7) Finally, in the interactive conversational scenarios, CMR allows users to refine queries through multi-turn dialogue, enabling more flexible and accurate searches across various domains.

Next, we will introduce composed multi-modal retrieval based on supervised, zero-shot, and semi-supervised learning in Sect.~\ref{SL-CMR}, \ref{ZSL-CMR}, and \ref{SSL-CMR}, respectively. Then, in Sect.~\ref{APPL}, we will introduce the CMR applications in different scenarios, including related benchmark datasets, evaluation metrics, and method performance. 
Finally, in Sect.~\ref{future} and Sect.~\ref{CONC}, we look forward to future research directions and conclude the paper, respectively.

\section{Supervised Learning-based CMR} \label{SL-CMR}

Existing SL-CMR can be categorized into three key dimensions: data, model architecture, and optimization objectives. \textbf{First}, as discussed in Section \ref{over}, the construction and annotation of multimodal triplet data pose a fundamental challenge in this field.  \textbf{Second}, in terms of model architecture, existing studies investigate various strategies for combining visual and linguistic modalities. These efforts emphasize the precise semantic modeling, focusing on retaining, modifying, and supplementing semantics during combination. The goal is to ensure that the combined representation captures the intended modifications and preserves the relevant context. \textbf{Finally}, the optimization objectives in the triplet data framework represent another critical aspect. Building on contrastive learning, recent advancements have introduced improvements to objective functions. In the following, we provide a detailed review of the progress made in these areas, highlighting key contributions and ongoing challenges.


\subsection{Data Construction}

The quantity and quality of data play a crucial role in deep learning, especially in light of the empirical evidence supporting scaling laws~\cite{kaplan2020scaling}. Unlike conventional cross-modal visual-language retrieval, where image-text pairs or video-text pairs can be readily collected from the web, CMR requires training data in the form of triplets, e.g., \{reference image, modified text, target image\}. In such triplets, the reference image and modified text jointly represent the user’s retrieval intent, with the text specifying flexible modification requests grounded in the rich visual objects and attributes of the reference image. The target image serves as the desired output of the retrieval process, with the relative text accurately describing the semantic differences between the reference and target images. However, in practice, triplets that capture such nuanced semantic differences are scarce, making construction of large-scale, high-quality triplet data a fundamental challenge in this field.

Early works primarily relied on manual annotation to construct triplet datasets. 
CSS \cite{vo2019composing} introduces a simple 3D scene rendering dataset. They first generate relative modification texts for reference images based on manually defined templates. These modifications target different object attributes such as color, shape, and size, and involve operations such as addition, removal, and alteration, for example, ``add a red cube". The CLEVR toolkit~\cite{johnson2017clevr} is then used to synthesize new target images, thereby constructing triplet data with relative semantic differences. CIRR \cite{liu2021image} is the first large-scale benchmark tailored for composed image retrieval with real-world photos. Derived from the NLVR$^2$ dataset \cite{suhr2019corpus}, CIRR includes 21,552 images and 36,554 triplets. The dataset emphasizes semantic diversity and minimizes false positives, with a single target image per query and an online evaluation protocol. CIRCO \cite{baldrati2023zero} introduces multiple ground truths per query (average of 4.53), which enhances annotation quality and reduces false negatives. Another earlier benchmark is Birds-to-Words \cite{forbes2019neural}, which contains 3,347 bird image pairs from iNaturalist, each accompanied by an average of 4.8 descriptive paragraphs detailing fine-grained differences. Though limited in size, it offers richer linguistic content per instance. MIT-States \cite{isola2015discovering} comprises 60K images labeled with 249 object nouns and 115 state adjectives, supporting adjective change scenarios (such as ``new camera", ``red tomato"), emphasizing the combination generalization ability.

In addition, representative benchmarks also include the fashion images domain. FashionIQ \cite{wu2021fashion} serves as a representative benchmark for composed fashion image retrieval. It comprises 77,684 images across three categories, i.e., dresses, shirts, and tops \& tees, including 60,272 triplets. In contrast, Fashion200k \cite{han2017automatic} contains over 200,000 product images sourced from online shopping platforms, with textual descriptions filtered and cleaned to retain 4,404 unique attributes for joint embedding. However, it lacks natural modification sentences, making it more suitable for attribute-based CMR. The Shoes dataset \cite{guo2018dialog} consists of 10,751 triplets from like.com, annotated with fine-grained relative descriptions.




Benchmarks in other domains, including videos~\cite{ventura2024covr, hummel2024egocvr, zhang2024localizing}, remote sensing images~\cite{psomas2024composed, wang2024scene}, person images~\cite{liu2025automaticsyntheticdatafinegrained}, sketch images~\cite{yu2016sketch, sangkloy2016sketchy, chowdhury2022fs, gao2020sketchycoco, hendrycks2021many}, multi-turn conversations~\cite{yuan2021conversational, pal2023fashionntm, wu2021deconfounded}, etc., can be referred to Sec.~\ref{APPL}.
Although the above datasets provide a feasible way to construct triplet data, the manual annotation process is time-consuming and laborious, which also leads to some key issues, such as the small data volume, single data domain, relatively simple text description, uncertainty, and noise in annotations. These problems restrict the further development of this field. Subsequent work has carried out a new paradigm (see Sec.~\ref{SSL-CMR}) for automatic data generation to alleviate this problem.

\subsection{Model Architecture }

For input reference images and modification texts, the core challenge of composed multi-modal retrieval lies in how to combine the two modalities, i.e., to represent a composite semantics that is similar to the reference image but adjusted in detail according to the modification text. Existing works can be categorized into three main aspects: (1) The most common approach is to design feature combiners for visual and language modalities, fusing the two modalities to achieve modified multimodal semantic representation; (2) Plug-and-play components, which offer broad compatibility across different existing methods and hold significant generalization value; (3) Non-explicit combiner methods that take an alternative path, implementing composed multimodal retrieval through novel architectures, providing important inspiration.

\begin{figure*}
  \centering
\includegraphics[width=1.0\textwidth]{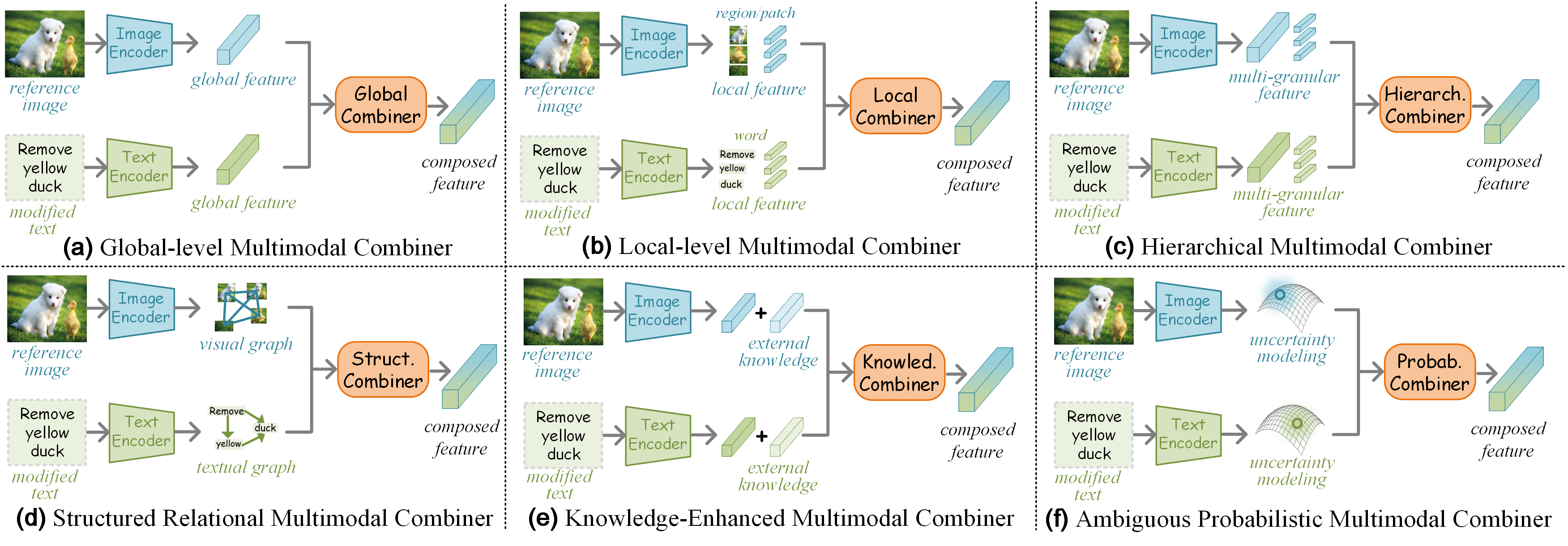}
  \caption{The existing multimodal combiner methods for CMR can be classified into 6 categories, progressing from simple to sophisticated structures.
  }
  \label{combiner}
\end{figure*}

\subsubsection{Design of Visual-Linguistic Combiner} \label{Combiner}
As illustrated in Fig.~\ref{combiner}, we classify existing combiner methods into the following 6 categories, ranging from simple to complex design structures: (1) Global-level combination based on coarse-grained modality features; (2) Local-level combination focusing on fine-grained modality features; (3) Hierarchical combination that incorporates multiple feature granularities; (4) Combination that models structured relationship information between modality samples; (5) Combination that leverages external knowledge for enhancement; (6) Combination that considers data polysemy by adopting ambiguity probability modeling.
Each combiner requires to solve two issues: the first is "where to see", i.e., to align the semantics between the modified text and the reference vision to find the content that needs to be modified, and the second is "how to change", i.e., to fuse multimodal information to modify the visual and language semantics, conveying more flexible retrieval intentions. Therefore, in the following, we will introduce the details of cross-modal alignment and fusion for each type of combiner.



\textbf{(1) Global-level Multimodal Combiner}: Global-level alignment occurs through interactions between image and text modalities exclusively at their global feature representations within a shared semantic space. Recent dual-stream models, including ViLBERT \cite{lu2019vilbert}, UniVL\cite{luo2020univl}, ERNIE-ViL \cite{ yu2021ernie}, have received more attention. They utilize separate visual and textual encoders to project images and texts into a shared space for semantic similarity measuring. The most representative work is CLIP~\cite{radford2021learning}, which has demonstrated remarkable capabilities. CLIP employs dedicated visual and textual encoders, learning semantic associations through contrastive optimization on 400 million image-text pairs.




In the CMR field, various methods have been proposed to effectively fuse global-level visual references and textual modifications into unified semantic representations, as illustrated in Fig.~\ref{combiner}(a). Early approaches focused on adaptive feature integration mechanisms. Vo et al.~\cite{vo2019composing} integrate gating mechanisms with residual connections, computing weighted combinations of original image representations and text-induced modifications to preserve visual content while enabling semantic adjustments. Dodds et al.~\cite{dodds2022training} employ residual attention fusion, deriving attention weights from textual features and applying them element-wise to image features for global semantic adaptation. Recent CLIP-based methods have gained prominence for their effectiveness. Baldrati et al.~\cite{baldrati2022effective,baldrati2022conditioned,baldrati2023composed} design Combiner networks that process concatenated CLIP features through linear layers and activations, later extending to normalized combinations including convex and learned mixtures. Lin et al.~\cite{lin2023clip} fine-tune CLIP encoders with fusion modules employing weighted summation, concatenation, and bilinear pooling to capture complex modification intents. In addition, advanced composition strategies explore specialized feature modeling approaches. Lee et al.~\cite{lee2021cosmo} propose CoSMo, disentangling textual features into content and style components that modulate spatial activation and distribution of image features through adaptive fusion. Anwaar et al.~\cite{anwaar2021compositional} develop complex-space transformations where text features apply rotational operations to image features under symmetry constraints. Xu et al.~\cite{xu2024align} introduce composition-decomposition frameworks with joint encoding and target image decoupling, employing attention mechanisms and bilinear pooling in closed-loop structures for enhanced feature interaction.

\textbf{(2) Local-level Multimodal Combiner}: Local-level feature alignment necessitates the capture of fine-grained visual-textual correspondences to enable precise cross-modal semantic measurement at the fragment level. Pioneering work, by Karpathy and Fei-Fei~\cite{karpathy2015deep}, first attempts to optimize the most similar region-word pairs for selecting matched semantics. The Stacked Cross Attention Network (SCAN)~\cite{lee2018stacked}, as a representative work, employs any fragment from one modality as a query to interact with all fragments from the complementary modality. 
SCAN inspires many variants, such as focal attention~\cite{BFAN, li2024improving}, iterative attention~\cite{IMRAM}, and negative-aware attention~\cite{zhang2022negative}. In recent years, pre-trained models based on the Transformer attention mechanism for fine-grained semantic alignment have also received great attention. This type of model integrates different modal information at an early stage, and because all modal information is interactively processed through only one branch, this type of model is also called a single-stream model. Representative works include Unicoder-VL~\cite{li2020unicoder}, ImageBERT~\cite{ qi2020imagebert}, UNITER~\cite{chen2020uniter}, BLIP2~\cite{li2023blip}, etc.

Current local feature-based CMR methods can be primarily categorized into two approaches: attention-based feature interaction methods \cite{hosseinzadeh2020composed, han2022fashionvil, zhang2021heterogeneous} that achieve fine-grained correspondence through cross-modal attention mechanisms; semantic enhancement and suppression methods \cite{pang2022heterogeneous, zhang2022composed, jandial2022sac, zhao2024neucore,wang2024negative} that focus on selective modulation of specific semantics via dynamic weights or kernels. While these approaches differ in their technical strategies, they all aim to improve the utilization of local features for more precise retrieval performance.
Attention-based feature interaction methods aim to construct fine-grained correspondences between visual and textual modalities by employing various forms of attention mechanisms. Hosseinzadeh and Wang \cite{hosseinzadeh2020composed} utilize a multi-layer bidirectional attention framework in which each layer computes both linguistically attended visual features and visually attended linguistic features, progressively enriching the joint representation. Han et al. \cite{han2022fashionvil} develop a dynamic, learnable attention module to establish region-word alignments and apply bilinear pooling to capture second-order interactions, thereby enhancing feature integration. Zhang et al. \cite{zhang2021heterogeneous} introduce Multi-modal Complementary Fusion and Cross-modal Guided Pooling, combining dual-attention-based projection of text into visual space, gated fusion, and modality-aware pooling to dynamically weight local features. These methods share the objective of improving cross-modal feature interaction through layered or dual-attention structures and enhanced fusion mechanisms.

Semantic enhancement and suppression methods, on the other hand, center around selectively enhancing or suppressing specific semantic features of the image based on textual information. 
Pang et al. \cite{pang2022heterogeneous} integrate relevant image features into an expanded text representation via a multi-layer perceptron and attention, enhancing understanding of short or incomplete queries. Zhang et al. \cite{zhang2022composed} suppress irrelevant image details while enhancing relevant ones through dynamic semantic weighting to align images with text better. Jandial et al. \cite{jandial2022sac} first detect salient image regions linked to text, then adjust these while preserving other areas, using LSTM and normalization for feature fusion. 
These methods are united by their focus on the selective enhancement or suppression of semantic features, using dynamic weights or kernels to modulate feature importance while maintaining the integrity of unmodified areas. Extending this line of research,
Zhao et al. \cite{zhao2024neucore} propose a concept-level visual-semantic alignment framework by concatenating feature tokens from both reference and target images and employing a Transformer for joint context modeling. An attention-based multiple-instance learning module then calculates alignment scores between visual concept representations and textual semantic embeddings, guiding the fusion process to ensure concept-level consistency in retrieval tasks. Wang et al. \cite{wang2024negative} introduce a threshold optimization strategy to refine the identification of positive and negative local feature correspondences. By dynamically adjusting decision thresholds, the model improves its capacity to discriminate between visually similar but semantically distinct instances, thereby reducing retrieval errors.

\textbf{(3) Hierarchical Multimodal Combiner}: This design is driven by the need to capture multi-granular visual and textual relationships, enabling a more comprehensive fusion of information across different levels of features. 
The alignment of semantic features of different granularities in visual language has been studied for a long time~\cite{dong2022hierarchical,ma2021hierarchical,zhu2025enhancing}.
Early work proposed to model the semantic alignment of global and local features of vision and language~\cite{wu2022global,wehrmann2020adaptive}. On this basis, to progressively mine the semantic alignment relationship between modalities, Chen et al.~\cite{IMRAM} proposed a serialized multi-level alignment model with iterative operation, Hu et al.~\cite{hu2019multi} employ a multi-layer CNN to capture the local correlations and
long-term dependencies between vision and language, and Ji et al.~\cite{SHAN} proposed a local-global cross-interaction alignment model.

Existing hierarchical multimodal combiners can be categorized into two primary types based on how hierarchical features are generated: encoder multi-layer strategies \cite{chen2020image, dodds2020modality, jandial2020trace, yang2023composed, zhang2023enhance, xu2021hierarchical, li2023multi, huang2023language} and global-local feature strategies \cite{wen2021comprehensive, liu2021image, goenka2022fashionvlp, wen2023self, zhang2022composed}.
The first type of approaches utilizes features extracted from different layers of encoders, such as CNNs or Transformers, allowing for a more detailed representation of image-text interactions.
Several works \cite{chen2020image, dodds2020modality, jandial2020trace, yang2023composed} utilize features from different CNN or Transformer layers, which are fused with textual features using attention mechanisms. For example, Chen et al. \cite{chen2020image} apply both self-attention and joint-attention to capture intra-modal and cross-modal relations. Similarly, Dodds et al. \cite{dodds2020modality} treat intermediate CNN features as visual tokens and align them with text via transformer-based attention, facilitating multi-granular fusion.
Others \cite{jandial2020trace, yang2023composed} explicitly construct multi-level visual pyramids aligned with text embeddings, which are then aggregated through specialized modules such as semantic feature transformation or hierarchical transformers. These designs support alignment across visual scales and improve visiolinguistic composition. A subset of methods \cite{zhang2023enhance, xu2021hierarchical, li2023multi, huang2023language} extend the above designs by incorporating spatial localization mechanisms. For instance, Zhang et al. \cite{zhang2023enhance} perform multi-layer localization alignment, while Xu et al. \cite{xu2021hierarchical} combine hierarchical vision features with global-local spatial alignment. Li et al. \cite{li2023multi} construct multi-grained representations for focused region modification, and Huang et al. \cite{huang2023language} employ language-guided masks to direct local feature adjustment. These approaches aim to better ground textual modifications in corresponding image regions.

The second type emphasizes a hierarchical approach to feature extraction, where local, regional, and global information is progressively integrated, ensuring a rich representation for multimodal retrieval tasks. 
Several methods \cite{wen2021comprehensive, wen2023self} employ distinct local and global composition modules, where fine-grained attribute alignment is combined with holistic visual-textual fusion, often facilitated by mutual learning components. Liu et al. \cite{liu2021image} utilize pre-trained vision-and-language models to encode local text and global image features, followed by transformer-based attention fusion, enabling abstract and broad semantic understanding.
In more specialized domains, models such as \cite{goenka2022fashionvlp} extract visual features at multiple spatial levels, including full images, object-centric crops, and region-of-interest segments, which are fused with textual feedback through vision-language transformers. Hybrid architectures \cite{zhang2022composed} integrate convolutional and transformer networks to jointly capture local textures and global scene structure, followed by feature fusion via residual or gating mechanisms for robust cross-modal matching.

\textbf{(4) Structured Relational Multimodal Combiner: }
Recent methods have significantly advanced composed multimodal retrieval by explicitly capturing and utilizing structural dependencies between visual and textual inputs. 
Leveraging structural dependencies can promote more accurate cross-modal semantic alignment. 
Early work introduced visual and linguistic scene graphs to enable context-aware alignment~\cite{wang2020cross, li2020visual, lee2019learning, liu2024semscene, pei2023scene}, learning modality-specific relations independently. A representative method, SGM by Wang et al.~\cite{wang2020cross}, designs visual and textual scene graph encoders that enrich node representations through neighborhood aggregation, thereby extracting cross-modal features at object- and relation-level for alignment interaction.
Moving beyond reliance on predefined scene graph extractors, the adoption of graph convolutional networks (GCNs)~\cite{kipf2016semi, velickovic2017graph} for non-Euclidean spatial data has enabled more flexible graph-based alignment strategies~\cite{GSMN, shi2022decoupled, long2022gradual, SGRAF, zhang2022show, ge2021structured, liu2022learning, wang2021wasserstein}. Liu et al.~\cite{GSMN} propose to model the graph structures of images and texts respectively, and use GCNs to construct fine-grained phrase-level alignment. Inspired by this, Diao et al. and Zhang et al. enhanced graph structure alignment from the perspectives of filtering and confidence~\cite{SGRAF,zhang2022show}, respectively, while works~\cite{wen2020learning, guo2023hgan} propose hierarchical graph structure interactions. In contrast to separate modeling modality graphs, multimodal graph structures~\cite{fu2023learning, li2024fast, zheng2024multimodal} that jointly represent both visual and textual elements have also been proposed, offering a unified framework to strengthen cross-modal interaction and fusion.

%


The structured relational multimodal combiner collectively emphasizes the importance of modeling hierarchical, complementary, and context-dependent relationships.
A line of work focuses on modeling visual and textual graphs~\cite{ding2023fine}. Jiang et al. \cite{jiang2024cala} introduce a dual-stream architecture with cross-attention to mine and align complementary information between modalities. Further emphasizing context-dependent relationships, Chen et al. \cite{chen2024spirit} incorporate structured style modeling to guide intra-image patch interactions in a text-conditioned manner, refining patch-level relationships.
Xu et al. \cite{xu2021hierarchical} propose a hierarchical composition framework that progressively models visual graphs
with entity, attribute, and relationship nodes, facilitating the gradual fusion of complex visual-textual associations. 
Li et al. \cite{shin2021rtic} enhance the flow of cross-modal information through residual learning with GCNs, effectively preserving and propagating modality-specific differences through structured residual connections. 
Zhang et al. \cite{zhang2022comprehensive} employ dynamic graph construction to model detailed semantic relationships between multimodal elements, supporting more flexible and context-sensitive retrieval.

Another line of work proposes to model the unified multimodal graph. 
Zhang et al.~\cite{zhang2020joint} construct a unified relational graph, where nodes represent visual elements enriched with textual attributes, and edges capture relationships among these attributes.
Nie et al.~\cite{nie2021conversational} propose a multimodal hierarchical graph neural network to effectively model conversational structures and multimodal contexts. They employ a graph attention network to dynamically aggregate information from different modalities for accurate user intent understanding.
Zhang et al.~\cite{zhang2021geometry} incorporate textual semantics into visual nodes, enabling each node to embed both visual features and corresponding textual information for enhanced multimodal representation.

\textbf{(5) Knowledge-Enhanced Multimodal Combiner: } 
By incorporating external knowledge, these approaches aim to enrich the semantic representation space, reinforce cross-modal alignment. Three types of knowledge are usually introduced: geometric information, dependency relations, and common sense knowledge.
Regional geometric information~\cite{song2021direction}, obtained by object detectors~\cite{ren2015faster,zhu2020deformable}, is typically used to enhance visual semantic features. Among them, Wang et al.~\cite{PFAN} propose an attention network based on position features, Liu et al. and Zhang et al. respectively propose geometric graph connection methods based on the relative positions of image regions~\cite{GSMN, zhang2022show}.
Dependency relations are typically obtained by off-the-shelf toolkits such as Spice~\cite{anderson2016spice}, Stanford CoreNLP~\cite{manning2014stanford}, or visual scene graph generation tools such as MSDN~\cite{li2017scene}  and NeuralMotifs~\cite{zellers2018neural}. Introducing structured dependency knowledge can significantly improve the alignment accuracy~\cite{guo2024visual}, as discussed in Sect.~\ref{Combiner}(4). 
Common-sense knowledge, including scene co-occurrence~\cite{shi2019knowledge}, corpus co-occurrence~\cite{CVSE}, and action information~\cite{li2022action}, is used to further enhance the accuracy of visual-language alignment.




Introducing external knowledge also improves the model’s capacity to interpret and reason over complex compositional queries. Zhang et al.~\cite{zhang2021geometry} propose a geometry-sensitive cross-modal reasoning network that integrates visual-semantic and spatial structural information through an inter-modal attention mechanism and a visual reasoning module guided by text. This method allows for explicit modeling of spatial relationships while enabling semantic extrapolation beyond the visual content of reference images. Zhang et al. ~\cite{zhang2024multimodal} design a compositional example mining strategy that constructs challenging negative samples by pairing mismatched image-text pairs and augmenting them with compositional variants. This technique enhances the model’s ability to learn discriminative representations by refining the decision boundary in the joint embedding space.
In a complementary direction, Bai et al.~\cite{bai2023sentence} propose a sentence-level prompting framework that decomposes textual descriptions into semantically meaningful constituents. These constituents are then used to generate structured prompts, which guide the model in locating corresponding visual content more precisely.

\textbf{(6) Ambiguous Probabilistic Multimodal Combiner: }
As shown in Fig.~\ref{combiner}(f), to address ambiguity and uncertainty problem, several approaches have been developed that incorporate probabilistic modeling. 
In vision-language alignment, a typical paradigm is to encode each input into a set of embeddings~\cite{PVSE,chun2021probabilistic,kim2023improving,10547196,zhang2025dh}. Song et al.~\cite{PVSE} propose to
regularize the learned embedding space by minimizing the
discrepancy using the maximum mean discrepancy.
Chun et al.~\cite{chun2021probabilistic} propose to use probabilistic embeddings to represent vision-language as probability distributions in a common embedding space.  Kim et al.~\cite{kim2023improving} propose a set prediction module and smooth-Chamfer similarity for set-based embeddings.
Zhang et al.~\cite{zhang2025dh} propose a novel set-embeddings-based method, which improves accuracy and efficiency from the perspective of dissecting key semantic dimensions within each subspace.


In composed multimodal retrieval, Xu et al.~\cite{xu2024set} introduced the Set of Diverse Queries with Uncertainty Regularization framework to address challenges arising from semantic polysemy and noisy supervision. This method extracts multiple deterministic embeddings using a vision-language encoder and encodes them into a distributional form to model uncertainty. A regularization module is employed to enable sampling-based probabilistic matching, thus establishing robust many-to-many correspondences between multimodal inputs and retrieval targets. Neculai et al.~\cite{neculai2022probabilistic} proposed the Multimodal Probabilistic Combiner, which extends the capacity of retrieval systems to handle an arbitrary number and types of query modalities.

\subsubsection{Design of Plug and Play Structure} 

Based on the key highlights of different approaches in achieving plug-and-play structures, we classify them into the following two categories: (1) modular design to enhance plug-and-play capability \cite{chen2020learning, kim2021dual, zhu2023amc, zhang2022comprehensive}; and (2) Training and inference strategies with plug-and-play characteristics \cite{liu2024bi, hou2024pseudo, pang2022heterogeneous, liucandidate}. 


Focusing on devising plug-and-play modules to enhance system flexibility and performance, Chen et al. \cite{chen2020learning} propose a Joint Visual Semantic Matching (JVSM) module that associates visual and textual information through a shared discriminative embedding space. Their four-module architecture, encompassing visual embedding, textual embedding, semantic projection, and composition modules, seamlessly integrates into existing retrieval systems, effectively handling image-text matching tasks, particularly excelling in complex fashion term retrieval. Meanwhile, Kim et al. \cite{kim2021dual} introduce a dual compositional module, which incorporates a correction network to complement traditional compositional networks by capturing differences between reference and target images. This dual mapping design not only generates composite features but also models image discrepancies, allowing it to be flexibly integrated into diverse retrieval systems. Yang et al. \cite{yang2021cross} propose a cross-modal joint prediction module, in which the modified text is treated as an implicit transformation between the query image and the target image. This module can be seamlessly integrated into existing methods to enhance the discriminability and robustness of visual and textual representations.

Besides, Zhu et al. \cite{zhu2023amc} present an adaptive multi-expert collaborative module, integrating multiple expert models focused on specific image regions or text segments, with an adaptive gating mechanism dynamically adjusting weights based on input. This approach facilitates more refined feature extraction and composition, improving the system's capability to handle complex and varied queries, demonstrating strong applicability as a plug-and-play component. Zhang et al. \cite{zhang2022comprehensive} develop a comprehensive relationship reasoning module, utilizing graph convolutional networks to analyze multiple relationships between images and texts, such as contextual and category relationships. This module improves the relevance and accuracy of retrieval results.

Focusing on plug-and-play design in training and inference strategies, Liu et al. \cite{liu2024bi} employ a bidirectional training strategy to optimize cross-modal interactions, improving model adaptability and overall retrieval accuracy, especially in few-shot scenarios. Hou et al. \cite{hou2024pseudo} construct pseudo-triplets using anchor samples to simulate semantic relationships and generate discriminative training instances, effectively alleviating data scarcity and demonstrating its potential as a lightweight plug-and-play component. Pang et al. \cite{pang2022heterogeneous} address heterogeneous feature fusion and cross-modal alignment by integrating multiple visual attributes via an MLP and applying contrastive loss to unify cross-modal semantics, enhancing the model's ability to understand complex compositional queries while maintaining modularity. Liu et al. \cite{liucandidate} present a dual multi-modal encoder for candidate set re-ranking in the inference stage, improving relevance and ranking quality with strong integration capabilities into retrieval systems.

\subsubsection{Design of Non-explicit Combiner Architecture}

Non-combinatorial structural design refers to methods that avoid directly fusing different modalities of data (such as images and texts) when processing multimodal data. Instead, these approaches achieve cross-modal information fusion and utilization through indirect means, aiming to circumvent the complexity and limitations that traditional combinatorial methods may introduce. 

Specifically, non-combinatorial structural design attempts to preserve the uniqueness of each modality's data through various strategies, and on this basis, conducts information interaction, enhancement, and complementation to achieve more effective information retrieval. Zhang et al.\cite{zhang2022tell} propose the Tell-Imagine-Search framework, which consists of three modules. The Tell module generates a detailed text description based on the input text and reference image; the Imagine module synthesizes an imaginary image according to this description; and the Search module uses the synthesized image to search for the most similar real image in the database. Subsequently, Delmas et al.\cite{delmas2022artemis} propose a novel approach to split the combined image retrieval task into two independent but complementary subtasks: explicit matching and implicit similarity. Explicit matching uses an attention mechanism to focus on processing visual elements explicitly described in the query text. Implicit similarity considers broader contextual similarities between images.



In addition, Yang et al.\cite{yang2023decompose} propose the decompose semantic shifts (FSS) method, which identifies key query elements that may cause semantic shifts. FSS explicitly decomposes the semantic change into two steps: from reference image to visual prototype and then from visual prototype to target image. This approach not only preserves key visual cues but also enriches the visual prototype with textual guidance. 
Wen et al.\cite{wen2024simple} transfer multimodal fusion to the raw data level, leveraging vision-language pre-training models to achieve multimodal encoding and cross-modal alignment. This framework resolves the embedding space shift problem that may arise from feature-level fusion, further validating the effectiveness of non-combinatorial structural design.

\subsection{Loss Optimization}

For the combined features of reference images and modified text, achieving constraints between these and the target images is a critical issue in the field. The loss optimization designs in existing methods mainly focus on three aspects: (1) fundamental contrastive learning constraints; (2) additional consistency constraints to refine details; and (3) joint optimization constraints that integrate multi-task learning.

\subsubsection{Based on Contrastive Learning}
The two most common contrastive learning losses are batch-based classification loss and soft triplet-based loss, both variants of Softmax Cross-Entropy Loss \cite{vo2019composing}. 
First, let us define some key variables. Assume we have a mini-batch of training triplet data containing $B$ queries, where each query composed of the reference image and modification text is denoted as $\psi_i$ and the target image is denoted as $\phi^+_i$. The negative samples can be randomly selected by $K$-1 target images of other queries from the same mini-batch, denoted as $(\phi^-_1, \phi^-_2, \ldots, \phi^-_{K-1} )$.
These negative and positive samples form a set $N_i$ for computing the loss function.
Below, we introduce these two losses and their relationships.

\textbf{Softmax Cross-Entropy Loss} \cite{vo2019composing}: The goal is to pull the "modified" query features closer to the target image features while pushing away dissimilar image features. Specifically, the loss function is formulated as:
\begin{equation} \label{E4}
 L = \frac{1}{MB} \sum_{i=1}^{B} \sum_{j=1}^{M} -\log \frac{\exp\{\kappa(\psi_i, \phi^+_i)\}}{{ \sum_{\phi_{i,j} \in N_i} \exp\{\kappa(\psi_i, \phi_{i,j})\} }},   
\end{equation}
where $\kappa$ is a similarity kernel function (e.g., the dot product or negative L2 distance). The maximum value of $M$ is $\binom{B}{K}$, but it is often selected as a smaller value for tractability. This loss ensures similar samples are close in the embedding space, while dissimilar ones are pushed apart.

\textbf{Batch-based Classification Loss} \cite{gidaris2018dynamic,goldberger2004neighbourhood,yang2023decompose,zhu2023amc}: When the number of negative samples $K$ equals the batch size $B$, the above loss simplifies to:
\begin{equation} \label{E5}
L = \frac{1}{B} \sum_{i=1}^{B} - \log \frac{\exp\{\kappa(\psi_i, \phi^+_i)\}}{{\sum_{j=1}^{B} \exp\{\kappa(\psi_i, \phi_{i,j})\} }}.
\end{equation}
In this setting, each query is compared against all other samples in the mini-batch, making the loss function more discriminative and accelerating convergence. However, it also increases the risk of overfitting.

\textbf{Soft Triplet-based Loss} \cite{liu2021image,hosseinzadeh2020composed,yang2021cross}: In the case of using the minimum value $K=2$,
each query has only one negative sample, the loss simplifies to:
\begin{equation} \label{E6}
L = \frac{1}{MB} \sum_{i=1}^{B} \sum_{j=1}^{M} \log{( 1 + \exp{( \kappa(\psi_i, \phi^-_{i,j}) - \kappa(\psi_i, \phi^+_i) )} )}.
\end{equation}
This form relaxes the strict distance constraints between positive and negative samples, making the model more stable and flexible for complex data distributions.
As demonstrated in \cite{vo2019composing,hosseinzadeh2020composed}, selecting an appropriate loss function allows the model to achieve better performance across different datasets.

\textbf{Triplet Ranking Loss} \cite{chen2020learning,chen2020image,zhang2021geometry}: Unlike soft triplet loss, which relaxes the distance constraint with a logarithmic function, triplet ranking loss uses a hard margin constraint. 
It constructs a triplet of an anchor (e.g., a multimodal composed query), a positive(e.g., a target image), and a negative sample (e.g., an unrelated image) to optimize the embedding space, ensuring similar samples are close and dissimilar ones are pushed apart.
The triplet loss function is formulated as:
\begin{equation} \label{E7}
L =  [\kappa(\psi_i, \phi^+_i) - \kappa(\psi_i, \phi^-_{i,j}) + \gamma ]_{+},
\end{equation}
where $[\cdot]_{+} = \max(0,\cdot)$, and $\gamma$ is a predefined margin. The loss occurs only when the distance to the positive sample exceeds that to the negative sample plus the margin. Soft triplet loss is more flexible and stable in complex data, while triplet ranking loss enforces stricter separation, making it suitable for clear sample distinctions. Though it converges faster, it may become unstable with complex or noisy data.

\textbf{Bi-directional Triplet Ranking Loss} \cite{chen2020learning}: Different from the conventional triplet loss which only constrains the positive combination $\psi^{+}_i$, it introduces the negative combination $\psi^{-}_i$, i.e., the reference image and modified text that are irrelevant to the target image, which can be expressed as:
\begin{equation} \label{E8}
\begin{aligned} L = &[\kappa(\psi^{+}_i, \phi^+_i) - \kappa(\psi^{-}_i, \phi^+_i) + \gamma]_+ + \\ & +[\kappa(\psi^{+}_i, \phi^+_i) - \kappa(\psi^{+}_i, \phi^-_{i,j}) + \gamma]_+.\end{aligned}
\end{equation}
Additionally, they also introduce a text-matching loss to optimize the alignment between the positive text $t^+_{i}$, i.e., the caption of the target image,  and the negative text $t^-_{i,j}$, i.e., the caption of the irrelevant target image. Specifically, the loss function is formulated as:
\begin{equation} \label{E9}
\begin{aligned} L = &[\kappa(\psi^{+}_i, t^+_{i}) - \kappa(\psi^{-}_i, t^+_{i}) + \gamma]_+ \\ & + [\kappa(\psi^{+}_i, t^+_{i}) - \kappa(\psi^{+}_i, t^-_{i,j}) + \gamma]_+.\end{aligned}
\end{equation}

The famous margin-based triplet ranking loss has many advanced variants in cross-modal retrieval. Wang et al.~\cite{wang2016learning} consider the external constraint loss that preserves the neighborhood structure in a single modality. There are also variants such as the ladder loss~\cite{zhou2020ladder}, the polynomial loss~\cite{wei2020universal}, and the adaptive offline quintuplet loss~\cite{chen2020adaptive}. 
In addition to designing the loss function, existing work also focuses on the negative sample mining strategies and the similarity metric function.


\textbf{Negative Sample Mining}: As a fundamental role in training, there are two common strategies: mini-batch hardest negative mining \cite{zhang2021geometry,schroff2015facenet,faghri2017vse} and mini-batch semi-hard negative mining \cite{chen2020learning,chen2020image}. The former only selects negative examples with the largest similarity to the combined query features, while the latter selects negative examples with a higher similarity to the combined query than the target image. It is validated that semi-hard mining is more stable and faster converging than the hardest mining~\cite{chen2020learning}.

Unlike simple selection, Feng et al.~\cite{feng2024improving} propose a method to expand positive and negative examples through MLLM and design a two-stage fine-tuning framework to optimize the representation space.
Zhang et al. \cite{zhang2024multimodal} propose a method to enhance multimodal fusion by replacing parts of the query's text to generate harder negative samples, improving model learning. They introduce two techniques for mining hard examples: (1) synthetic composition examples, which create negative samples by replacing the text while keeping the image part unchanged, forming a new sample as \( f = r + t' \), where \( t' \) is from another query. (2) augmented synthetic composition examples, which replace only parts of the text, generating even harder-to-distinguish negatives. A mask vector \( s \) randomly determines which dimensions of the text to retain or replace, with a Bernoulli distribution governing the decision. The new text embedding is \( \tilde{t}' = s \odot t + (1 - s) \odot t' \). They also introduce a dynamic replacement ratio, sampling \( p \) from a Beta distribution for more diverse hard examples.

To optimize the use of these hard examples, they propose an improved loss function:
\begin{equation} \label{E10}
L = \frac{1}{B} \sum_{i=1}^{B} -\log  \frac{\exp(\kappa(\psi_i, \phi^+_i))}{\exp(\kappa(\psi_i, \phi^+_i)) + \sum_{j=1}^{B-1} \exp(\kappa(\phi^+_i, f_j))}.
\end{equation}
This formula ensures that the query \( \psi_i \) has high similarity with the target image \( \phi^+_i \) and low similarity with other synthetic composed negative samples \(f_j \),  enhancing the model's discriminative power.

\textbf{Similarity Metric Function}: Conventional contrastive learning losses typically employ Euclidean distance or cosine similarity as metrics for measuring the distance between samples \cite{vo2019composing}. However, these methods consider only the absolute distance or angular relationship between samples, neglecting the combined effect of both factors. To address this limitation, Zhang et al. \cite{zhang2022composed} propose the Triangle Area (TA) as a novel metric for sample distance measurement. Specifically, as illustrated in Figure A, for an anchor sample \(a\), a positive sample \(c\), and a negative sample \(b\), the distances between the samples are represented by the areas of triangles \(Oab\) and \(Oac\), where \(O\) denotes the origin. TA is defined as:
\begin{equation} \label{E11}
\text{TA}(\psi_i, \varphi_i) = \frac{1}{2}|\psi_i||\phi_i| \sqrt{1 - \left( \frac{\psi_i \cdot \phi_i}{|\psi_i||\phi_i|} \right)^2}. 
\end{equation}


TA has two key benefits: (1) It considers both the distance and angle between samples, letting the model optimize both without manual weight tuning. (2) For small datasets, it uses squared distance to slow convergence and prevent overfitting; for large datasets, it uses area-based distance to avoid exaggerating differences, leading to faster training.

Beyond TA, researchers have also proposed many novel similarity metric functions in cross-modal alignment, including the feature block weighting similarity \cite{GSMN}, the cross-modal feature mapping similarity \cite{SGRAF}, the cross-modal feature dimensional dependency modeling similarity \cite{zhang2023enhanced}, the conditional similarity based on feature decoupling \cite{veit2017conditional}, the dimensionally interpretable similarity \cite{chen2023stair}, and the learnable similarity via enhanced measure-units~\cite{zhang2023unlocking}.


\subsubsection{Consistency Constraint}

To ensure that visual and linguistic information can be effectively represented and mutually complementary within a shared space, researchers have proposed methods based on consistency constraints. These methods aim to capture the complementary information between images and text, enabling them to jointly express query intentions and accurately reflect user modification intentions, thereby generating more relevant retrieval results.

Works \cite{jandial2020trace,jandial2022sac,anwaar2021compositional} focus on consistency loss for image and text reconstruction, aiming to construct a robust and meaningful joint representation space. Within this space, the model attempts to reconstruct the original input images and texts as faithfully as possible, learning their associations and semantic relationships to improve retrieval accuracy. Additionally, works \cite{wen2021comprehensive} and \cite{yang2023composed} delve into the reconstruction consistency between global and local composite features. For instance, Wen et al.\cite{wen2021comprehensive} utilize fine-grained local and global combination modules to achieve multimodal integration and introduce an inter-enhancement module to promote mutual improvement between local and global features. It enhances knowledge transfer across different feature levels using KL divergence loss and feature-level $L_2$-norm loss to optimize the consistency between local and global features. This approach not only improves the expressiveness of local features but also reinforces the synergy between the global and local combination modules.
Pang et al. \cite{pang2022heterogeneous} and Zhang et al. \cite{zhang2021heterogeneous} propose the Relative Caption-aware Consistency (RCC) loss to bridge the semantic gap between image and text modalities. RCC guides cross-modal alignment by minimizing the divergence between generated and ground-truth modification texts.

Several studies \cite{zhang2024collaborative,wen2023target,yang2023decompose,zhu2023amc} introduce consistency constraints across multi-branch or multi-expert networks to enhance model robustness and generalization. Zhang et al. \cite{zhang2024collaborative} propose a consensus network with diverse combiners generating complementary image-text embeddings, regularized by KL divergence to encourage agreement and reduce annotation noise sensitivity. Wen et al. \cite{wen2023target} employ a teacher-student framework, where the student mimics the teacher’s target-query reasoning, guided by KL consistency to improve multimodal query alignment. Yang et al. \cite{yang2023decompose} develop a Semantic Shift Network that decomposes text into upgradation and degradation steps, using KL constraints to align positive samples with targets and increase separation from negatives. Zhu et al. \cite{zhu2023amc} design an adaptive multi-expert network that enforces consistency among expert branches focusing on distinct visual features. Collectively, these approaches leverage consistency losses to promote mutual learning, reduce bias, and improve cross-modal representation.

\subsubsection{Multi-task Joint Loss}

These works aim to enhance the effectiveness of multimodal compositional learning through carefully designed multi-task joint loss functions. They typically incorporate some form of knowledge distillation, contrastive learning, or adversarial learning, along with task-specific loss functions such as classification loss or triplet ranking loss.

The FAME-ViL and FashionViL models by Han et al. \cite{han2023fame,han2022fashionvil} use distillation loss to transfer knowledge from single-task teacher models to multi-task student models, enhancing generalization and preventing negative transfer. FAME-ViL \cite{han2023fame} combines distillation with mutual information maximization, guiding learning with teacher model outputs and improving parameter efficiency. FashionViL \cite{han2022fashionvil} uses contrastive learning, classification loss, and distillation loss, combining these to boost task-specific performance and foster synergy across tasks, improving retrieval accuracy and robustness.

Some works introduce adversarial learning mechanisms, leveraging generative adversarial networks to capture finer-grained semantic information. The multi-order adversarial network by Fu et al. \cite{fu2021multi} introduces an adversarial learning mechanism. In addition to adversarial losses, MAN combines triplet ranking loss to enhance the retrieval network's performance, ensuring that the synthesized features better reflect the semantic information of text modifications. This comprehensive loss design effectively integrates adversarial learning with triplet loss, boosting the model's performance in multimodal compositional retrieval tasks.
Similarly, VISTA by Zhou et al. \cite{zhou2024vista} combines cross-modal contrastive learning and bidirectional generative adversarial networks, aligning semantic information across modalities through adversarial training, improving both feature consistency and generation quality.

\section{Zero-Shot Learning-based CMR} \label{ZSL-CMR}

Without relying on triplet annotations,  existing methods focus on simulating composed retrieval based solely on image-text pairs or unlabeled images, establishing a zero-shot learning mechanism. Popular paradigms can be divided into three categories: (1) using images as input, designing a learning framework for visual-to-language transformation to combine modified text as query to search target images at test time; (2) using text as input, simulating the visual-to-language transformation process, and replacing text input with visual image at test time, which offers higher training efficiency; 
(3) using linear interpolation of visual and textual features.
The following sections will introduce progress in each of these areas.


\subsection{Image-side Transformation}
\begin{figure}
  \centering
\includegraphics[width=0.5\textwidth]{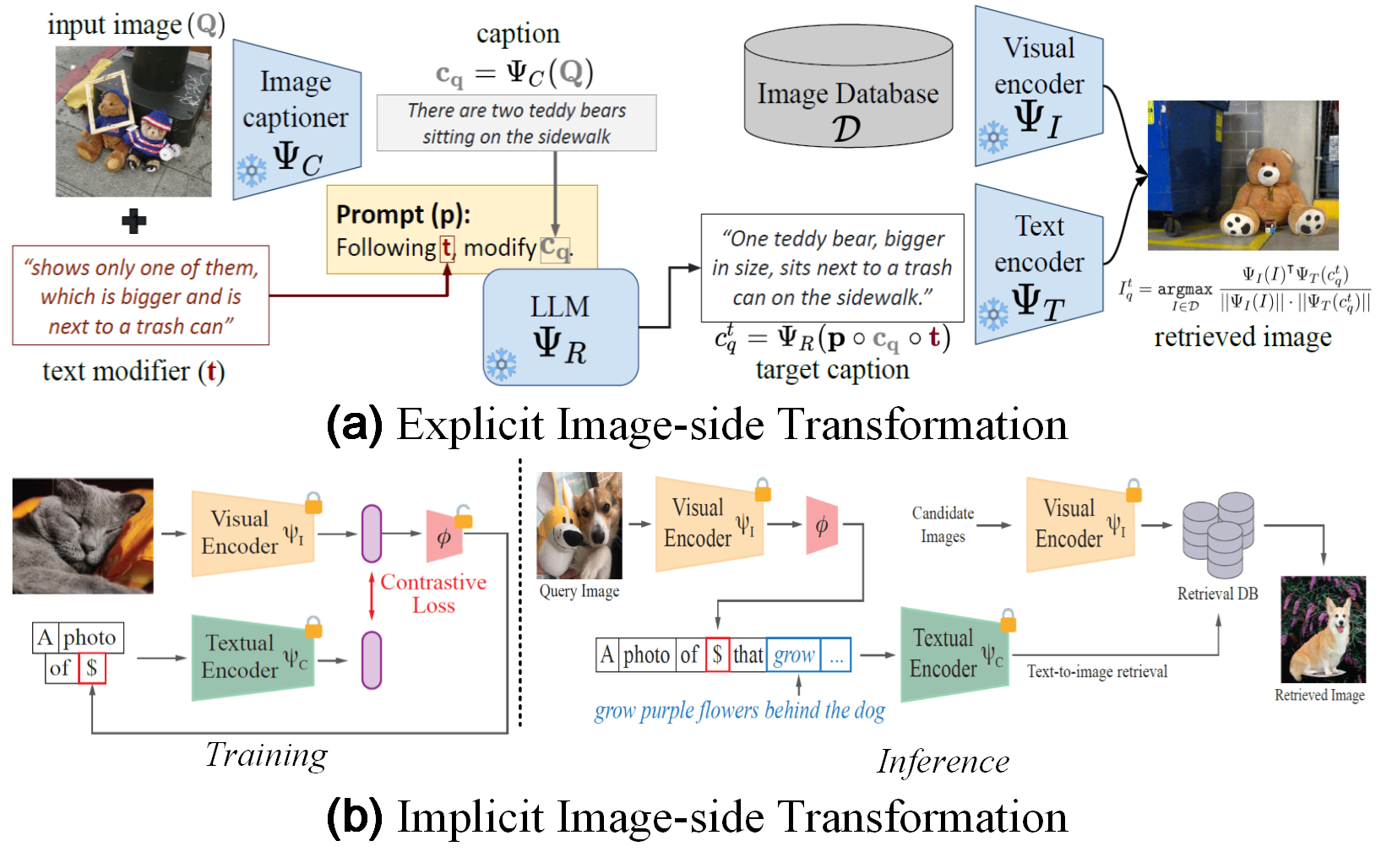}
  \caption{Explicit and  Implicit Visual Transformation.  Sub-figures (a) and (b) are from \cite{karthik2023vision} and \cite{gu2024language}, respectively. 
  }
  \label{ZS-CMR-1}
\end{figure}

In existing zero-shot composed multi-modal retrieval, as shown in Fig.~\ref{ZS-CMR-1}, the process of image-to-visual transformation can be categorized into explicit and implicit approaches. The explicit approach directly converts input images into textual descriptions, typically using pre-trained captioning models due to their training-free nature, but it is limited by the performance of these captioning models. The implicit approach, on the other hand, transforms input images into implicit feature vectors, usually encoded in a word vector space. This method is learnable and more flexible, allowing for a more sophisticated design tailored to the specific characteristics of the task.

\subsubsection{Explicit Visual Transformation (Training-Free)}
Existing explicit visual transformation methods \cite{karthik2023vision, sun2023training, yang2024ldre,yang2024semantic} utilize pre-trained visual-language models (VLMs)  and large language models (LLMs) to achieve zero-shot CMR in a training-free manner.

Karthik et al. \cite{karthik2023vision} propose a CIReVL framework, which generates captions for reference images and uses LLMs to recompose these captions based on the textual modifications of the target. This approach avoids the need for additional adaptation resources, supports flexible model component replacements, and allows user-level interventions by refining captions post-hoc, achieving competitive results. Sun et al. \cite{sun2023training} introduce a two-stage ranking method to address modality gaps and ambiguous requirements from the reference image. The first stage converts composed image-text queries into text-only queries for global retrieval, while the second stage re-ranks top-K results by extracting and evaluating local attributes from the modified instructions. This method combines global and local information to outperform other training-free approaches on open-domain datasets. Yang et al. \cite{yang2024ldre} develop LDRE, which employs dense caption generation for reference images and uses LLMs for divergent reasoning to create a range of potential target image captions. An ensemble strategy integrates these candidates to better align with the query intent, effectively handling complex or ambiguous queries. 
Yang et al. \cite{yang2024semantic} propose SEIZE, which generates diverse edited captions for the reference image by combining multiple captions from a pre-trained captioning model with relative text through LLM-based reasoning, enabling training-free ZS-CMR.
Despite their differences in visual transformation strategies, all methods rely on pre-trained models and language-driven reasoning to bridge the gap between modalities.

\subsubsection{Implicit Visual Transformation}
Existing works on implicit visual transformation for zero-shot CMR can be categorized into three main groups: single pseudo-word methods \cite{saito2023pic2word, baldrati2023zero, agnolucci2024isearle, tang2024context, tang2025missing, wang2025mapping,tang2024denoise}, multiple pseudo-word methods \cite{lin2024fine, du2024image2sentence,suo2024knowledge, chen2025data}.


Single pseudo-word methods focus on mapping an image into a single pseudo-word token for textual representation. Saito et al. \cite{saito2023pic2word} propose Pic2Word, which maps CLIP visual features to a single pseudo-word token via a cycle contrastive loss, showing the powerful generalization across various ZS-CMR tasks using only image-text data. 
Similarly, Baldrati et al. \cite{baldrati2023zero} introduce SEARLE, an optimization-based textual inversion to generate pseudo-word tokens and use knowledge distillation to train a Textual Inversion Network. SEARLE achieves efficient inference and improves performance. 
Expanding SEARLE, the iSEARLE proposed by Agnolucci et al. \cite{agnolucci2024isearle} further improves the retrieval accuracy by improving robustness with Gaussian noise, adding regularization to ensure dense token representations, and incorporating hard negative sampling to capture fine-grained details. Then, Context-I2W \cite{tang2024context} employs a context-dependent mapping strategy, combining an Intent View Selector for dynamic view selection and a Visual Target Extractor for contextually relevant pseudo-word tokens generation, achieving significant performance gains.
On this basis, Tang et al. \cite{tang2024denoise} further propose Denoise-I2W, which integrates a learnable attention-based denoising module to suppress intention-irrelevant visual regions during image-to-word mapping, and leverages contrastive learning to enhance semantic consistency, achieving notable performance gains on multiple CIR benchmarks.
Wang et al. \cite{wang2025mapping} propose a framework, TSCIR, that generates pseudo-word tokens by modeling complex semantic structures such as attribute-object and action-target pairs through a structured text encoder and semantic alignment loss, improving performance in fine-grained ZS-CMR tasks.
Lastly, Tang et al. \cite{tang2025missing} propose MTIP-WM, which predicts missing target-relevant semantics using a world model trained on pseudo-triplets, generating enriched pseudo-word tokens that capture both observed and inferred attributes, and then combines these tokens with manipulation text for robust contrastive learning.

Multiple pseudo-word methods extend representation by mapping images to multiple pseudo-word tokens to better capture fine-grained details. Lin et al.  \cite{lin2024fine} introduce a fine-grained textual inversion network, FTI4CIR, where an image is represented by a subject-oriented pseudo-word token and several attribute-oriented pseudo-word tokens, aligned with real-word tokens using semantic regularization. This approach enhances the expressiveness of image content. Similarly, Du et al. \cite{du2024image2sentence} propose an asymmetric ZS-CMR framework ISA, which employs an adaptive token learner to map images into sentence-like representations with discriminative visual information, and combines global contrastive distillation and local alignment regularization to improve retrieval accuracy and efficiency, particularly in resource-constrained environments.
Suo et al. \cite{suo2024knowledge} introduce the Knowledge-Enhanced Dual-stream ZS-CMR (KEDs) framework to address the limitation of existing methods that overlook fine-grained attribute information. KEDs enhances pseudo-word tokens by integrating external knowledge from a database, which captures detailed attributes like color, object count, and layout. This approach leverages external knowledge to significantly improve the accuracy and domain-specific performance of ZS-CMR.
Chen et al. \cite{chen2025data} introduce a Data-efficient Generalization (DeG) framework for ZS-CMR, addressing modality and distribution shift challenges. The framework comprises two key components: the Textual Supplement (TS) module, which refines pseudo-word tokens by enhancing their semantic representation, and the Semantic Set (S-Set), which mitigates overfitting by leveraging the zero-shot capabilities of VLMs, improving the generalization.
Recently, Li et al. \cite{li2025rethinking} propose MOA, a object-aware pseudo-word learning framework that identifies query-relevant objects through a multi-object recognizer and a noun-guided filtering strategy, and then maps the screened objects into multiple pseudo-word tokens for precise ZS-CMsR. This simple yet effective design achieves competitive results on multiple benchmarks, significantly outperforming prior methods.

\subsection{Text-side Transformation}
The ZS-CMR methods for image-side transformation rely on pre-defined simple prompts (e.g., “a photo of [\$]”), which leads to a very homogeneous input to text encoders and cannot fully learn the subtleties of natural language variations. This limitation becomes more significant, especially when trying to scale up the model, as larger models require more diverse data to capture more complex semantic information.

\begin{figure}
  \centering
\includegraphics[width=0.5\textwidth]{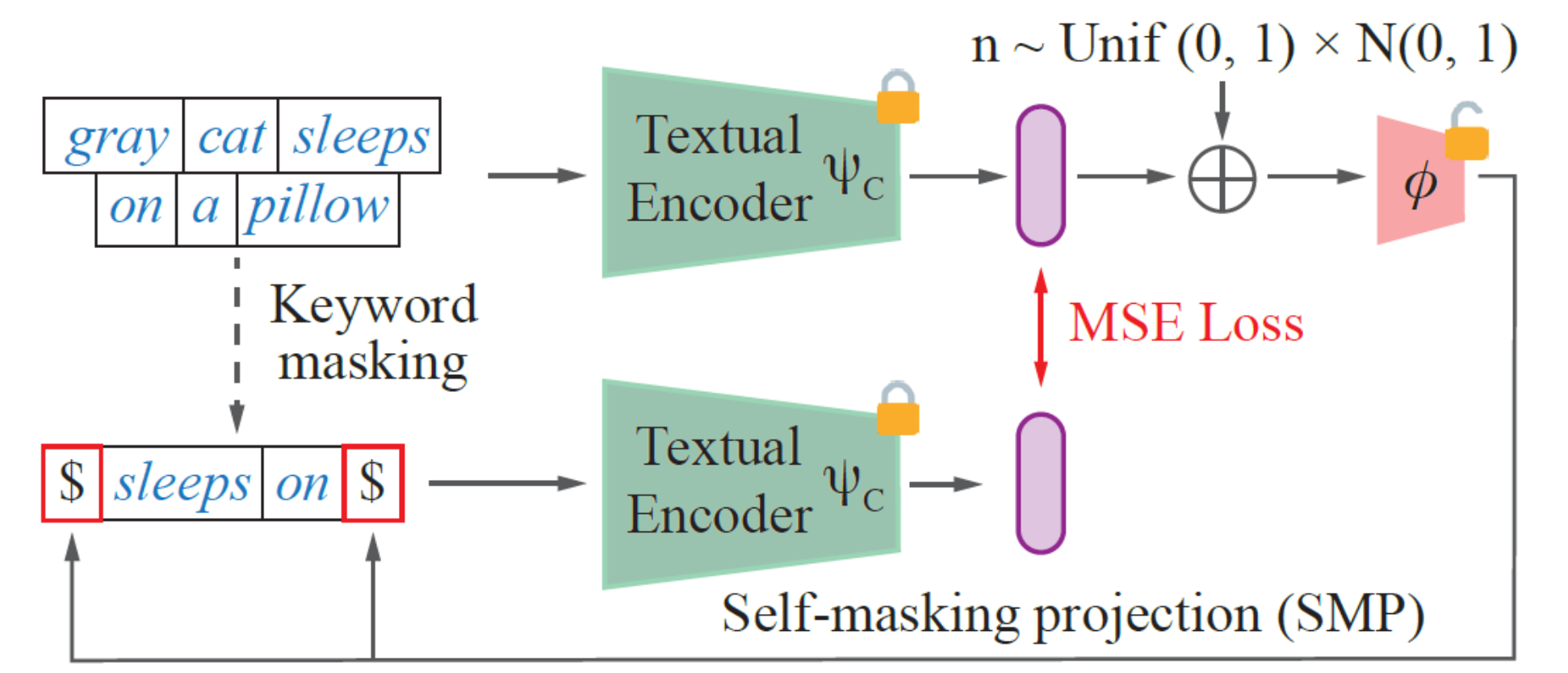}
  \caption{Text-side  Transformation.  Figure is from \cite{gu2024language}. 
  }
  \label{ZS-CMR-2}
\end{figure}

Recently, text-side transformations for CMR have been proposed to address the above limitations. 
As shown in Fig.~\ref{ZS-CMR-2}, the LinCIR proposed by Gu et al. \cite{gu2024language} leverages a self-supervised mechanism called self-masking projection to reduce dependency on triplet datasets and minimize modality gaps. This involves generating augmented texts through keyword masking and adding noise to textual embeddings, ensuring consistency between original and augmented representations. 
Building on this, Li et al. \cite{li2024motadual} propose MoTaDual, a multimodal-task dual alignment framework that uses LLMs (e.g., LLaMA, GPT-4) to generate diverse modification instructions and target texts, and applies prompt tuning to a lightweight textual inversion network, effectively addressing both modality and task discrepancies and achieving strong performance across four ZS-CMR benchmarks.
Other methods also propose text-side enhancement strategies. Byun et al. \cite{byun2024reducing} introduce RTD, a plug-and-play training scheme that enhances text encoders through target-anchored contrastive learning, refined batch sampling with hard negatives, and a concatenation scheme, effectively addressing task discrepancies without requiring additional fine-tuning data. 
Thawakar et al. \cite{thawakar2024composed} propose CoVR, which uses rich language descriptions to encode query-specific contexts and learns discriminative visual-textual embeddings for accurate video retrieval, achieving SOTA results. Li et al. \cite{liimproving} focus on multimodal compositional learning for LLMs, introducing tasks like Multimodal-Context Captioning and Retrieval to guide frozen language models in synthesizing multimodal information. 

\subsection{Linear Interpolation}
\begin{figure}
  \centering
\includegraphics[width=0.47\textwidth]{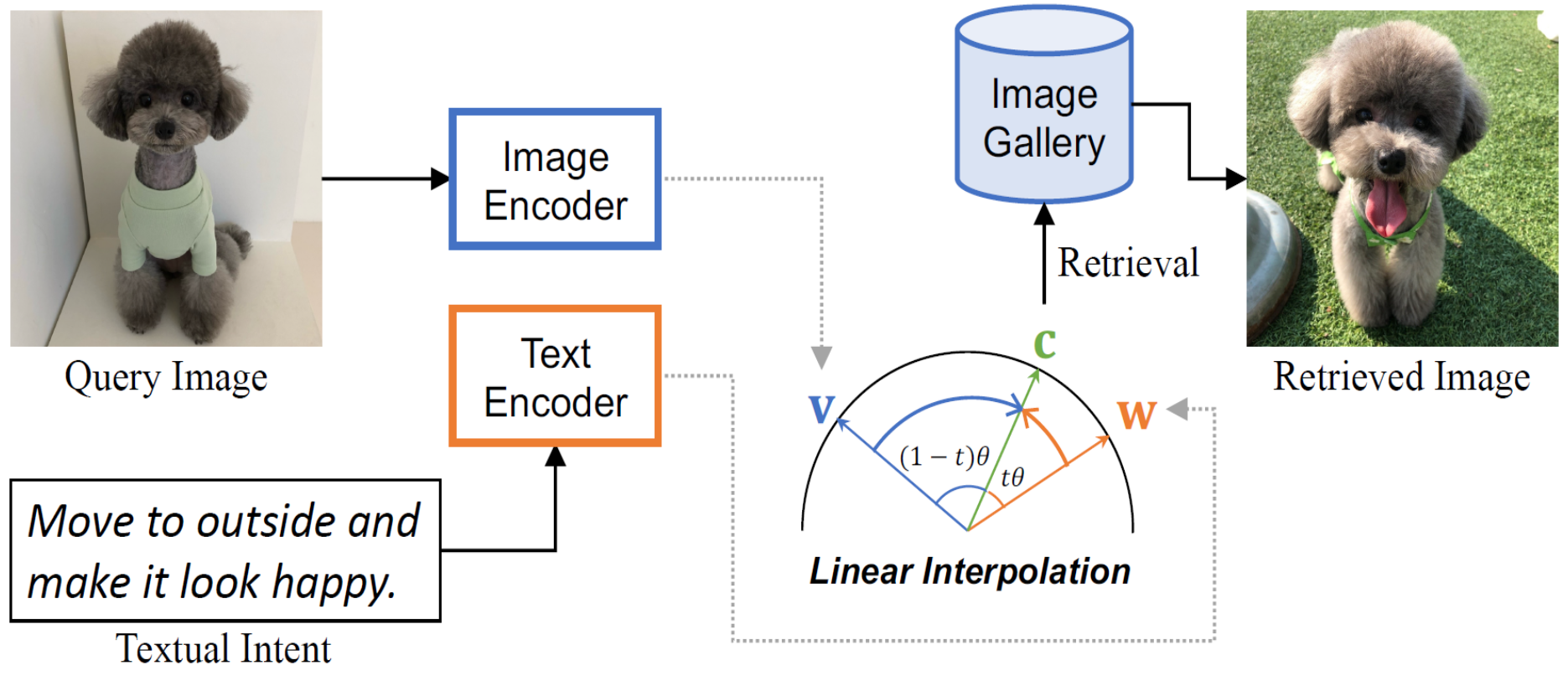}
  \caption{Linear Interpolation for ZS-CMR.  Figure is from \cite{jang2024spherical}. 
  }
  \label{ZS-CMR-3}
\end{figure}

As shown in Fig.~\ref{ZS-CMR-3}, Other approaches focus on improving the integration of image and text representations for ZS-CMR. Early work ALIGN proposed by Jia et al.~\cite{jia2021scaling} attempts to directly use the linear image+text features as multimodal queries to retrieve the intended images.
The Slerp proposed by Jang et al. \cite{jang2024spherical} employs Spherical Linear Interpolation for direct merging of image and text representations and Text-Anchored Tuning (TAT) to refine the image encoder, narrowing the modality gap and preserving image representation integrity. 
Jang et al. further \cite{jangtext} propose Slerp++, which introduces semantics-aware dynamic interpolation weights to adaptively control the fusion of image and text features in high-dimensional space, and combines it with text-anchored tuning to keep language representations stable, enabling more accurate composed embeddings for ZS-CMR.
Chen et al. \cite{chen2023pretrain} propose Masked Tuning to bridge the gap between pre-trained models and ZS-CMR tasks by masking image patches to generate triplets (masked image, text, image).
During inference, the model uses a linearly weighted combination of image and text features as a multimodal query, where the weight is a predefined mask rate to alleviate the distribution shift between masked images in pre-training and complete images in inference.

\section{Semi-Supervised Learning-based CMR} \label{SSL-CMR}





Early supervised works primarily relied on manual annotation to construct triplet datasets. 
While these efforts provided a foundation for dataset construction, the manual annotation process is time-consuming and labor-intensive, leading to several significant limitations:
(1) The data scale and domains are limited, constrained by the human effort required for annotation, impairing generalizability; 
(2) The modified text is frequently simplistic, lacking the complexity required for real-world applications; 
(3) Annotations may have noise and uncertainty, which becomes a bottleneck for further advancements in the CMR field.


To mitigate these issues, the semi-supervised learning-based CMR (SSL-CMR) paradigm is proposed, which generates pseudo-triplets via image and text generation techniques. These approaches aim to improve the quantity of triplet data while having high-quality triplet annotations by minimizing the introduction of noise and uncertainty. These advancements allow SSL-CMR models to perform well and achieve high accuracy, improving model robustness and offering new solutions to the data bottleneck in CMR tasks. 
In the following, we highlight related key works and their technical innovations.

\subsection{Automatic Data Construction}


Works in this research line can be divided into two major categories: (1) generating language content only~\cite{liu2023zero,hou2024pseudo,zhang2024multimodal,liu2024bi,jiang2024hycir,jang2024visual,levy2024data,zhou2025scale}; and (2) generating language and visual content simultaneously~\cite{zhang2022tell, gu2023compodiff, zhou2024vista,liu2025automaticsyntheticdatafinegrained}.

%


\subsubsection{Generating Language Content}
The augmentation of triplet data through text generation can be categorized into two main approaches: (1) generating text from a single image and (2) generating text describing differences between two images.

In the first category, Liu et al. \cite{liu2023zero} propose a pipeline leveraging large-scale image-text datasets (e.g., Laion-COCO) to synthesize relative descriptions. They use predefined templates and LLMs such as ChatGPT to edit image captions, forming triplets by matching generated text with candidate image captions. This yields two datasets: Laion-CIR-Template and Laion-CIR-LLM, each with approximately 16K triplets. Template-based descriptions demonstrate superior performance in fine-grained retrieval tasks. Hou et al. \cite{hou2024pseudo} introduce a masked training method, where patches of the original image are masked to create a reference image. Captions generated for the original (target) image are used as modification text, forming pseudo triplets. Zhang et al. \cite{zhang2024multimodal} focus on generating challenging negative examples. They modified query sentences or replaced reference images with semantically mismatched alternatives to create pseudo triplets with subtle differences. This improved the model's ability to distinguish examples and learn a better metric space for complex retrieval tasks.
Liu et al. \cite{liu2024bi} introduce bidirectional training, where the model is trained on both forward (reference image + modification text to target image) and backward (target image + reverse modification text to reference image) queries. This expands data volume and strengthens multimodal fusion capabilities.
Jawade et al. \cite{jawade2025scot} propose SCOT, which generates pseudo-triplets by masking parts of the image to create reference images, with captions for the target images used as modification text. These pseudo-triplets are then used for training, where a self-training process with a teacher-student framework is applied to generate pseudo-labels, and multi-factor fusion and attention mechanisms are used to enhance the model's robustness.
Duan et al. \cite{duan2025scaling} introduce a prompt-driven ZS-CMR method, which reduces the reliance on annotated triplets and textual data. By clustering images to generate pseudo-labels and using LLMs to create diverse textual modification instructions, this approach enables the model to perform composed retrieval using only image data.

In the second category, Jiang et al. \cite{jiang2024hycir} present a synthetic labeling framework that selects visually similar image pairs, generates differential captions using models like BLIP and LLMs, and filters outputs for semantic quality. Levy et al. \cite{levy2024data} introduce LaSCo, a large-scale dataset formed by selecting image pairs that yield different answers to the same question, followed by GPT-3-generated transition text. LaSCo significantly enlarges the training set and improves model performance. Jang et al. \cite{jang2024visual} propose the Visual Delta Generator, which identifies suitable image pairs from auxiliary datasets and generates textual descriptions of visual differences to construct pseudo triplets. 
The CoLLM proposed by Huynh et al. \cite{huynh2025collm} generates pseudo triplets by finding in-batch nearest neighbor images for each target in large-scale image-text pairs, interpolating reference features via Slerp, and creating modification texts with predefined templates and LLMs based on the captions. 
Noting that describing the differences between two images has been a long-standing research topic of interest~\cite{
tu2023neighborhood, yue2023i3n, black2024vixen, tu2023adaptive,jhamtani2018learning}.


\subsubsection{Generating Language and Visual Content}
Some works not only generate text but also generate images to further augment the triple data. Zhang et al. \cite{zhang2022tell} propose CTI-IR, a GAN-based network that jointly trains generative and retrieval models to handle image generation and retrieval tasks. The model uses global-local collaborative discriminators to ensure semantic consistency between the generated image and the modification text by capturing both overall and fine-grained differences. 
Gu et al. \cite{gu2023compodiff} introduce a diffusion-based model for constructing a large-scale synthetic dataset called SynthTriplets18M. By leveraging existing image-caption datasets (e.g., COYO 700M, LAION-2B-en-aesthetic), they first generate diverse modified texts by replacing object terms in the original captions or editing operation via LLMs. Then, a latent diffusion model is used to generate high-quality target images based on the reference image and modification text. This approach enhances the model's ability to handle diverse modifications and introduces capabilities like handling negative text and image mask conditions.
Zhou et al. \cite{zhou2024vista} propose one strategy, generating multi-modal triplets from large-scale image-text datasets. They use LLMs (GPT 3.5) to generate modification texts, which can be used to change certain objects, colors, positions, etc, and then employ stable diffusion to create target images, ensuring alignment between the modified text and the generated image.

Liu et al. ~\cite{liu2025automaticsyntheticdatafinegrained}
propose an automated pipeline for a large-scale, synthetic, composed person retrieval dataset. First, a LLM generates textual quadruples that include reference and target image descriptions along with relative captions detailing appearance changes. Second, a fine-tuned generative model produces identity-consistent image pairs from these textual descriptions, with each pair split into reference and target images. 
Tu et al. \cite{tu2025multimodal} propose MRA-CIR, which generates pseudo-triplets by selecting semantically related image pairs and creating modification texts with a multimodal reasoning agent combining GNN-based visual reasoning, Transformer-based textual reasoning, and attention-based cross-modal fusion for effective compositional learning in ZS-CMR.
Recently, Li et al. \cite{li2025imagine} propose an Imagine-and-Seek paradigm IP-CIR, where a diffusion model generates potential target image representations from reference images and modification texts, and these generated images serve as intermediate supervision to train the retrieval model, enhancing cross-modal alignment without real target images.
Building on this, Wang et al. \cite{wang2025generative} design a generative framework that directly synthesizes target image representations from reference images and modification texts, jointly training a diffusion model with a contrastive matching module to enable effective retrieval without predefined triplets or pseudo-word tokens.

\subsection{Noise/Uncertainty in Data}

In addition to the volume of data, ensuring data quality is a critical focus in existing research. High-quality triplet data can significantly enhance model training effectiveness by mitigating the impact of noisy data, thereby improving retrieval performance. Data uncertainty and noise primarily manifest in two forms: (1) Content noise, where erroneous content is incorrectly deemed semantically relevant \cite{jiang2024hycir,liu2023zero,gu2023compodiff,
lin2023clip,hou2024pseudo,levy2024data}. 
For instance, the target image does not fully align with the retrieval intent specified by the modification text. This issue is particularly pronounced in automatically generated data. (2) Annotation uncertainty, where the target image in the triplet is not the sole positive example, as other images in the dataset may also satisfy the retrieval intent defined by the reference image and modification text \cite{wen2023target,zhang2024collaborative,chen2023ranking,chen2022composed,baldrati2023zero,dodds2022training,li2025learning}.
Existing research has addressed these challenges through a series of studies. Below, we review and summarize these efforts in detail.

\subsubsection{Content Noise}
Existing methods primarily employ threshold-based filtering strategies to ensure semantic relevance and consistency. These strategies focus on measuring the similarity between reference images, modified texts, and target images to align with the intended modification intention. 
Jiang et al. \cite{jiang2024hycir} and Liu et al. \cite{liu2023zero} both utilize semantic similarity filtering to ensure consistency between generated modification texts and image pairs. Jiang et al. \cite{jiang2024hycir} screen modification texts within the language space, retaining triplets with high semantic similarity to both reference and target images. Similarly, Liu et al. \cite{liu2023zero} generate modification instructions using a LLM, followed by sentence transformation to produce edited descriptions. They apply a similarity-based filter to select triplets with high semantic relevance between edited descriptions and reference images.
Gu et al. \cite{gu2023compodiff} design a CLIP-based filtering strategy, calculating CLIP similarity between reference and generated target images, as well as between original and modified texts, and applying thresholds to ensure visual and semantic consistency. They introduce directional CLIP similarity to evaluate alignment between image and text modifications and verify that generated images reflect specified keywords or instructions.

Lin et al. \cite{lin2023clip} introduce a similarity-based data augmentation method. Specifically, they utilize CLIP's image encoder to extract features from all images in the training set, and then compute the cosine similarity between the reference or target images and other images in the training set. Based on the similarity scores, several of the most similar images are selected as new reference or target images, thereby generating pseudo-triplets. To ensure quality, the generated pseudo-triplets select the most similar images, constrained by predefined similarity thresholds and a limit on the number of pseudo-triplets per image to control quality and prevent overfitting.
Hou et al. \cite{hou2024pseudo} introduce a random sampling strategy based on the 3-$\sigma$ rule to select challenging samples for fine-tuning. They generate pseudo-modification texts for unlabeled image pairs, calculate distance metrics, and define a candidate range using the 3-$\sigma$ rule. Randomly selected samples from this range mitigate noise while ensuring sufficient challenge, enhancing model robustness.
Levy et al. \cite{levy2024data} develop an analytical tool to detect redundant noise in queries by evaluating the contribution of different modalities. Queries with minimal impact from either image or text are identified as containing redundant information. By identifying and filtering out triplets that include such redundant information, the quality of the generated data can be enhanced. 
Liu et al. ~\cite{liu2025automaticsyntheticdatafinegrained}
use a MLLM to evaluate and filter the generated triplets based on image quality, identity consistency, text-image alignment, and caption quality, retaining only high-quality examples that meet strict criteria.

\subsubsection{Annotation Uncertainty}
To address the issue of annotation uncertainty in triplets, where a single query may correspond to multiple valid target images, existing methods focus on mitigating false positives and enhancing semantic diversity. These approaches can be categorized into two main strategies: uncertainty-aware modeling and constructing evaluation datasets with multiple ground-truth labels.

Some works focus on modeling uncertainty in data.
Wen et al. \cite{wen2023target} introduce a matching regularization module based on target similarity distribution. It dynamically adjusts match scores by measuring similarities between the ground-truth target and candidate images within a mini-batch, allowing for partial matches rather than strictly treating all non-targets as negatives. Zhang et al. \cite{zhang2024collaborative} propose the Consensus Network (Css-Net), which utilizes four diverse compositors to generate robust image-text embeddings. By applying Kullback-Leibler divergence loss, Css-Net encourages consistent outputs across compositors, mitigating individual biases. Chen et al. \cite{chen2023ranking} present a rank-aware uncertainty framework with three modules: in-sample uncertainty models image features as Gaussian distributions; cross-sample uncertainty identifies shared positives across queries; and distribution regularization aligns source-target features. Chen et al. \cite{chen2022composed} further refine uncertainty modeling by introducing feature-space perturbations to simulate multi-granularity queries. Their model adjusts between one-to-one and one-to-many matching based on query precision via an uncertainty regularization module, thereby reducing false positives and enhancing retrieval accuracy.


Some studies have highlighted that uncertainty in data can become noise during model training, such as image-text pairs that are labeled as mismatched but are actually matched.
To address this problem, researchers have proposed improved methods~\cite{biten2022image,feng2023learning,qin2022deep,wang2022point,li2022unified,huang2021learning,yang2023bicro}.
Huang et al.~\cite{huang2021learning} propose Noisy Correspondence Rectifier, which divides data into two types, correct and noisy, based on the memory effect of neural networks, and then corrects the correspondence through an adaptive prediction model in a cooperative teaching manner.
Biten et al.~\cite{biten2022image} propose a new strategy that defines a semantic adaptive boundary using the image captioning metric CIDEr and improves and optimizes it in the standard triple ranking loss.
Yang et al.~\cite{yang2023bicro} proposed estimating soft labels for noisy annotations to reflect their true correspondence.
Recently, Li et al.~\cite{li2025learning} propose pseudo-text enhancement for triple noisy annotations, which transforms the noise into a visual difference modeling problem, bridges the gap between real modifications and annotated text by generating adapters, and introduces learnable task-oriented prompts to replace reference images to construct independent queries and reduce the impact of visually irrelevant noise.

To better reflect real-world ambiguity, similar to the expansion of the dataset by supplementing the missing positive sample associations in the image-text pair~\cite{chun2022eccv}, new datasets provide multiple valid targets per query. Baldrati et al. \cite{baldrati2023zero} create the CIRCO dataset, which annotates multiple relevant images for each query. They retrieve the top 100 candidates using retrieval models, filter to the 50 most visually similar images, and annotate all plausible matches, yielding an average of 4.53 ground truths per query. This reduces false negatives and improves evaluation robustness. Similarly, the CFQ dataset \cite{dodds2022training} offers multi-label annotations in the fashion domain. For each query, annotators label several target images from the same category as positive or negative based on their alignment with the textual description, providing a more comprehensive benchmark for fashion-oriented image retrieval.

\section{Applications} \label{APPL}

\begin{figure*}
  \centering
\includegraphics[width=1.0\textwidth]{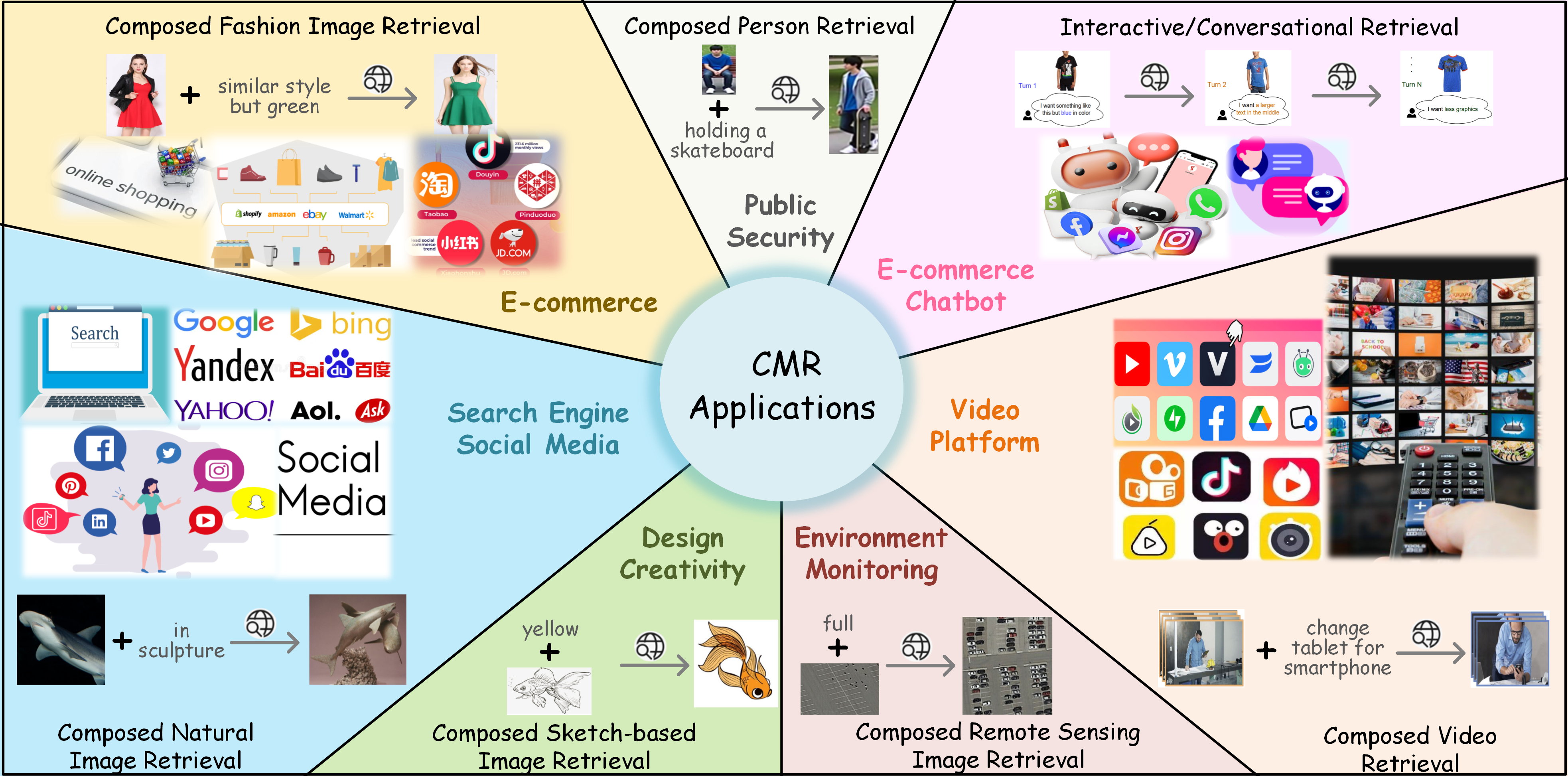}
  \caption{Existing composed multi-modal retrieval (CMR) applications can be categorized based on specific scenarios, including fashion images, natural images, videos, remote sensing images, person re-identification, skeletal images, and interactive conversation..
  }
  \label{applications}
\end{figure*}

At the application level, as shown in Fig.~\ref{applications}, we categorize existing composed multimodal retrieval applications based on specific scenarios, including fashion images, natural images, videos, remote sensing images, person re-identification, skeletal images, and interactive conversation. The specific contexts and challenges faced by each task in these different applications are thoroughly summarized, with the hope of inspiring exploration in other yet-to-be-explored scenarios.

To assess the effectiveness of composed multimodal retrieval models across various applications, several standard evaluation metrics are commonly employed. Among them, \textbf{Recall@K} is the most widely used. It measures the proportion of relevant items retrieved among the top-K results. Due to the presence of potential false negatives in annotations, high values of K (e.g., 10, 50) are often used to mitigate incomplete labeling effects. We report average Recall@1, Recall@10, and Recall@50 following standard practice. Another commonly used metric is \textbf{Mean Average Precision at K (mAP@K)}, which accounts for both precision and the rank of relevant items within the top-K results. It is particularly helpful when evaluating models that must not only retrieve relevant samples but also rank them effectively. These metrics provide a comprehensive view of both retrieval accuracy and ranking quality, and they are used selectively across different datasets and tasks according to their characteristics.

\subsection{Composed Fashion Image Retrieval}
Traditional fashion image retrieval primarily relies on simple image search or keyword-based search. However, these methods often fail to meet user needs when searching for specific fashion items with complex attributes such as color, style, and material. The main goal of Composed Fashion Image Retrieval (CFIR) is to achieve more accurate and personalized fashion searches by combining both images and texts. 
CFIR has broad applications in e-commerce. By combining image-based search with textual refinement, it improves search efficiency and helps users locate products that better match their preferences. For example, users can input a reference image (e.g., a photo of a clothing item) along with a textual modification (e.g., "long-sleeve version").
This leads to higher customer satisfaction and greater user engagement. For retailers and brands, CFIR supports personalized product recommendations, fashion trend analysis, and more effective inventory management.


\subsubsection{Benchmark Datasets}
The commonly used datasets related to the retrieval of composed fashion images can be summarized as follows:

\textbf{FashionIQ} \cite{wu2021fashion} consists of 77,684 images of fashion products across three categories: dresses, shirts, and tops\&tees. A subset of 49,464 images is annotated with side information derived from product descriptions, including various attributes. Additionally, 60,272 image pairs are annotated with relative captions, which provide natural language descriptions of the differences between reference and target images. These captions serve as modification text for retrieving target images. 


\textbf{Fashion200k} \cite{han2017automatic} is collected from online shopping websites. The authors crawl over 300,000 product images along with their descriptions, removing those with descriptions containing fewer than four words, resulting in more than 200,000 images. They divide the dataset into 172,049 images for training, 12,164 for validation, and 25,331 for testing. For text cleaning, they eliminate stop words, symbols, and words that occur fewer than five times. The remaining words are treated as attributes, giving a total of 4,404 attributes to train the joint embedding.

\textbf{Shoes} \cite{guo2018dialog} is collected from the Internet, specifically from like.com, a shopping website that aggregates product data from a wide range of e-commerce sources.  They crawl 10,751 pairs of shoe images with relative expressions that describe fine-grained visual differences. Typically, 10,000 samples are used for training, and 4,658 samples are used for evaluation. 


There are also some datasets that are not commonly used, including CFQ \cite{dodds2022training}, UT-Zap50k \cite{yu2014fine, yu2017semantic}, FashionGen \cite{rostamzadeh2018fashion}, FACAD \cite{yang2020fashion}, and PolyvoreOutfits \cite{vasileva2018learning}.

%


\subsubsection{Results and Analysis}
We evaluate the performance of various methods on three widely used benchmarks for composed fashion image retrieval: \textbf{FashionIQ}, \textbf{Fashion200K}, and \textbf{Shoes}, covering diverse types of textual modifications and attribute granularities. For the \textbf{FashionIQ} dataset, we follow two commonly adopted evaluation protocols: \textit{Original Split} and \textit{VAL Split}, as shown in Table~\ref{tab:fashioniq} and Table~\ref{tab:fashioniq_val}, respectively. The VAL Split, introduced in early CMR studies, constructs a reduced candidate pool within the validation set, resulting in a narrower search space and relatively simpler task. In contrast, the Original Split leverages the full candidate pool provided by the dataset, substantially increasing the retrieval difficulty and typically leading to lower Recall@K scores. In recent years, the Original Split has been increasingly adopted as a more rigorous benchmark for evaluating model generalization under large-scale open-domain settings.

Under the Original Split, early supervised methods typically employed ResNet + RNN architectures. For instance, JSVM only achieved an average R@50 of 26.6\%. With the emergence of large-scale vision-language pretraining, models such as CLIP and BLIP have been progressively introduced into compositional retrieval tasks. CLIP4CIR2 and BLIP4CIR2 achieved average R@50 scores of 64.2\% and 73.1\%, respectively, demonstrating significant performance improvements. More recent methods, such as DQU-CIR and SDQUR, adopt non-explicit fusion architectures. Specifically, DQU-CIR performs modality fusion at the data level, while SDQUR leverages BLIP2’s Q-Former for diverse query generation and uncertainty modeling. These models achieve average R@50 scores of 75.6\% and 76.4\%, respectively. These results highlight the importance of backbone strength and deep fusion mechanisms in advancing compositional retrieval performance.

In the \textbf{zero-shot learning (ZSL-CMR)} setting, models do not involve task-specific training and rely entirely on the inherent cross-modal alignment capabilities of VLMs. For example, PrediCIR, using a ViT-G/14 backbone, achieves an average R@10 of 47.2\%, comparable to or even outperforming several early supervised models. Other approaches such as FTI4CIR (pseudo-token based), LinCIR (self-supervised in the text domain), Slerp (spherical interpolation), and CIReVL (LLM-based retrieval) explore different composition paradigms, contributing to the growing diversity and maturity of the ZSL-CMR paradigm.

Under the \textbf{semi-supervised learning (SSL-CMR)} setting, methods such as VDG and CASE generate large-scale pseudo-triplets by using VLMs and LLMs to produce fine-grained textual differences for visually similar image pairs. Built on BLIP backbones, VDG and CASE achieve average R@10 scores of 50.85\% and 48.79\%, respectively, outperforming all existing ZSL methods and many supervised baselines. These results suggest that high-quality, automatically constructed pseudo-supervision can provide rich and effective learning signals. Furthermore, methods like CompoDiff extend this paradigm by directly generating target images via diffusion models, enabling more complex compositional scene synthesis.

For the \textbf{Fashion200K} and \textbf{Shoes} datasets (Table~\ref{tab:f200k_shoes_comparison}), despite differing task emphases, we observe similar trends to FashionIQ. Fashion200K features broader category diversity; the R@50 improved from 63.8\% (TIRG) to 87.8\% (DQU-CIR). In the Shoes dataset, which requires finer attribute matching, performance increased from 75.8\% (VAL) to 88.3\% (SDQUR). These results indicate that stronger vision backbones (e.g., ViT, BLIP2), coupled with more expressive multimodal fusion modules (e.g., Q-Former, set-level alignment), consistently lead to improved retrieval accuracy.


\begin{table*}[htp]
\centering
\scriptsize
\setlength{\tabcolsep}{4.5pt}  
\caption{Comparison of Different Methods on FashionIQ Dataset (Original Split).}
\label{tab:fashioniq}
\newcolumntype{C}[1]{>{\centering\arraybackslash}p{#1}}
\begin{tabular}{@{}p{4.5cm}C{2.5cm}C{2cm}*{8}{S[table-format=2.2,round-mode=places,round-precision=2]}@{}}
\toprule\toprule  
\multirow{2}{*}{Method} & \multicolumn{2}{c}{Backbone} & \multicolumn{2}{c}{Dresses} & \multicolumn{2}{c}{Shirts} & \multicolumn{2}{c}{Tops\&Tees} & \multicolumn{2}{c}{Avg} \\
\cmidrule(lr){2-3} \cmidrule(lr){4-5} \cmidrule(lr){6-7} \cmidrule(lr){8-9} \cmidrule(lr){10-11}
& Visual & Textual & {R@10} & {R@50} & {R@10} & {R@50} & {R@10} & {R@50} & {R@10} & {R@50} \\
\midrule
\multicolumn{11}{l}{{\textbf{Supervised Learning-based CMR (SL-CMR)}}} \\[2pt]
JSVM\pubinfo{20'ECCV} \cite{chen2020learning} & MobileNet & LSTM & 10.70 & 25.90 & 12.00 & 27.10 & 13.00 & 26.90 & 11.90 & 26.63 \\
TRACE\pubinfo{21'AAAI} \cite{jandial2020trace} & ResNet50 & GRU & 22.70 & 44.91 & 20.80 & 40.80 & 24.22 & 49.80 & 22.57 & 46.19 \\
ComposeAE\pubinfo{21'WACV} \cite{anwaar2021compositional} & ResNet18 & BERT & 10.77 & 28.29 & 9.96 & 25.14 & 12.74 & 30.79 & 11.80 & 29.40 \\
CoSMo\pubinfo{21'CVPR} \cite{lee2021cosmo} & ResNet50 & LSTM & 21.39 & 44.45 & 16.90 & 37.49 & 21.32 & 46.02 & 19.87 & 42.62 \\
FashionVLP\pubinfo{22'CVPR} \cite{goenka2022fashionvlp} & ResNet50 & BERT & 26.77 & 53.2 & 22.67 & 46.22 & 28.51 & 57.47 & 25.98 & 52.30 \\
Combiner\pubinfo{22'CVPR-D} \cite{baldrati2022effective} & CLIP(RN50) & CLIP & 31.63 & 56.67 & 36.36 & 51.86 & 38.19 & 62.42 & 35.39 & 59.03 \\
MultiColSAP\pubinfo{22'TMM} \cite{zhang2023enhance} & ResNet50 & LSTM & 30.74 & 55.92 & 26.05 & 50.64 & 34.42 & 61.14 & 30.4 & 55.9 \\
CLIP4Cir1\pubinfo{22'CVPR-W} \cite{baldrati2022conditioned} & CLIP(RN50) & CLIP & 33.81 & 59.40 & 39.99 & 60.45 & 41.41 & 65.37 & 38.32 & 61.74 \\
Progressive\pubinfo{22'SIGIR} \cite{zhao2022progressive} & ResNet50 & BERT & 29.00 & 53.94 & 35.43 & 58.88 & 39.16 & 64.56 & 34.53 & 59.13 \\
Progressive\pubinfo{22'SIGIR} \cite{zhao2022progressive} & CLIP(ViT-B/32) & BERT & 33.60 & 58.90 & 39.45 & 61.78 & 43.96 & 68.33 & 39.02 & 63.00 \\
ARTEMIS\pubinfo{22'ICLR} \cite{delmas2022artemis} & ResNet50 & BiGRU & 25.68 & 51.05 & 21.57 & 44.13 & 28.59 & 55.06 & 25.28 & 50.08 \\
ComqueryFormer\pubinfo{23'TMM} \cite{xu2023multi} & Swin-T & BERT & 28.85 & 55.38 & 25.64 & 50.22 & 33.61 & 60.48 & 29.37 & 55.36 \\
TransAgg(LaionCIR)\pubinfo{23'BMVC} \cite{liu2023zero} & BLIP & BLIP & 27.67 & 49.38 & 32.83 & 52.31 & 35.70 & 58.08 & 32.07 & 53.26 \\
CLIP-CD\pubinfo{23'AI} \cite{lin2023clip} & CLIP(RN50) & CLIP & 37.68 & 62.62 & 42.44 & 63.74 & 45.33 & 67.72 & 41.82 & 64.79 \\
CLIP4Cir2\pubinfo{23'TOMM} \cite{baldrati2023composed} & CLIP(RN50) & CLIP & 37.67 & 63.16 & 39.87 & 60.84 & 44.88 & 68.59 & 40.80 & 64.20 \\
BLIP4CIR\pubinfo{24'WACV} \cite{liu2024bi} & BLIP & BLIP & 40.65 & 66.34 & 37.49 & 60.06 & 43.60 & 67.77 & 39.34 & 63.85 \\
BLIP4CIR+Bi\pubinfo{24'WACV} \cite{liu2024bi} & BLIP & BLIP & 42.09 & 67.33 & 41.76 & 64.28 & 46.61 & 70.32 & 43.49 & 67.31 \\
SPIRIT\pubinfo{24'TOMM} \cite{chen2024spirit} & CLIP(RN50) & CLIP & 39.86 & 64.30 & 47.67 & 71.10 & 44.11 & 65.60 & 43.88 & 67.20 \\
DQU-CIR\pubinfo{24'SIGIR} \cite{wen2024simple} & CLIP(ViT-H/14) & {--} & 51.90 & 74.37 & 53.57 & 73.21 & 58.48 & 79.23 & 54.65 & 75.60 \\
FashionViL\pubinfo{24'ECCV} \cite{han2022fashionvil}  & ResNet50 & BERT & 33.47 & 59.94 & 25.17 & 50.39 & 34.98 & 60.79 & 31.21 & 57.04 \\
BLIP4CIR2\pubinfo{24'TMLR} \cite{liucandidate} & BLIP(ViT-B) & BLIP & 43.78 & 71.34 & 50.15 & 71.25 & 55.23 & 76.80 & 51.17 & 73.13 \\
CompoDiff\pubinfo{23'TMLR} \cite{gu2023compodiff} & {CLIP(ViT-L/14)} & {--} & 35.53 & 49.56 & 40.88 & 53.06 & 41.15 & 54.12 & 39.05 & 52.34 \\
CompoDiff\pubinfo{23'TMLR} \cite{gu2023compodiff} & {CLIP(ViT-G/14)} & {--} & 38.39 & 51.03 & 41.68 & 56.02 & 45.70 & 57.32 & 39.81 & 51.90 \\
SDQUR\pubinfo{24'TCSVT} \cite{xu2024set} & BLIP2 & {--} & 49.93 & 73.33 & 56.87 & 76.50 & 59.66 & 79.25 & 55.49 & 76.36 \\
AlRet\pubinfo{24'TMM} \cite{xu2024align} & CLIP & LSTM & 35.75 & 60.56 & 37.02 & 60.05 & 42.25 & 67.52 & 38.20 & 62.82 \\
NSFSE\pubinfo{24'TMM} \cite{wang2024negative} & ResNet50 & BiGRU & 29.62 & 54.41 & 22.96 & 45.93 & 31.08 & 57.01 & 27.84 & 52.39 \\
NSFSE\pubinfo{24'TMM} \cite{wang2024negative} & ResNet152 & BiGRU & 31.12 & 55.73 & 24.58 & 45.85 & 31.93 & 58.37 & 29.17 & 53.24 \\
MANME\pubinfo{24'TCSVT} \cite{li2023multi} & ResNet50 & BiGRU & 31.26 & 57.66 & 26.37 & 47.94 & 36.33 & 59.31 & 29.95 & 54.90 \\
\midrule
\multicolumn{11}{l}{{\textbf{Zero-Shot Learning-based CMR (ZSL-CMR)}}} \\[2pt]
PALAVAR\pubinfo{22'ECCV} \cite{cohen2022my} & \multicolumn{2}{c}{CLIP(ViT-B/32)} & 17.25 & 35.94 & 21.49 & 37.05 & 20.55 & 38.76 & 19.76 & 37.25 \\
Pic2Word\pubinfo{23'CVPR} \cite{saito2023pic2word} & \multicolumn{2}{c}{CLIP(ViT-L/14)} & 20.00 & 40.20 & 26.20 & 43.60 & 27.90 & 47.40 & 24.70 & 43.70 \\
SEARLE\pubinfo{23'ICCV} \cite{baldrati2023zero} & \multicolumn{2}{c}{CLIP(ViT-B/16)} & 18.54 & 38.51 & 24.44 & 41.61 & 25.70 & 46.46 & 22.89 & 42.53 \\
SEARLE-XL\pubinfo{23'ICCV} \cite{baldrati2023zero} & \multicolumn{2}{c}{CLIP(ViT-L/14)} & 20.48 & 43.13 & 26.86 & 45.58 & 29.32 & 49.97 & 25.56 & 46.23 \\
MTCIR\pubinfo{23'Arxiv} \cite{chen2023pretrain} & \multicolumn{2}{c}{CLIP(ViT-L/14)} & 28.11 & 51.12 & 38.63 & 58.51 & 39.42 & 62.68 & 35.39 & 57.44 \\
KEDs\pubinfo{24'CVPR} \cite{suo2024knowledge} & \multicolumn{2}{c}{CLIP(ViT-L/14)} & 21.7 & 43.8 & 28.9 & 48.0 & 29.9 & 51.9 & 26.8 & 47.9 \\
PM\pubinfo{24'Arxiv} \cite{zhang2024zero} & \multicolumn{2}{c}{CLIP(ViT-L/14)} & 27.1 & 43.8 & 21.4 & 41.7 & 28.9 & 47.3 & 25.8 & 44.2 \\
FTI4CIR\pubinfo{24'SIGIR} \cite{lin2024fine} & \multicolumn{2}{c}{CLIP(ViT-L/14)} & 24.39 & 47.84 & 31.35 & 50.59 & 32.43 & 54.21 & 29.39 & 50.88 \\
CIReVL\pubinfo{24'ICLR} \cite{karthik2023vision}  & \multicolumn{2}{c}{CLIP(ViT-L/14)} & 24.79 & 44.76 & 29.49 & 47.40 & 31.36 & 53.65 & 28.55 & 48.57 \\
CIReVL\pubinfo{24'ICLR} \cite{karthik2023vision}  & \multicolumn{2}{c}{CLIP(ViT-G/14)} & 27.07 & 49.53 & 33.71 & 51.42 & 35.80 & 56.14 & 32.19 & 52.36 \\
LinCIR\pubinfo{24'CVPR} \cite{gu2024language} & \multicolumn{2}{c}{CLIP(ViT-L/14)} & 20.92 & 42.44 & 29.10 & 46.81 & 28.81 & 50.18 & 26.28 & 46.49 \\
LinCIR\pubinfo{24'CVPR} \cite{gu2024language} & \multicolumn{2}{c}{CLIP(ViT-G/14)} & 38.08 & 60.88 & 46.76 & 65.11 & 50.48 & 71.09 & 45.11 & 65.69 \\
iSEARLE-XL\pubinfo{24'Arxiv} \cite{agnolucci2024isearle} & \multicolumn{2}{c}{CLIP(ViT-L/14)} & 22.51 & 46.36 & 28.75 & 47.84 & 31.31 & 52.68 & 27.52 & 48.96 \\
RTD(SEARLE)\pubinfo{24'Arxiv} \cite{byun2024reducing} & \multicolumn{2}{c}{CLIP(ViT-B/32)} & 20.72 & 43.13 & 26.69 & 44.31 & 26.67 & 48.75 & 24.7 & 45.4 \\
RTD((LinCIR)\pubinfo{24'Arxiv} \cite{byun2024reducing} & \multicolumn{2}{c}{CLIP(ViT-L/14)} & 24.49 & 48.24 & 32.83 & 50.44 & 33.4 & 54.56 & 30.24 & 51.08 \\
ISA(Sym)\pubinfo{24'ICLR} \cite{du2024image2sentence}& \multicolumn{2}{c}{BLIP} & 24.69 & 43.88 & 30.79 & 50.05 & 33.91 & 53.65 & 29.79 & 49.19 \\
ISA(Asy)\pubinfo{24'ICLR} \cite{du2024image2sentence} & \multicolumn{2}{c}{EfficientNet+BLIP} & 25.33 & 46.26 & 30.03 & 48.58 & 33.45 & 53.8 & 29.6 & 49.54 \\
LDRE\pubinfo{24'SIGIR} \cite{yang2024ldre} & \multicolumn{2}{c}{CLIP(ViT-L/14)} & 22.93 & 46.76 & 31.04 & 51.22 & 31.57 & 53.64 & 28.51 & 50.54 \\
LDRE\pubinfo{24'SIGIR} \cite{yang2024ldre} & \multicolumn{2}{c}{CLIP(ViT-G/14)} & 26.11 & 51.12 & 35.94 & 58.58 & 35.42 & 56.67 & 32.49 & 55.46 \\
Context-I2W\pubinfo{24'AAAI} \cite{tang2024context} & \multicolumn{2}{c}{CLIP(ViT-L/14)} & 23.1 & 45.3 & 29.7 & 48.6 & 30.6 & 52.9 & 27.8 & 48.9 \\
Slerp+TAT\pubinfo{24'ECCV}  \cite{jang2024spherical}& \multicolumn{2}{c}{CLIP(ViT-L/14)} & 23.35 & 45.12 & 29.94 & 46.47 & 31.97 & 51.20 & 28.32 & 47.00 \\
Slerp+TAT\pubinfo{24'ECCV}  \cite{jang2024spherical}& \multicolumn{2}{c}{BLIP(ViT-L/16)} & 29.15 & 50.62 & 32.14 & 51.62 & 37.02 & 57.73 & 32.77 & 53.32 \\
Denoise-I2W\pubinfo{24'Arxiv}~\cite{tang2024denoise} & \multicolumn{2}{c}{CLIP(ViT-L/14)} & 24.4 & 47.8 & 30.9 & 49.8 & 31.6 & 54.1 & 29.0 & 50.6 \\
MoTaDual(LinCIR)\pubinfo{24'Arxiv}~\cite{li2024motadual} & \multicolumn{2}{c}{CLIP-L} & {--} & {--} & {--} & {--} & {--} & {--} & 28.94 & 49.43 \\
InstructCIR1\pubinfo{24'Arxiv}~\cite{zhong2024compositional} & \multicolumn{2}{c}{CLIP(ViT-L/14)} & 28.15 & 49.38 & 32.24 & 52.11 & 37.26 & 56.13 & 32.55 & 52.54 \\
DeG\pubinfo{25'Arxiv} \cite{chen2025data} & \multicolumn{2}{c}{CLIP(ViT-L/14)} & 24.4 & 46.5 & 30.7 & 50.3 & 31.6 & 52.0 & 28.9 & 49.6 \\
Slerp+\pubinfo{24'Arxiv} \cite{jangtext} & \multicolumn{2}{c}{BLIP} & 31.78 & 54.05 & 37.73 & 56.82 & 41.36 & 62.37 & 36.96 & 57.74 \\
InstructCIR2\pubinfo{25'Arxiv} \cite{duan2025scaling} & \multicolumn{2}{c}{CLIP(ViT-L/14)} & {--} & {--} & {--} & {--} & {--} & {--} & 37.32 & 56.84 \\
TSCIR\pubinfo{25'Arxiv} \cite{wang2025mapping} & \multicolumn{2}{c}{CLIP(ViT-L/14)} & 24.14 & 46.80 & 31.01 & 50.05 & 32.94 & 54.26 & 29.37 & 50.37 \\
SEIZE\pubinfo{24'ACMMM} \cite{yang2024semantic} & \multicolumn{2}{c}{CLIP(ViT-L/14)} & 30.93 & 50.76 & 33.04 & 53.22 & 35.57 & 58.64 & 33.18 & 54.21 \\
SEIZE\pubinfo{24'ACMMM} \cite{yang2024semantic} & \multicolumn{2}{c}{CLIP(ViT-G/14)} & 39.61 & 61.02 & 43.60 & 65.42 & 45.94 & 71.12 & 43.05 & 65.85 \\
PrediCIR\pubinfo{25'CVPR} \cite{tang2025missing} & \multicolumn{2}{c}{CLIP(ViT-L/14)} & 25.4 & 49.5 & 31.8 & 52.0 & 33.1 & 55.4 & 30.1 & 52.3 \\
PrediCIR\pubinfo{25'CVPR} \cite{tang2025missing} & \multicolumn{2}{c}{CLIP(ViT-G/14)} & 39.7 & 62.4 & 48.2 & 67.4 & 53.7 & 73.6 & 47.2 & 67.8 \\
CoLLM\pubinfo{25'CVPR} \cite{huynh2025collm} & \multicolumn{2}{c}{CLIP(ViT-L/14)} & {--} & {--} & {--} & {--} & {--} & {--} & 30.1 & 49.5 \\
\midrule
\multicolumn{11}{l}{{\textbf{Semi-Supervised Learning-based CMR (SSL-CMR)}}} \\[2pt]
CompoDiff(ST18M)\pubinfo{23'TMLR} \cite{gu2023compodiff} & \multicolumn{2}{c}{CLIP(ViT-L/14)} & 32.24 & 46.27 & 37.69 & 49.08 & 38.12 & 50.57 & 36.02 & 48.64 \\
PTG(+SPRC)\pubinfo{24'Arxiv} \cite{hou2024pseudo} & \multicolumn{2}{c}{--} & {--} & {--} & {--} & {--} & {--} & {--} & 31.10 & 51.90 \\
CASE\pubinfo{24'AAAI} \cite{levy2024data} & \multicolumn{2}{c}{BLIP}& 47.44 & 69.36 & 48.48 & 70.23 & 50.18 & 72.24 & 48.79 & 70.68 \\
HyCIR(CC3M+syn)\pubinfo{24'Arxiv} \cite{jiang2024hycir} & \multicolumn{2}{c}{CLIP} & 19.98 & 40.8 & 27.62 & 44.94 & 28.14 & 47.67 & 25.25 & 44.47 \\
HyCIR(CC3M+syn)\pubinfo{24'Arxiv} \cite{jiang2024hycir} & \multicolumn{2}{c}{BLIP} & 18.88 & 34.5 & 22.52 & 37.58 & 22.13 & 40.33 & 21.18 & 37.47 \\
VDG(+DF'se)\pubinfo{24'CVPR} \cite{jang2024visual} & \multicolumn{2}{c}{BLIP} & 47.10 & 69.10 & 49.95 & 69.96 & 53.90 & 74.35 & 50.32 & 71.14 \\
VDG(+FIQ'se)\pubinfo{24'CVPR} \cite{jang2024visual} & \multicolumn{2}{c}{BLIP}  & 47.89 & 69.81 & 51.36 & 71.08 & 53.29 & 74.65 & 50.85 & 71.85 \\
SCOT\pubinfo{25'Arxiv} \cite{jawade2025scot} & \multicolumn{2}{c}{BLIP(ViT-L/16)} & 26.42 & 49.23 & 30.91 & 49.65 & 34.72 & 55.12 & 30.68 & 51.33 \\
SCOT\pubinfo{25'Arxiv} \cite{jawade2025scot} & \multicolumn{2}{c}{BLIP2(ViT-G/14)} & 32.78 & 55.91 & 41.42 & 61.09 & 41.15 & 63.10 & 38.45 & 60.03 \\
InstructCIR2\pubinfo{25'Arxiv} \cite{duan2025scaling} & \multicolumn{2}{c}{CLIP(ViT-L/14)} & {--} & {--} & {--} & {--} & {--} & {--} & 49.03 & 70.96 \\
TSCIR\pubinfo{25'Arxiv} \cite{wang2025mapping} & \multicolumn{2}{c}{CLIP(ViT-L/14)} & 27.22 & 50.87 & 33.71 & 53.43 & 34.73 & 57.22 & 31.88 & 53.84 \\
MRA-CIR\pubinfo{25'Arxiv} \cite{tu2025multimodal} & \multicolumn{2}{c}{BLIP2(ViT-L/14)} & 31.87 & 54.23 & 40.43 & 60.20 & 41.25 & 62.51 & 37.85 & 58.98 \\
IP-CIR(LDRE)\pubinfo{25'CVPR} \cite{li2025imagine} & \multicolumn{2}{c}{CLIP(ViT-G/14)} & 39.02 & 61.03 & 48.04 & 66.68 & 50.18 & 71.14 & 45.74 & 66.28 \\
CIG-XL(SEARLE)\pubinfo{25'CVPR} \cite{wang2025generative} & \multicolumn{2}{c}{CLIP(ViT-B/32)} & 17.74 & 39.86 & 24.58 & 41.41 & 25.65 & 46.35 & 22.99 & 42.54 \\
CIG-XL(LinCIR)\pubinfo{25'CVPR} \cite{wang2025generative} & \multicolumn{2}{c}{CLIP(ViT-L/14)} & 21.27 & 43.98 & 28.66 & 47.20 & 29.83 & 50.28 & 26.59 & 47.15 \\
CoLLM\pubinfo{25'CVPR} \cite{huynh2025collm} & \multicolumn{2}{c}{CLIP(ViT-L/14)} & {--} & {--} & {--} & {--} & {--} & {--} & 32.9 & 54.2 \\
CoLLM\pubinfo{25'CVPR} \cite{huynh2025collm} & \multicolumn{2}{c}{BLIP(ViT-L/16)} & {--} & {--} & {--} & {--} & {--} & {--} & 39.1 & 60.7  \\
\bottomrule\bottomrule
\end{tabular}
\end{table*}

{\small
\begin{table*}[htp]
\centering
\scriptsize
\setlength{\tabcolsep}{4pt}
\caption{Comparison of Different Methods on FashionIQ Dataset (Val Split)}
\label{tab:fashioniq_val}
\newcolumntype{C}[1]{>{\centering\arraybackslash}p{#1}}
\begin{tabular}{@{}p{4cm}C{2.5cm}C{1.5cm}*{8}{S[table-format=2.2,round-mode=places,round-precision=2]}@{}}
\toprule\toprule
\multirow{2}{*}{Method} & \multicolumn{2}{c}{Backbone} & \multicolumn{2}{c}{Dresses} & \multicolumn{2}{c}{Shirts} & \multicolumn{2}{c}{Tops\&Tees} & \multicolumn{2}{c}{Avg} \\
\cmidrule(lr){2-3} \cmidrule(lr){4-5} \cmidrule(lr){6-7} \cmidrule(lr){8-9} \cmidrule(lr){10-11}
& Visual & Textual & {R@10} & {R@50} & {R@10} & {R@50} & {R@10} & {R@50} & {R@10} & {R@50} \\
\midrule
\multicolumn{11}{l}{{\textbf{Supervised Learning-based CMR (SL-CMR)}}} \\[2pt]
TIRG\pubinfo{18'CVPR}  \cite{vo2019composing} & ResNet18 & LSTM & 14.87 & 34.66 & 18.26 & 37.89 & 19.08 & 39.62 & 17.40 & 37.39 \\
VAL\pubinfo{20'CVPR} \cite{chen2020image} & ResNet50 & LSTM & 22.53 & 44.00 & 22.38 & 44.15 & 27.53 & 51.68 & 24.15 & 46.61 \\
MAAF\pubinfo{20'Arxiv} \cite{dodds2020modality} & ResNet50 & LSTM & 23.8 & 48.6 & 21.3 & 44.2 & 27.9 & 53.6 & 24.3 & 48.8 \\
TRACE\pubinfo{21'AAAI} \cite{jandial2020trace} & ResNet50 & GRU & 26.52 & 51.01 & 28.02 & 51.86 & 32.7 & 61.23 & 29.08 & 54.7 \\
RTIC\pubinfo{21'Arxiv} \cite{shin2021rtic} & ResNet50 & LSTM & 19.4 & 43.51 & 16.93 & 38.36 & 21.58 & 47.88 & 19.3 & 43.25 \\
DCNet\pubinfo{21'AAAI} \cite{kim2021dual} & ResNet50 & Glove & 28.95 & 56.07 & 23.95 & 47.30 & 30.44 & 58.29 & 27.78 & 53.89 \\
CoSMo\pubinfo{21'CVPR} \cite{lee2021cosmo} & ResNet50 & LSTM & 25.64 & 50.3 & 24.90 & 49.18 & 29.21 & 57.46 & 26.58 & 52.31 \\
HFCA\pubinfo{21'ACMMM} \cite{zhang2021heterogeneous} & ResNet50 & LSTM & 26.2 & 51.2 & 22.4 & 46.0 & 29.7 & 56.4 & 26.1 & 51.2 \\
CLVC-Net\pubinfo{21'SIGIR} \cite{wen2021comprehensive} & ResNet50 & LSTM & 29.85 & 56.47 & 28.75 & 54.76 & 33.50 & 64.00 & 30.70 & 58.41 \\
CIRPLANT\pubinfo{21'ICCV} \cite{liu2021image} & ResNet152 & LSTM & 14.38 & 34.66 & 13.64 & 33.56 & 16.44 & 38.34 & 14.82 & 35.52 \\
CIRPLANT(OSCAR)\pubinfo{21'ICCV} \cite{liu2021image} & ResNet152 & LSTM & 17.45 & 40.41 & 17.53 & 38.81 & 21.64 & 45.38 & 18.87 & 41.53 \\
JPM(VAL+MSE)\pubinfo{21'ACMMM} \cite{yang2021cross} & ResNet18 & LSTM & 21.27 & 43.12 & 21.88 & 43.3 & 25.81 & 50.27 & 22.98 & 45.59 \\
JPM(VAL+Tri)\pubinfo{21'ACMMM} \cite{yang2021cross} & ResNet18 & LSTM & 21.38 & 45.15 & 22.81 & 45.18 & 27.78 & 51.70 & 23.99 & 47.34 \\
FashionVLP\pubinfo{22'CVPR} \cite{goenka2022fashionvlp} & ResNet50 & BERT & 32.42 & 60.29 & 31.89 & 58.44 & 38.51 & 68.79 & 34.27 & 62.51 \\
SAC\pubinfo{22'WACV} \cite{jandial2022sac} & ResNet50 & GRU & 26.52 & 51.01 & 28.02 & 51.86 & 32.70 & 61.23 & 29.08 & 54.70 \\
ERR\pubinfo{22'TIP} \cite{zhang2022composed} & ResNet50 & LSTM & 30.02 & 55.44 & 25.32 & 49.87 & 33.20 & 60.34 & 29.51 & 55.22 \\
CRR\pubinfo{22'ACMMM} \cite{zhang2022comprehensive} & ResNet101 & BiGRU & 30.41 & 57.11 & 30.73 & 58.02 & 33.67 & 64.48 & 31.60 & 59.87 \\
Progressive\pubinfo{22'SIGIR} \cite{zhao2022progressive} & ResNet50 & BERT & 33.22 & 59.99 & 46.17 & 68.79 & 46.46 & 73.84 & 41.98 & 67.54 \\
Progressive\pubinfo{22'SIGIR} \cite{zhao2022progressive} & CLIP(ViT-B/32) & BERT & 38.18 & 64.50 & 48.63 & 71.54 & 52.32 & 76.90 & 46.37 & 70.98 \\
MCR\pubinfo{22'TMM} \cite{pang2022heterogeneous} & ResNet50 & LSTM & 26.20 & 51.20 & 22.40 & 46.00 & 29.70 & 56.40 & 26.10 & 51.20 \\
NEUCORE\pubinfo{23'NIPS-W} \cite{zhao2024neucore} & ResNet & BiGRU & 27.00 & 53.79 & 22.84 & 45.00 & 29.63 & 56.65 & 26.45 & 51.75 \\
ARTEMIS\pubinfo{22'ICLR} \cite{delmas2022artemis} & ResNet50 & LSTM & 27.34 & 51.71 & 21.05 & 49.87 & 27.34 & 44.18 & 24.43 & 48.59 \\
ARTEMIS\pubinfo{22'ICLR} \cite{delmas2022artemis} & ResNet50 & BiGRU & 27.16 & 52.40 & 21.78 & 43.64 & 29.20 & 54.83 & 26.05 & 50.29 \\
AMC\pubinfo{23'TOMM} \cite{zhu2023amc} & ResNet50 & LSTM & 31.73 & 59.25 & 36.21 & 66.60 & 30.67 & 59.08 & 32.87 & 61.64 \\
ComqueryFormer\pubinfo{23'TMM} \cite{xu2023multi} & Swin-T & BERT & 33.86 & 61.08 & 35.57 & 62.19 & 42.07 & 69.30 & 37.17 & 64.19 \\
RankUn\pubinfo{23'Arxiv} \cite{chen2023ranking} & CLIP(RN50) & CLIP & 34.8 & 60.22 & 45.21 & 69.06 & 47.68 & 74.85 & 42.5 & 68.04 \\
CRN\pubinfo{23'TIP} \cite{yang2023composed} & Swin-T(base) & LSTM & 30.24 & 57.61 & 29.83 & 55.54 & 33.91 & 64.04 & 31.36 & 59.06 \\
CRN\pubinfo{23'TIP} \cite{yang2023composed} & Swin-T(Large) & LSTM & 32.67 & 59.30 & 30.27 & 56.97 & 37.74 & 65.94 & 33.56 & 60.74 \\
SSN\pubinfo{23'Arxiv} \cite{yang2023decompose} \cite{yang2023decompose} & CLIP(ViT-B/32) & CLIP & 34.36 & 60.78 & 38.13 & 61.83 & 44.26 & 69.025 & 38.92 & 63.89 \\
TG-CIR\pubinfo{23'ACMMM} \cite{wen2023target} & CLIP(ViT-B/16) & CLIP & 45.22 & 69.66 & 52.60 & 72.52 & 56.14 & 77.10 & 51.32 & 73.09 \\
MCEMiner\pubinfo{24'TIP} \cite{zhang2024multimodal} & ResNet50 & LSTM & 33.23 & 59.16 & 26.15 & 50.87 & 33.83 & 61.40 & 31.07 & 57.14 \\
CaLa-CLIP4Cir\pubinfo{24'SIGIR} \cite{jiang2024cala} & {--} & {--} & 32.93 & 56.82 & 39.20 & 60.13 & 39.16 & 63.83 & 37.10 & 60.26 \\
CaLa-ARTEMIS\pubinfo{24'SIGIR} \cite{jiang2024cala} & {--} & {--} & 40.13 & 66.88 & 46.86 & 67.28 & 49.87 & 74.11 & 45.62 & 69.42 \\
CaLa-BLIP2cir\pubinfo{24'SIGIR} \cite{jiang2024cala} & {--} & {--} & 42.38 & 66.08 & 46.76 & 68.16 & 50.93 & 73.42 & 46.69 & 69.22 \\
CSS-Net\pubinfo{24'KBS} \cite{zhang2024collaborative}& ResNet50 & RoBERTa & 33.65 & 63.16 & 35.96 & 61.96 & 42.65 & 70.70 & 37.42 & 65.27 \\
SPIRIT\pubinfo{24'TOMM} \cite{chen2024spirit} & CLIP(RN50) & CLIP & 43.83 & 68.86 & 52.50 & 74.19 & 56.60 & 79.25 & 50.98 & 74.10 \\
LIMN\pubinfo{24'TPAMI} \cite{wen2023self} & ResNet50 & LSTM & 35.6 & 62.37 & 34.69 & 59.81 & 40.64 & 68.33 & 36.98 & 63.50 \\
DQU-CIR\pubinfo{24'SIGIR} \cite{wen2024simple} & CLIP(ViT-H/14) & {--} & 57.63 & 78.56 & 62.14 & 80.38 & 66.15 & 85.73 & 61.97 & 81.56 \\
Uncertainty\pubinfo{24'ICLR} \cite{chen2022composed} & ResNet50 & RoBERTa & 30.60 & 57.46 & 31.54 & 58.29 & 37.37 & 68.41 & 33.17 & 61.39 \\
Uncertainty(CLVC-Net)\pubinfo{24'ICLR} & ResNet50 & RoBERTa & 31.25 & 58.35 & 31.69 & 60.65 & 39.82 & 71.07 & 34.25 & 63.36 \\
Uncertainty(CLIP4Cir)\pubinfo{24'ICLR} & ResNet50 & RoBERTa & 32.61 & 61.34 & 33.23 & 62.55 & 41.40 & 72.51 & 35.75 & 65.47 \\
SDQUR\pubinfo{24'TCSVT} \cite{xu2024set} & BLIP2 & {--} & 56.87 & 76.50 & 49.93 & 73.33 & 59.66 & 79.25 & 55.49 & 76.36 \\
AlRet\pubinfo{24'TMM} \cite{xu2024align} & ResNet50 & LSTM & 30.19 & 58.80 & 29.39 & 55.69 & 37.66 & 64.97 & 32.26 & 59.76 \\
SPRC\pubinfo{24'ICLR} \cite{bai2023sentence}  & BLIP2(ViT-L) & {--} & 49.18 & 72.43 & 55.64 & 73.89 & 59.35 & 78.58 & 54.92 & 74.97 \\
\midrule
\multicolumn{11}{l}{{\textbf{Zero-Shot Learning-based CMR (ZSL-CMR)}}} \\[2pt]
Progressive(stage1)\pubinfo{22'SIGIR} \cite{zhao2022progressive} & ResNet50 & BERT & 5.75 & 13.04 & 11.73 & 21.98 & 11.22 & 21.93 & 9.57 & 18.98 \\
\bottomrule\bottomrule
\end{tabular}
\end{table*}
}

{\small
\begin{table*}[htp]
\centering
\scriptsize
\setlength{\tabcolsep}{5pt}
\caption{Comparison of Different Methods on Fashion200K \& Shoes Datasets.}
\label{tab:f200k_shoes_comparison}
\newcolumntype{C}[1]{>{\centering\arraybackslash}p{#1}}
\begin{tabular}{@{}p{4cm}C{2cm}C{3cm}*{8}{S[table-format=2.2,round-mode=places,round-precision=2]}@{}}
\toprule\toprule
\multirow{2}{*}{Method} & \multicolumn{2}{c}{Backbone} & \multicolumn{4}{c}{Fashion200K} & \multicolumn{4}{c}{Shoes} \\
\cmidrule(lr){2-3} \cmidrule(lr){4-7} \cmidrule(lr){8-11}
& Visual & Textual & {R@1} & {R@10} & {R@50} & {Avg} & {R@1} & {R@10} & {R@50} & {Avg} \\
\midrule
\multicolumn{11}{l}{{\textbf{Supervised Learning-based CMR (SL-CMR)}}} \\[2pt]
TIRG\pubinfo{18'CVPR}  \cite{vo2019composing} & ResNet18 & LSTM & 14.1 & 42.5 & 63.8 & 40.1 & {--} & {--} & {--} & {--} \\
VAL\pubinfo{20'CVPR} \cite{chen2020image} & MobileNet\&ResNet50 & LSTM & 22.9 & 50.8 & 72.7 & 48.8 & 17.18 & 51.52 & 75.83 & 48.18 \\
LBF(small)\pubinfo{20'CVPR} \cite{hosseinzadeh2020composed} & Faster-RCNN & {--} & 16.26 & 46.9 & 71.73 & 44.96 & {--} & {--} & {--} & {--} \\
LBF(big)\pubinfo{20'CVPR} \cite{hosseinzadeh2020composed} & Faster-RCNN & {--} & 17.8 & 48.35 & 68.5 & 44.83 & {--} & {--} & {--} & {--} \\
VAL\pubinfo{20'CVPR} \cite{chen2020image} & MobileNet\&ResNet50 & LSTM & 22.9 & 50.8 & 72.7 & 48.8 & 17.18 & 51.52 & 75.83 & 48.18 \\
JSVM\pubinfo{20'ECCV} \cite{chen2020learning} & MobileNet & LSTM & 19.0 & 52.1 & 70.0 & 47.0 & {--} & {--} & {--} & {--} \\
TRACE\pubinfo{21'AAAI} \cite{jandial2020trace} & ResNet50 & GRU & {--} & {--} & {--} & {--} & {--} & 18.5 & 51.73 & 77.28 \\
MAAF\pubinfo{20'Arxiv} \cite{dodds2020modality} & ResNet50 & LSTM & 18.94 & {--} & {--} & {--} & {--} & {--} & {--} & {--} \\
GSCMR\pubinfo{21'TIP}\cite{zhang2021geometry} & Faster-RCNN & BiGRU & 21.57&52.84&70.12&48.18& {--} & {--} & {--} & {--}\\
MAN\pubinfo{21'ICASSP} \cite{fu2021multi} & ResNet18 & BERT & 17.1 & 47.9 & 68.1 & 44.37 & {--} & {--} & {--} & {--} \\
MAN\pubinfo{21'ICASSP} \cite{fu2021multi} & MobileNet & BERT & 22.3 & 54.5 & 74.1 & 50.3 & {--} & {--} & {--} & {--} \\
RTIC\pubinfo{21'Arxiv} \cite{shin2021rtic} & ResNet50 & LSTM & {--} & {--} & {--} & {--} & {--} & {--} & 43.66 & 72.11 \\
ComposeAE\pubinfo{21'WACV} \cite{anwaar2021compositional} & ResNet18 & BERT & 22.8 & 55.3 & 73.4 & 50.5 & {--} & {--} & {--} & {--} \\
DCNet\pubinfo{21'AAAI} \cite{kim2021dual} & ResNet50 & Glove & {--} & 46.89 & 67.56 & {--} & {--} & 53.82 & 79.33 & {--} \\
CoSMo\pubinfo{21'CVPR} \cite{lee2021cosmo} & ResNet18\&ResNet50 & LSTM & 23.3 & 50.4 & 69.3 & 47.7 & 16.72 & 48.36 & 75.64 & 46.91 \\
HFCA\pubinfo{21'ACMMM} \cite{zhang2021heterogeneous} & ResNet50 & LSTM & 18.24 & 49.41 & 69.37 & 45.67 & 17.85 & 50.95 & 77.24 & 48.68 \\
CLVC-Net\pubinfo{21'SIGIR} \cite{wen2021comprehensive} & ResNet50 & LSTM & 22.6 & 53.0 & 72.2 & 49.3 & 17.64 & 54.39 & 79.47 & 50.50 \\
HCL\pubinfo{21'MMAsia} \cite{xu2021hierarchical} & ResNet18 & LSTM & 23.48 & 54.03 & 73.71 & 50.41 & {--} & {--} & {--} & {--} \\
JPM(TIRG+MSE)\pubinfo{21'ACMMM} \cite{yang2021cross} & ResNet18 & LSTM & 19.8 & 46.5 & 66.6 & 44.3 & {--} & {--} & {--} & {--} \\
JPM(TIRG+Tri)\pubinfo{21'ACMMM} \cite{yang2021cross} & ResNet18 & LSTM & 17.7 & 44.7 & 64.5 & 42.3 & {--} & {--} & {--} & {--} \\
FashionVLP\pubinfo{22'CVPR} \cite{goenka2022fashionvlp} & ResNet18\&ResNet50 & BERT & {--} & 49.9 & 70.5 & 60.2 & 49.08 & 77.32 & 63.20 & 63.20 \\
SAC\pubinfo{22'WACV} \cite{jandial2022sac} & ResNet50 & GRU & {--} & {--} & {--} & {--} & 18.5 & 51.73 & 77.28 & 49.17 \\
ERR\pubinfo{22'TIP} \cite{zhang2022composed} & ResNet50 & LSTM & {--} & 50.88 & 70.60 & {--} & 19.87 & 55.96 & 79.58 & 51.80 \\
TriArea\pubinfo{22'Sci.Rep.} \cite{zhang2022composed} & ResNet18+TF & LSTM & 17.7 & 46.8 & 66.2 & 43.6 & {--} & {--} & {--} & {--} \\
CRR\pubinfo{22'ACMMM} \cite{zhang2022comprehensive} & ResNet101 & GRU\&BiGRU & 24.85 & 56.41 & 73.56 & 51.61 & 19.41 & 56.38 & 79.92 & 51.90 \\
MultiColSAP\pubinfo{22'TMM} \cite{zhang2023enhance} & ResNet50 & LSTM & {--} & 51.06 & 70.13 & 60.60 & 19.54 & 65.39 & 79.47 & 54.80 \\
MCR\pubinfo{22'TMM} \cite{pang2022heterogeneous} & ResNet50 & LSTM & 18.24 & 49.41 & 69.37 & 45.67 & {--} & {--} & {--} & {--} \\
Progressive\pubinfo{22'SIGIR} \cite{zhao2022progressive} & ResNet50 & BERT & {--} & {--} & {--} & {--} & 19.53 & 55.65 & 80.58 & 51.92 \\
Progressive\pubinfo{22'SIGIR} \cite{zhao2022progressive} & CLIP(ViT-B/32) & BERT & {--} & {--} & {--} & {--} & 22.88 & 58.53 & 84.16 & 55.29 \\
ComqueryFormer\pubinfo{23'TMM} \cite{xu2023multi} & Swin-T & BERT & {--} & 52.2 & 72.2 & 62.2 & {--} & {--} & {--} & {--} \\
NEUCORE\pubinfo{23'NIPS-W} \cite{zhao2024neucore} & ResNet & BiGRU & {--} & {--} & {--} & {--} & {--} & 19.76 & 55.48 & 80.75 \\
ARTEMIS\pubinfo{22'ICLR} \cite{delmas2022artemis} & ResNet50 & LSTM & {--} & {--} & {--} & {--} & 17.60 & 51.05 & 76.85 & 48.50 \\
ARTEMIS\pubinfo{22'ICLR} \cite{delmas2022artemis} & ResNet50 & BiGRU & {--} & {--} & {--} & {--} & 18.72 & 53.11 & 79.31 & 50.38 \\
AMC\pubinfo{23'TOMM} \cite{zhu2023amc} & ResNet50 & LSTM & {--} & {--} & {--} & {--} & {--} & 19.99 & 56.89 & 79.27 \\
CRN\pubinfo{23'TIP} \cite{yang2023composed} & Swin-T(base) & LSTM & 17.32 & 54.15 & 79.34 & 50.27 & 53.3 & 73.3 & 63.3 & 63.3 \\
CRN\pubinfo{23'TIP} \cite{yang2023composed} & Swin-T(Large) & LSTM & 18.92 & 54.55 & 80.04 & 51.17 & 53.5 & 74.5 & 64.0 & 64.0 \\
TG-CIR\pubinfo{23'ACMMM} \cite{wen2023target} & CLIP(ViT-B/16) & CLIP & {--} & {--} & {--} & {--} & {--} & {--} & {--} & 58.05 \\
MCEMiner\pubinfo{24'TIP} \cite{zhang2024multimodal} & ResNet18\&ResNet50 & LSTM & 26.82 & 56.76 & 76.91 & 53.50 & 19.10 & 55.37 & 79.57 & 51.35 \\
CSS-Net\pubinfo{24'KBS} \cite{zhang2024collaborative}& RoBERTa & RoBERTa & 22.2 & 50.5 & 69.7 & 47.5 & 20.13 & 56.81 & 81.32 & 52.75 \\
SPIRIT\pubinfo{24'TOMM} \cite{chen2024spirit} & CLIP(RN50) & CLIP & {--} & 55.2 & 73.6 & {--} & 56.90 & 81.49 & 69.19 & 69.19 \\
DQU-CIR\pubinfo{24'SIGIR} \cite{wen2024simple} & CLIP(ViT-H/14) & {--} & 36.8 & 67.9 & 87.8 & 64.1 & 31.47 & 69.19 & 88.52 & 63.06 \\
LIMN\pubinfo{24'TPAMI} \cite{wen2023self} & ResNet50 & LSTM & {--} & {--} & {--} & {--} & {--} & 57.30 & 82.7 & {--} \\
Uncertainty\pubinfo{24'ICLR} \cite{chen2022composed} & ResNet50 & RoBERTa & 21.8 & 52.1 & 70.2 & 48.0 & 18.41 & 53.63 & 79.84 & 50.63 \\
SDQUR\pubinfo{24'TCSVT} \cite{xu2024set} & BLIP2 & {--} & {--} & {--} & {--} & {--} & 30.14 & 68.30 & 88.30 & 62.25 \\
AlRet\pubinfo{24'TMM} \cite{xu2024align} & ResNet18\&ResNet50 & LSTM & 24.42 & 53.93 & 73.25 & 53.53 & 18.13 & 53.98 & 78.81 & 50.31 \\
AlRet\pubinfo{24'TMM} \cite{xu2024align} & CLIP & LSTM & {--} & {--} & {--} & {--} & 21.02 & 55.72 & 80.77 & 52.50 \\
NSFSE\pubinfo{24'TMM} \cite{wang2024negative} & ResNet50 & BiGRU & 25.3 & 53.8 & 73.5 & 50.87 & {--} & {--} & {--} & {--} \\
NSFSE\pubinfo{24'TMM} \cite{wang2024negative} & ResNet152 & BiGRU & 24.9 & 54.3 & 73.4 & 50.87 & {--} & {--} & {--} & {--} \\
MANME\pubinfo{24'TCSVT} \cite{li2023multi} & ResNet50 & BiGRU & 23.0 & 57.9 & 75.3 & 52.0 & 20.73 & 55.96 & 80.98 & 52.56 \\
\bottomrule\bottomrule
\end{tabular}
\end{table*}
}

\begin{table}[htp]
\centering
\scriptsize 
\setlength{\tabcolsep}{5pt}
\caption{Comparison of Different Methods on CSS Dataset}
\label{tab:css_performance}
\newcolumntype{C}[1]{>{\centering\arraybackslash}p{#1}}
\begin{tabular}{@{}p{3.5cm}C{1cm}C{0.5cm}*{2}{S[table-format=2.2,round-mode=places,round-precision=2]}@{}}
\toprule\toprule
\multirow{2}{*}{Methods} & \multicolumn{2}{c}{Backbone} & \multicolumn{2}{c}{Recall@1} \\
\cmidrule(lr){2-3} \cmidrule(lr){4-5}
& Visual & Textual & {3D-to-3D }& {2D-to-3D}\\
\midrule
\multicolumn{4}{l}{{\textbf{Supervised Learning-based CMR (SL-CMR)}}} \\[2pt]
TIRG\pubinfo{18'CVPR}  \cite{vo2019composing} & ResNet18 & LSTM & 73.70 & 46.60\\
LBF(small)\pubinfo{20'CVPR} \cite{hosseinzadeh2020composed} & Faster-RCNN & {--} & 67.26 & 50.31 \\
LBF(big)\pubinfo{20'CVPR} \cite{hosseinzadeh2020composed} & Faster-RCNN & {--} & 79.20 & 55.69 \\
MAAF\pubinfo{20'Arxiv} \cite{dodds2020modality} & ResNet50 & LSTM & 87.80 & {--}\\
GSCMR\pubinfo{21'TIP} \cite{zhang2021geometry} & Faster-RCNN & BiGRU & 81.81 & 58.74 \\
MAN\pubinfo{21'ICASSP} \cite{fu2021multi} & ResNet18 & BERT & 79.60 &  {--} \\
MAN\pubinfo{21'ICASSP} \cite{fu2021multi} & MobileNet & BERT & 80.40 &  {--} \\
HCL\pubinfo{21'MMAsia} \cite{xu2021hierarchical} & ResNet18 & LSTM & 81.59 & 58.65\\
JPM(TIRG+MSE)\pubinfo{21'MM} \cite{yang2021cross} & ResNet18 & LSTM & 83.80  & {--}\\
JPM(TIRG+Tri)\pubinfo{21'MM} \cite{yang2021cross} & ResNet18 & LSTM & 83.20  & {--}\\
CRR\pubinfo{22'ACMMM} \cite{zhang2022comprehensive} & ResNet101 & BiGRU & 85.84  & {--}\\
\bottomrule\bottomrule
\end{tabular}
\end{table}

\subsection{Composed Natural Image Retrieval}
Traditional image retrieval systems typically rely on either visual or textual inputs alone. However, single-modality approaches are limited in their ability to represent complex queries involving multiple attributes or concepts. Composed Natural Image Retrieval (CNIR) addresses this limitation by integrating both textual descriptions and image content, enabling systems to better understand abstract and nuanced user requirements. For example, users can upload an image of a favorite landscape along with a description such as “same location but in Autumn,” and the CNIR system will analyze both modalities to retrieve semantically relevant images. By enabling more accurate and personalized search experiences, CNIR significantly improves efficiency and user satisfaction. It also opens new opportunities for innovative services and the broader development of advanced image retrieval technologies.

\subsubsection{Benchmark Datasets}

\textbf{CSS} \cite{vo2019composing} is a dataset close to the earliest, which consists of 32K queries (16K for training and 16K for test), such as 'make yellow sphere small' that serve as modification text for the images synthesized in a 3-by-3 grid scene. Although it is a relatively simple dataset, CSS has the benefit of facilitating carefully controlled experiments.

\textbf{CIRR} \cite{liu2021image} is the first released dataset for the natural domain. It consists of a total of 36554 triples derived from 21552 real-life images from the popular natural language reasoning dataset NLVR$^{2}$ \cite{suhr2019corpus}. Each triplet consists of real-life images and human-generated modified sentences, arranged in an 80\%, 10\%, 10\% split between the train/validate/test. This dataset encompasses rich object interactions, which addresses the issues of overly narrow domains and the high number of false negatives in the FashionIQ \cite{wu2021fashion}. Each query image in the validation and test sets has only one target image, and the test is evaluated on an online platform. 

\textbf{CIRCO} \cite{baldrati2023zero} is the first dataset for CIR with multiple ground truths collected from the COCO 2017 unlabeled set~\cite{lin2014microsoft}. It consists of a total of 1020 queries, 220 for the validation set, and 800 for the test set, with an average of 4.53 ground truths per query. In addition, it uses all the 120K images of COCO as the index set, thus providing significantly more distractors than the 2K images of the CIRR test set. Compared to CIRR \cite{liu2021image}, CIRCO employs a two-phase annotation strategy to ensure higher quality, reduced false negatives, and the availability of multiple ground truths. Thus, it is also extremely challenging. The test is similarly evaluated on an online platform.

\textbf{MIT-States} \cite{isola2015discovering} has 60k images, and each comes with an object/noun label and a state/adjective label (such as “red tomato” or “new camera”).  For nouns, it selected words that refer to physical objects, materials, and scenes. For adjectives, it selected words that refer to specific physical transformations. Then, for each adjective, if a clear antonym exists, it is paired with another antonym adjective in the list. Finally, there are 249 nouns and 115 adjectives; on average, each noun is only modified by 9 adjectives it affords.  

\textbf{Birds-to-Words} \cite{forbes2019neural} consists of images of birds from iNaturalist combined with human-annotated paragraphs to describe the difference between these pairs of images. This dataset is characterized by “long” natural language queries, with every one of 3,347 image pairs having on average 4.8 paragraphs, each describing the differences between the pair of birds in an average of 31.38 words. Birds-to-Words provides richer text descriptions in each example than any of the other datasets in the current study, although the number of examples is small.

The following are representative datasets constructed through data generation. 

\textbf{LaSCo} \cite{levy2024data} has 10 times more queries, 2 times more unique tokens, and 17 times more corpus images than the CIRR dataset \cite{liu2021image}. It contains over 389K triplets of query image, modification text, and target image. The training image corpus has 81,653 images and the validation corpus has 39,826 images. The dataset has 13,488 unique language tokens and an average text length of 30.70 tokens per query. Analysis of the dataset shows it has significantly less bias towards a single modality for retrieval compared to previous datasets.

\textbf{Laion-CIR-Combined} \cite{liu2023zero} consists of two sub-datasets, Laion-CIR-Template and Laion-CIR-LLM, which are constructed by two methods of modifying captions (i.e., using templates or LLMs). Specifically, for one image-caption sample, they revise its caption and use the resulting edited caption as a query to retrieve an image with a similar caption as the target image. Both datasets contain around 16K triplets. In addition, by combining the two approaches, a 32K dataset.

\textbf{SynthTriplets18M} \cite{gu2023compodiff} is a vast set of high-quality 18M triplet datasets synthesized using large-scale generative models such as OPT \cite{zhang2022opt} and Stable Diffusion \cite{brooks2023instructpix2pix,hertz2022prompt}. Specifically, they use two strategies: the keyword-based generation (11.4M), and the LLM-based generation (7.4M). SynthTriplets18M is over 500 times larger than existing datasets and covers a diverse range of conditioning cases.

\textbf{Good4cir}~\cite{kolouju2025good4cir} generates two datasets via a three-stage synthetic pipeline: \textbf{CIRR-R} rewrites CIRR captions to produce 199,350 training triplets (28,225 pairs) and 22,620 validation triplets (4,184 pairs). Captions avoid simplistic edits (e.g., "add a red ball") and instead offer fine-grained object modifications. \textbf{Hotel-CIR}: Mines hotel image pairs using perceptual hashing and CLIP embeddings, yielding 415,447 training triplets (65,364 pairs), 13,298 validation triplets (2,092 pairs), and 13,178 test triplets (2,069 pairs + 5 distractors/pair). Generates compound instructions (e.g., "add a vase, change curtains, remove painting").

There are also some datasets that are not commonly
used, including Spot-the-Diff \cite{jhamtani2018learning} and GeneCIS \cite{vaze2023genecis}.

\subsubsection{Results and Analysis}

We evaluate the performance of compositional image retrieval methods on several benchmarks in the natural image domain. Compared to fashion retrieval, natural image tasks involve greater semantic diversity, scene complexity, and more open-ended language descriptions, posing distinct challenges for generalization and robustness.

The \textbf{CSS dataset} (Table~\ref{tab:css_performance}) is one of the earliest benchmarks, constructed using a 3D rendering engine to generate synthetic scenes with precise control over object attributes, viewpoints, and modifications. This design allows for targeted evaluation of a model’s capacity to learn compositional concepts. For instance, MAAF achieves a Recall@1 of 87.8\% on the 3D-to-3D subtask, demonstrating strong performance under idealized conditions. However, the simplicity of the synthetic scenes limits the dataset’s ability to reflect real-world visual complexity. Consequently, CSS is now rarely used to evaluate model performance in natural image settings.

In contrast, the \textbf{CIRR dataset} (Table~\ref{tab:cirr_comparison}) has emerged as the most widely adopted benchmark for natural image retrieval with compositional queries. It emphasizes fine-grained discrimination within visually similar scenes or categories, and performance is evaluated using Recall@K (K = 1, 5, 10). Early supervised methods, such as CIRPLANT, which employed a ResNet+LSTM architecture, achieved only 15.18\% Recall@1. The introduction of CLIP significantly improved retrieval performance; Combiner and CLIP4Cir achieved Recall@1 scores approaching 40\%. More recent models such as SDQUR and ConText-CIR, built on stronger vision-language backbones (e.g., BLIP/BLIP2) and advanced fusion mechanisms, further elevated Recall@1 to 53--55\%. Notably, SDQUR reaches 79.47\% on Recall\_subset@1, illustrating its capability to resolve fine-grained semantic ambiguities. Semi-supervised approaches such as VDG and CASE also show strong performance, leveraging large-scale pseudo-triplets to achieve Recall@1 of 50.96\% and 48.00\%, respectively, outperforming most fully supervised baselines. These results indicate that automatically constructed training data, when of sufficiently high quality, can achieve performance comparable to or exceeding that of limited human annotations in practical settings. In the zero-shot setting, models like SEIZE and PrediCIR, built on ViT-G/14, reach approximately 38\% Recall@1, showing promising results without task-specific supervision, though still behind the best SL and SSL models.

The \textbf{CIRCO dataset} (Table~\ref{tab:circo_comparison}) introduces a more challenging setup by associating each query with multiple valid target images and expanding the retrieval pool to the full COCO dataset (approximately 120K images). Evaluation is based on mAP@1, 10, 25, 50, measuring robustness to ambiguity and scalability in open-world retrieval. Results indicate a strong correlation between model capacity and performance. For example, CIReVL’s mAP@5 improves from 14.94\% (ViT-B/32) to 26.77\% (ViT-G/14), while LDRE increases from 23.35\% (ViT-L/14) to 31.12\% (ViT-G/14). This suggests that larger models generalize better to diverse target sets. Semi-supervised methods such as ConText-CIR (mAP@50 34.72\%) and IP-CIR (mAP@50 38.03\%) benefit from enhanced training paradigms and data augmentation strategies. Despite these gains, the absolute performance on CIRCO remains modest, with IP-CIR representing the current best at 38.03\% mAP@50. This highlights ongoing challenges in fine-grained discrimination and semantic ambiguity resolution in large-scale retrieval.

In addition, Table~\ref{tab:mit_bird_comparison} presents results on two specialized datasets: \textbf{MIT-States}, which focuses on object-state composition, and \textbf{Birds-to-Words}, which targets long-form descriptive query understanding. Although these datasets are less frequently used in recent work, they remain valuable for evaluating a model’s capacity to capture nuanced semantic shifts and process complex textual inputs.

Overall, while method trends in natural image retrieval share some similarities with the fashion domain, the focus shifts more toward handling semantic complexity, open-set diversity, and one-to-many relationships. The performance on CIRR and CIRCO reveals that modeling ambiguity, leveraging high-quality supervision, and improving multimodal fusion remain critical areas for future research.

\begin{table*}[htp]
\centering
\scriptsize
\setlength{\tabcolsep}{4pt}
\caption{Comparison of Different Methods on CIRR Dataset}
\label{tab:cirr_comparison}
\newcolumntype{C}[1]{>{\centering\arraybackslash}p{#1}}
\begin{tabular}{@{}p{4.5cm}C{3cm}C{2cm}*{3}{S[table-format=2.2,round-mode=places,round-precision=2]}*{3}{S[table-format=2.2,round-mode=places,round-precision=2]}@{}}
\toprule\toprule
\multirow{2}{*}{Method} & \multicolumn{2}{c}{Backbone} & \multicolumn{3}{c}{Recall@K} & \multicolumn{3}{c}{Recall\textsubscript{subset}@K} \\
\cmidrule(lr){2-3} \cmidrule(lr){4-6} \cmidrule(lr){7-9}
& Visual & Textual & {K=1} & {K=5} & {K=10} & {K=1} & {K=2} & {K=3} \\
\midrule
\multicolumn{9}{l}{{\textbf{Supervised Learning-based CMR (SL-CMR)}}} \\[2pt]
CIRPLANT\pubinfo{21'ICCV} \cite{liu2021image} & ResNet152 & LSTM & 15.18 & 43.36 & 60.48 & 33.81 & 56.99 & 75.40 \\
CIRPLANT(OSCAR)\pubinfo{21'ICCV} \cite{liu2021image} & ResNet152 & LSTM & 19.55 & 52.55 & 68.39 & 39.20 & 63.03 & 79.49 \\
Combiner\pubinfo{22'CVPR-D} \cite{baldrati2022effective} & CLIP(RN50) & CLIP & 33.59 & 65.35 & 77.35 & 62.39 & 81.81 & 92.02 \\
CLIP4Cir1\pubinfo{22'CVPR-W} \cite{baldrati2022conditioned} & CLIP(RN50) & CLIP & 38.53 & 69.98 & 81.86 & 68.19 & 85.64 & 94.17 \\
NEUCORE\pubinfo{23'NIPS-W} \cite{zhao2024neucore} & ResNet & BiGRU & 18.46 & 49.40 & 63.57 & 44.27 & 67.06 & 78.92 \\
ComqueryFormer\pubinfo{23'TMM} \cite{xu2023multi} & Swin-T & BERT & 25.76 & 61.76 & 75.90 & 51.86 & 76.26 & 89.25 \\
ARTEMIS\pubinfo{22'ICLR} \cite{delmas2022artemis} & ResNet50 & BiGRU & 16.96 & 46.10 & 61.31 & 39.99 & 62.20 & 75.67 \\
RankUn\pubinfo{23'Arxiv} \cite{chen2023ranking} & CLIP(RN50) & CLIP & 32.24 & 66.63 & 79.23 & 61.25 & 81.33 & 92.02 \\
CLIP4Cir2\pubinfo{23'TOMM} \cite{baldrati2023composed} & CLIP(RN50) & CLIP & 42.05 & 76.13 & 86.51 & 70.15 & 87.18 & 94.40 \\
SSN\pubinfo{23'Arxiv} \cite{yang2023decompose} \cite{yang2023decompose} & CLIP(ViT-B/32) & CLIP & 43.91 & 77.25 & 86.48 & 71.76 & 88.63 & 95.54 \\
TG-CIR\pubinfo{23'ACMMM} \cite{wen2023target} & CLIP(ViT-B/16) & CLIP & 45.25 & 78.29 & 87.16 & 72.84 & 89.25 & 94.13 \\
MCEMiner\pubinfo{24'TIP} \cite{zhang2024multimodal} & ResNet152 & LSTM & 17.48 & 46.13 & 62.17 & {--} & {--} & {--} \\
CaLa-CLIP4Cir\pubinfo{24'SIGIR} \cite{jiang2024cala} & {--} & {--} & 35.37 & 68.89 & 80.04 & 66.68 & 84.65 & 93.42 \\
CaLa-ARTEMIS\pubinfo{24'SIGIR} \cite{jiang2024cala} & {--} & {--} & 47.37 & 79.33 & 88.17 & 76.02 & 90.29 & 96.19 \\
CaLa-BLIP2cir\pubinfo{24'SIGIR} \cite{jiang2024cala} & {--} & {--} & 49.11 & 81.21 & 89.59 & 76.27 & 91.04 & 96.46 \\
BLIP4CIR\pubinfo{24'WACV} \cite{liu2024bi} & BLIP & BLIP & 40.17 & 71.81 & 83.18 & 72.34 & 88.70 & 95.23 \\
BLIP4CIR+Bi\pubinfo{24'WACV} \cite{liu2024bi} & BLIP & BLIP & 40.15 & 73.08 & 83.88 & 72.10 & 88.27 & 95.93 \\
SPIRIT\pubinfo{24'TOMM} \cite{chen2024spirit} & CLIP(RN50) & CLIP & 40.23 & 75.10 & 84.16 & 73.74 & 89.60 & 95.93 \\
DQU-CIR\pubinfo{24'SIGIR} \cite{wen2024simple} & CLIP(ViT-H/14) & {--} & 46.22 & 78.17 & 87.64 & 70.92 & 87.69 & 94.68 \\
BLIP4CIR2\pubinfo{24'TMLR} \cite{liucandidate} & BLIP(ViT-B) & BLIP & 44.70 & 76.59 & 86.43 & 75.02 & 89.92 & 95.64 \\
CompoDiff(ST18M \& FT)\pubinfo{23'TMLR} \cite{gu2023compodiff} & CLIP(ViT-L/14) & CLIP & 21.30 & 55.01 & 72.62 & 58.82 & 77.60 & 88.37 \\
CompoDiff(ST18M \& FT)\pubinfo{23'TMLR} \cite{gu2023compodiff} & CLIP(ViT-G/14) & CLIP & 32.39 & 57.61 & 77.25 & 67.88 & 85.29 & 94.07 \\
SDQUR\pubinfo{24'TCSVT} \cite{xu2024set} & BLIP2 & {--} & 53.13 & 83.16 & 90.60 & 79.47 & 91.74 & 96.63 \\
NSFSE\pubinfo{24'TMM} \cite{wang2024negative} & ResNet152 & BiGRU & 20.70 & 52.50 & 67.96 & 44.20 & 65.53 & 78.50 \\
MANME\pubinfo{24'TCSVT} \cite{li2023multi} & ResNet50 & BiGRU & 18.27 & 48.02 & 63.23 & 42.43 & 64.89 & 77.93 \\
SPRC\pubinfo{24'ICLR} \cite{bai2023sentence}  & BLIP2(ViT-L) & {--} & 51.96 & 82.12 & 89.74 & 80.65 & 92.31 & 96.60 \\
\midrule
\multicolumn{9}{l}{{\textbf{Zero-Shot Learning-based CMR (ZSL-CMR)}}} \\[2pt]
PALAVAR\pubinfo{22'ECCV} \cite{cohen2022my} & \multicolumn{2}{c}{CLIP(ViT-B/32)} & 16.62 & 43.49 & 58.51 & {--} & {--} & {--} \\
Pic2Word\pubinfo{23'CVPR} \cite{saito2023pic2word} & \multicolumn{2}{c}{CLIP(ViT-L/14)} & 23.90 & 51.70 & 65.30 & {--} & {--} & {--} \\
SEARLE\pubinfo{23'ICCV} \cite{baldrati2023zero} & \multicolumn{2}{c}{CLIP(ViT-B/16)} & 24.00 & 53.42 & 66.82 & 54.89 & 76.60 & 88.19 \\
SEARLE-XL-OTI\pubinfo{23'ICCV} \cite{baldrati2023zero} & \multicolumn{2}{c}{CLIP(ViT-L/14)} & 24.87 & 52.31 & 66.29 & 53.80 & 74.31 & 86.94 \\
SEARLE-XL\pubinfo{23'ICCV} \cite{baldrati2023zero} & \multicolumn{2}{c}{CLIP(ViT-L/14)} & 24.24 & 52.48 &  66.29 & 53.76 & 75.01 & 88.19 \\
MTCIR\pubinfo{23'Arxiv} \cite{chen2023pretrain} & \multicolumn{2}{c}{CLIP(ViT-L/14)} & 25.52 & 54.58 & 67.59 & 55.64 & 77.54 & 89.47 \\
KEDs\pubinfo{24'CVPR} \cite{suo2024knowledge} & \multicolumn{2}{c}{CLIP(ViT-L/14)} & 26.40 & 54.80 & 67.20 & {--} & {--} & {--} \\
PM\pubinfo{24'Arxiv} \cite{zhang2024zero} & \multicolumn{2}{c}{CLIP(ViT-L/14)} & 26.10 & 55.20 & 67.50 & 56.00 & 76.60 & 88.00 \\
FTI4CIR\pubinfo{24'SIGIR} \cite{lin2024fine} & \multicolumn{2}{c}{CLIP(ViT-L/14)} & 25.90 & 55.61 & 67.66 & 55.21 & 75.88 & 87.98 \\
Denoise-I2W\pubinfo{24'Arxiv} \cite{tang2024denoise} & \multicolumn{2}{c}{CLIP(ViT-L/14)} & 26.9 & 57.2 & 69.8 & 90.6 & {--} & {--} \\
MoTaDual(LinCIR)\pubinfo{24'Arxiv} \cite{li2024motadual} & \multicolumn{2}{c}{CLIP(ViT-L/14)} & 27.28 & 56.39 & {--} & {--} & {--} & {--} \\
MoTaDual(LinCIR)\pubinfo{24'Arxiv} \cite{li2024motadual} & \multicolumn{2}{c}{CLIP(ViT-G/14)} & 38.10 & 68.94 & {--} & {--} & {--} & {--} \\
InstructCIR1\pubinfo{24'Arxiv} \cite{zhong2024compositional} & \multicolumn{2}{c}{CLIP(ViT-L/14)} & 35.18 & 65.12 & 77.61 & {--} & 67.54 & 84.77 \\
DeG\pubinfo{25'Arxiv} \cite{chen2025data} & \multicolumn{2}{c}{CLIP(ViT-L/14)} & 26.8 & 55.0 & 67.7 & {--} & {--} & {--} \\
Slerp+\pubinfo{24'Arxiv} \cite{jangtext} & \multicolumn{2}{c}{BLIP} & 39.74 & 67.74 & 77.40 & 91.55 & 70.65 & 86.72 \\
TSCIR\pubinfo{25'Arxiv} \cite{wang2025mapping} & \multicolumn{2}{c}{CLIP(ViT-L/14)} & 26.10 & 55.15 & 68.66 & 90.06 & {--} & {--} \\
SEIZE\pubinfo{24'ACMMM} \cite{yang2024semantic} & \multicolumn{2}{c}{CLIP(ViT-L/14)} & 28.65 & 57.16 & 69.23 & {--} & 66.22 & 84.05 \\
SEIZE\pubinfo{24'ACMMM} \cite{yang2024semantic} & \multicolumn{2}{c}{CLIP(ViT-G/14)} & 38.87 & 69.42 & 79.42 & {--} & 74.15 & 89.23 \\
PrediCIR\pubinfo{25'CVPR} \cite{tang2025missing} & \multicolumn{2}{c}{CLIP(ViT-L/14)} & 27.2 & 57.0 & 70.2 & {--} & {--} & {--} \\
PrediCIR\pubinfo{25'CVPR} \cite{tang2025missing} & \multicolumn{2}{c}{CLIP(ViT-G/14)} & 37.0 & 66.1 & 77.9 & {--} & {--} & {--} \\
CoLLM\pubinfo{25'CVPR} \cite{huynh2025collm} & \multicolumn{2}{c}{CLIP(ViT-L/14)} & 29.7 & 72.8 & 91.5 & {--} & {--} & {--} \\
\midrule
\multicolumn{9}{l}{{\textbf{Semi-Supervised Learning-based CMR (SSL-CMR)}}} \\[2pt]
MCL(OPT-6.7B)\pubinfo{21'ICML} \cite{liimproving} & \multicolumn{2}{c}{--} & 24.15 & 55.98 & 90.92 & 59.52 & {--} & {--} \\
MCL(Llama2-7B)\pubinfo{21'ICML} \cite{liimproving} & \multicolumn{2}{c}{--} & 26.22 & 56.84 & 91.35 & 61.45 & {--} & {--} \\
CompoDiff(ST18M)\pubinfo{23'TMLR} \cite{gu2023compodiff} & \multicolumn{2}{c}{CLIP(ViT-L/14)} & 18.24 & 53.14 & 70.82 & 57.42 & 77.10 & 87.90 \\
CompoDiff(ST18M)\pubinfo{23'TMLR} \cite{gu2023compodiff} & \multicolumn{2}{c}{CLIP(ViT-G/14)} & 26.71 & 55.14 & 74.52 & 64.54 & 82.39 & 91.81 \\
TransAgg\pubinfo{23'BMVC} \cite{liu2023zero} & \multicolumn{2}{c}{BLIP} & 38.10 & 68.42 & 93.51 & {--} & {--} & {--} \\
CASE\pubinfo{24'AAAI} \cite{levy2024data} & \multicolumn{2}{c}{BLIP(ViT)+BERT} & 48.00 & 79.11 & 87.25 & 75.88 & 90.58 & 96.00 \\
VISTA\pubinfo{24'ACL} \cite{zhou2024vista} & \multicolumn{2}{c}{--} & {--} & 76.10 & {--} & 75.70 & {--} & {--} \\
HyCIR(CC3M+synthetic)\pubinfo{24'Arxiv} \cite{jiang2024hycir} & \multicolumn{2}{c}{CLIP} & 25.08 & 53.49 & 67.03 & 53.83 & 75.06 & 87.18 \\
HyCIR(CC3M+synthetic)\pubinfo{24'Arxiv} \cite{jiang2024hycir} & \multicolumn{2}{c}{BLIP} & 38.28 & 69.03 & 79.71 & 66.79 & 84.79 & 93.06 \\
VDG(Human+COCO'se)\pubinfo{24'CVPR} \cite{jang2024visual} & \multicolumn{2}{c}{BLIP} & 49.37 & 78.12 & 85.52 & 76.68 & 90.46 & 96.05 \\
VDG(Human+NLVR2'se)\pubinfo{24'CVPR} \cite{jang2024visual} & \multicolumn{2}{c}{BLIP} & 50.96 & 80.15 & 86.86 & 77.45 & 90.65 & 96.10 \\
PTG(SPRC+PTG)\pubinfo{24'Arxiv} \cite{hou2024pseudo} & \multicolumn{2}{c}{--} & 36.40 & 66.10 & {--} & {--} & {--} & {--} \\
SCOT\pubinfo{25'Arxiv} \cite{jawade2025scot} & \multicolumn{2}{c}{BLIP(ViT-L/16)} & 36.31 & 66.19 & 77.37 & 92.96 & 64.73 & 83.20 \\
SCOT\pubinfo{25'Arxiv} \cite{jawade2025scot} & \multicolumn{2}{c}{BLIP2(ViT-G/14)} & 36.82 & 64.34 & 74.48 & 93.42 & 75.73 & 88.70 \\
InstructCIR2\pubinfo{25'Arxiv} \cite{duan2025scaling} & \multicolumn{2}{c}{CLIP(ViT-L/14)} & 50.70 & 81.61 & 98.27 & {--} & 76.10 & {--} \\
TSCIR\pubinfo{25'Arxiv} \cite{wang2025mapping} & \multicolumn{2}{c}{CLIP(ViT-L/14)} & 29.16 & 59.33 & 71.88 & 91.52 & {--} & {--} \\
MRA-CIR\pubinfo{25'Arxiv} \cite{tu2025multimodal} & \multicolumn{2}{c}{BLIP2(ViT-L/14)} & 37.98 & 67.45 & 78.07 & 93.98 & 64.17 & 83.01 \\
IP-CIR(LDRE)\pubinfo{25'CVPR} \cite{li2025imagine} & \multicolumn{2}{c}{CLIP(ViT-L/14)} & 29.76 & 58.82 & 71.21 & 90.41 & 62.48 & 81.64 \\
IP-CIR(LDRE)\pubinfo{25'CVPR} \cite{li2025imagine} & \multicolumn{2}{c}{CLIP(ViT-G/14)} & 39.25 & 70.07 & 80.00 & 94.89 & 69.95 & 86.87 \\
CIG-XL(SEARLE)\pubinfo{25'CVPR} \cite{wang2025generative} & \multicolumn{2}{c}{CLIP(ViT-B/32)} & 24.75 & 54.36 & 67.81 & 90.58 & 56.24 & 77.18 \\
CIG-XL(LinCIR)\pubinfo{25'CVPR} \cite{wang2025generative} & \multicolumn{2}{c}{CLIP(ViT-L/14)} & 25.06 & 53.69 & 66.99 & 89.01 & 55.78 & 76.63 \\
ConText-CIR\pubinfo{25'CVPR} \cite{wang2025generative} & \multicolumn{2}{c}{CLIP(ViT-L/14)} & 52.65 & 83.27 & 89.51 & 98.87 & 80.32 & 92.13 \\
ConText-CIR\pubinfo{25'CVPR} \cite{wang2025generative} & \multicolumn{2}{c}{CLIP(ViT-H/14)} & 55.24 & 84.85 & 90.75 & 98.82 & 82.96 & 93.12 \\
CoLLM\pubinfo{25'CVPR} \cite{huynh2025collm} & \multicolumn{2}{c}{CLIP(ViT-L/14)} & 34.7 & {--} & {--} & 77.0 & 93.1 & {--} \\

\bottomrule\bottomrule
\end{tabular}
\end{table*}


\begin{table}[htp]
\centering
\scriptsize
\setlength{\tabcolsep}{2pt} 
\caption{Comparison of Different Methods on CIRCO Dataset.}
\label{tab:circo_comparison}
\newcolumntype{C}[1]{>{\centering\arraybackslash}p{#1}}
\begin{tabular}{@{}p{3.7cm}C{1cm}C{0.4cm}*{4}{S[table-format=2.2,round-mode=places,round-precision=2]}@{}}
\toprule\toprule
\multirow{2}{*}{Method} & \multicolumn{2}{c}{Backbone} & \multicolumn{4}{c}{CIRCO mAP@} \\
\cmidrule(lr){2-3} \cmidrule(lr){4-7}
& Visual & Textual & {5} & {10} & {25} & {50} \\
\midrule
\multicolumn{7}{l}{{\textbf{Zero-Shot Learning-based CMR (ZSL-CMR)}}} \\[2pt]
PALAVAR\pubinfo{22'ECCV} \cite{cohen2022my} & \multicolumn{2}{c}{CLIP(ViT-B/32)} & 4.61 & 5.32 & 6.33 & 6.80 \\
MTCIR\pubinfo{23'Arxiv} \cite{chen2023pretrain} & \multicolumn{2}{c}{CLIP(ViT-L/14)} & 10.36 & 11.63 & 12.95 & 13.67 \\
Pic2Word\pubinfo{23'CVPR} \cite{saito2023pic2word} & \multicolumn{2}{c}{CLIP(ViT-L/14)} & 8.72 & 9.51 & 10.64 & 11.29 \\
SEARLE-XL-OTI\pubinfo{23'ICCV} \cite{baldrati2023zero} & \multicolumn{2}{c}{CLIP(ViT-L/14)} & 10.18 & 11.03 & 12.72 & 13.67 \\
SEARLE-XL\pubinfo{23'ICCV} \cite{baldrati2023zero} & \multicolumn{2}{c}{CLIP(ViT-L/14)} & 11.68 & 12.73 & 14.33 & 15.12 \\
FTI4CIR\pubinfo{24'SIGIR} \cite{lin2024fine} & \multicolumn{2}{c}{CLIP(ViT-L/14)} & 15.05 & 16.32 & 18.06 & 19.05 \\
CIReVL\pubinfo{24'ICLR} \cite{karthik2023vision}  & \multicolumn{2}{c}{CLIP(ViT-B/32)} & 14.94 & 15.42 & 17.00 & 17.82 \\
CIReVL\pubinfo{24'ICLR} \cite{karthik2023vision}  & \multicolumn{2}{c}{CLIP(ViT-L/14)} & 18.57 & 19.01 & 20.89 & 21.80 \\
CIReVL\pubinfo{24'ICLR} \cite{karthik2023vision}  & \multicolumn{2}{c}{CLIP(ViT-G/14)} & 26.77 & 27.59 & 29.96 & 31.03 \\
LinCIR\pubinfo{24'CVPR} \cite{gu2024language} & \multicolumn{2}{c}{CLIP(ViT-L/14)} & 12.59 & 13.58 & 15.00 & 15.85 \\
LinCIR\pubinfo{24'CVPR} \cite{gu2024language} & \multicolumn{2}{c}{CLIP(ViT-H/14)} & 17.60 & 18.52 & 20.46 & 21.39 \\
LinCIR\pubinfo{24'CVPR} \cite{gu2024language} & \multicolumn{2}{c}{CLIP(ViT-G/14)} & 19.71 & 21.01 & 23.13 & 24.18 \\
iSEARLE-XL-OTI\pubinfo{24'Arxiv} \cite{agnolucci2024isearle} & \multicolumn{2}{c}{CLIP(ViT-L/14)} & 11.31 & 12.67 & 14.46 & 15.34 \\
iSEARLE-XL\pubinfo{24'Arxiv} \cite{agnolucci2024isearle} & \multicolumn{2}{c}{CLIP(ViT-L/14)} & 12.50 & 13.61 & 15.36 & 16.25 \\
RTD(SEARLE)\pubinfo{24'Arxiv} \cite{byun2024reducing} & \multicolumn{2}{c}{CLIP(ViT-B/32)} & 11.26 & 12.11 & 13.63 & 14.37 \\
RTD(LinCIR)\pubinfo{24'Arxiv} & \multicolumn{2}{c}{CLIP(ViT-B/32)} & 8.94 & 9.35 & 10.57 & 11.21 \\
RTD(SEARLE)\pubinfo{24'Arxiv} \cite{byun2024reducing} & \multicolumn{2}{c}{CLIP(ViT-L/14)} & 16.53 & 17.89 & 19.77 & 20.68 \\
RTD(LinCIR)\pubinfo{24'Arxiv} \cite{byun2024reducing} & \multicolumn{2}{c}{CLIP(ViT-L/14)} & 17.11 & 18.11 & 20.06 & 21.01 \\
ISA(Sym)\pubinfo{24'ICLR} \cite{du2024image2sentence}& \multicolumn{2}{c}{BLIP} & 9.67 & 10.32 & 11.26 & 11.61 \\
ISA(Asy)\pubinfo{24'ICLR} \cite{du2024image2sentence} & \multicolumn{2}{c}{EfficientNet+BLIP} & 11.33 & 12.25 & 13.42 & 13.97 \\
Slerp\pubinfo{24'ECCV} \cite{jang2024spherical} & \multicolumn{2}{c}{CLIP(ViT-B/32)} & 6.35 & 7.11 & 8.12 & 8.75 \\
Slerp+TAT\pubinfo{24'ECCV}  \cite{jang2024spherical}& \multicolumn{2}{c}{CLIP(ViT-B/32)} & 9.34 & 10.26 & 11.65 & 12.33 \\
Slerp+TAT\pubinfo{24'ECCV}  \cite{jang2024spherical}& \multicolumn{2}{c}{CLIP(ViT-L/14)} & 18.46 & 19.41 & 21.43 & 22.41 \\
Slerp+TAT\pubinfo{24'ECCV}  \cite{jang2024spherical}& \multicolumn{2}{c}{BLIP(ViT-L/16)} & 17.84 & 18.44 & 20.24 & 21.07 \\
LDRE\pubinfo{24'SIGIR} \cite{yang2024ldre} & \multicolumn{2}{c}{CLIP(ViT-B/32)} & 17.96 & 18.32 & 20.21 & 21.11 \\
LDRE\pubinfo{24'SIGIR} \cite{yang2024ldre} & \multicolumn{2}{c}{CLIP(ViT-L/14)} & 23.35 & 24.03 & 26.44 & 27.50 \\
LDRE\pubinfo{24'SIGIR} \cite{yang2024ldre} & \multicolumn{2}{c}{CLIP(ViT-G/14)} & 31.12 & 32.24 & 34.95 & 36.03 \\
HyCIR(Pic2Word)\pubinfo{24'Arxiv} \cite{jiang2024hycir} & \multicolumn{2}{c}{CLIP} & 14.12 & 15.02 & 16.72 & 17.56 \\
HyCIR(Pic2Word)\pubinfo{24'Arxiv} \cite{jiang2024hycir} & \multicolumn{2}{c}{BLIP} & 18.91 & 19.67 & 21.58 & 22.49 \\
MoTaDual\pubinfo{24'Arxiv} \cite{li2024motadual} & \multicolumn{2}{c}{CLIP(ViT-L/14)} & 20.42 & 21.62 & {--} & {--} \\
InstructCIR1\pubinfo{24'Arxiv} \cite{zhong2024compositional} & \multicolumn{2}{c}{CLIP(ViT-L/14)} & 22.32 & 23.80 & 26.25 & 27.32 \\
DeG\pubinfo{25'Arxiv} \cite{chen2025data} & \multicolumn{2}{c}{CLIP(ViT-L/14)} & 13.7 & 14.9 & 16.8 & 17.7 \\
InstructCIR2\pubinfo{25'Arxiv} \cite{duan2025scaling} & \multicolumn{2}{c}{CLIP(ViT-L/14)} & 10.98 & 12.94 & 13.84 & 15.62 \\
TSCIR\pubinfo{25'Arxiv} \cite{wang2025mapping} & \multicolumn{2}{c}{CLIP(ViT-L/14)} & 14.79 & 15.15 & 16.92 & 19.00 \\
SEIZE\pubinfo{24'ACMMM} \cite{yang2024semantic} & \multicolumn{2}{c}{CLIP(ViT-B/32)} & 19.04 & 19.64 & 21.55 & 22.49 \\
SEIZE\pubinfo{24'ACMMM} \cite{yang2024semantic} & \multicolumn{2}{c}{CLIP(ViT-L/14)} & 24.98 & 25.82 & 28.24 & 29.35 \\
PrediCIR\pubinfo{25'CVPR} \cite{tang2025missing} & \multicolumn{2}{c}{CLIP(ViT-L/14)} & 15.7 & 17.1 & 18.6 & 19.3 \\
CoLLM\pubinfo{25'CVPR} \cite{huynh2025collm} & \multicolumn{2}{c}{CLIP(ViT-L/14)} & 20.3 & 20.8 & {--} & 23.4 \\
\midrule
\multicolumn{7}{l}{{\textbf{Semi-Supervised Learning-based CMR (SSL-CMR)}}} \\[2pt]
MCL(OPT-6.7B)\pubinfo{21'ICML} \cite{liimproving} & \multicolumn{2}{c}{--} & 15.14 & 16.13 & 17.88 & 18.82 \\
MCL(Llama2-7B)\pubinfo{21'ICML} \cite{liimproving} & \multicolumn{2}{c}{--} & 17.67 & 18.86 & 20.80 & 21.68 \\
CompoDiff\pubinfo{23'TMLR} \cite{gu2023compodiff} & \multicolumn{2}{c}{CLIP(ViT-L/14)} & 12.55 & 13.36 & 15.83 & 16.43 \\
CompoDiff\pubinfo{23'TMLR} \cite{gu2023compodiff} & \multicolumn{2}{c}{CLIP(ViT-G/14)} & 15.33 & 17.71 & 19.45 & 21.01 \\
TSCIR\pubinfo{25'Arxiv} \cite{wang2025mapping} & \multicolumn{2}{c}{CLIP(ViT-L/14)} & 18.37 & 19.55 & 21.64 & 22.71 \\
MRA-CIR\pubinfo{25'Arxiv} \cite{tu2025multimodal} & \multicolumn{2}{c}{BLIP2(ViT-L/14)} & 27.14 & 28.85 & 31.54 & 32.63 \\
CAT-LLM\pubinfo{25'CVPR} \cite{wang2025generative} & \multicolumn{2}{c}{CLIP(ViT-L/14)} & 15.00 & 15.73 & 17.51 & 18.45 \\
CAT-LLM\pubinfo{25'CVPR} \cite{wang2025generative} & \multicolumn{2}{c}{CLIP(ViT-B/16)} & 13.95 & 14.47 & 16.00 & 16.74 \\
IP-CIR(LDRE)\pubinfo{25'CVPR} \cite{li2025imagine} & \multicolumn{2}{c}{CLIP(ViT-L/14)} & 26.43 & 27.41 & 29.87 & 31.07 \\
IP-CIR(LDRE)\pubinfo{25'CVPR} \cite{li2025imagine} & \multicolumn{2}{c}{CLIP(ViT-G/14)} & 32.75 & 34.26 & 36.86 & 38.03 \\
CIG-XL(SEARLE)\pubinfo{25'CVPR} \cite{wang2025generative} & \multicolumn{2}{c}{CLIP(ViT-B/32)} & 10.3 & 10.79 & 12.12 & 12.76 \\
CIG-XL(LinCIR)\pubinfo{25'CVPR} \cite{wang2025generative} & \multicolumn{2}{c}{CLIP(ViT-L/14)} & 12.97 & 13.64 & 15.14 & 16.01 \\
ConText-CIR\pubinfo{25'CVPR} \cite{wang2025generative} & \multicolumn{2}{c}{CLIP(ViT-L/14)} & 30.05 & 30.53 & 34.79 & 34.72 \\
\bottomrule\bottomrule
\end{tabular}
\end{table}

\begin{table*}[htp]
\centering
\scriptsize
\setlength{\tabcolsep}{5pt}
\caption{Performance Comparison of Different Methods on MIT-States and Birds-to-Words Datasets.}
\label{tab:mit_bird_comparison}
\newcolumntype{C}[1]{>{\centering\arraybackslash}p{#1}}
\begin{tabular}{@{}p{3cm}C{2cm}C{2cm}*{6}{S[table-format=2.2,round-mode=places,round-precision=2]}@{}}
\toprule\toprule
\multirow{2}{*}{Method} & \multicolumn{2}{c}{Backbone} & \multicolumn{4}{c}{MIT-States} & \multicolumn{2}{c}{Birds-to-Words} \\
\cmidrule(lr){2-3} \cmidrule(lr){4-7} \cmidrule(lr){8-9}
& Visual & Textual & {R@1} & {R@5} & {R@10} & {Avg} & {R@10} & {R@50} \\
\midrule
\multicolumn{9}{l}{{\textbf{Supervised Learning-based CMR (SL-CMR)}}} \\[2pt]
TIRG\pubinfo{18'CVPR}  \cite{vo2019composing} & ResNet18 & LSTM & 12.20 & 31.90 & 43.10 & 29.07 & {--} & {--} \\
LBF(small)\pubinfo{20'CVPR} \cite{hosseinzadeh2020composed} & Faster-RCNN & {--} & 14.72 & 35.3 & 46.56 & 32.19 & {--} & {--} \\
LBF(big)\pubinfo{20'CVPR} \cite{hosseinzadeh2020composed} & Faster-RCNN & {--} & 14.72 & 35.3 & 46.56 & 32.19 & {--} & {--} \\
MAAF\pubinfo{20'Arxiv} \cite{dodds2020modality} & ResNet50 & LSTM & 12.70 & 32.60 & 44.80 & 30.03 & 34.75 & 66.29 \\
TRACE\pubinfo{21'AAAI} \cite{jandial2020trace} & ResNet50 & GRU & {--} & {--} & {--} & {--} & 19.56 & 45.24 \\
GSCMR\pubinfo{21'TIP}\cite{zhang2021geometry} & Faster-RCNN & BiGRU & 17.28 & 36.45 & 47.04 & 33.59 & {--} & {--}\\
ComposeAE\pubinfo{21'WACV} \cite{anwaar2021compositional} & ResNet18 & BERT & 13.90 & 35.30 & 47.90 & 32.37 & {--} & {--} \\
MAN\pubinfo{21'ICASSP} \cite{fu2021multi} & ResNet18 & BERT & 13.90 & 35.30 & 47.90 & 31.53 & {--} & {--} \\
MAN\pubinfo{21'ICASSP} \cite{fu2021multi} & MobileNet & BERT & 15.60 & 36.70 & 47.70 &  33.33 & {--} & {--} \\
RTIC\pubinfo{21'Arxiv} \cite{shin2021rtic} & ResNet50 & LSTM & {--} & {--} & {--} & {--} & 37.40 & 66.97 \\
HFCA\pubinfo{21'ACMMM} \cite{zhang2021heterogeneous} & ResNet50 & LSTM & 14.30 & 35.36 & 47.12 & 32.26 & {--} & {--} \\
HCL\pubinfo{21'MMAsia} \cite{xu2021hierarchical} & ResNet18 & LSTM & 15.22 & 35.95 & 46.71 & 32.63 & {--} & {--} \\
CRR\pubinfo{22'ACMMM} \cite{zhang2022comprehensive} & ResNet101 & GRU & 17.71 & 37.16 & 47.83 & 34.23 & {--} & {--} \\
TriArea\pubinfo{22'Sci.Rep.} \cite{zhang2022composed} & ResNet18+TF & LSTM & 13.20 & 33.30 & 44.30 & 30.27 & {--} & {--} \\
SAC\pubinfo{22'WACV} \cite{jandial2022sac} & ResNet50 & GRU & {--} & {--} & {--} & {--} & 19.56 & 45.24 \\
LIMN\pubinfo{24'TPAMI} \cite{wen2023self} & ResNet50 & LSTM & {--} & {--} & {--} & {--} & 40.34 & 66.18 \\
\bottomrule\bottomrule
\end{tabular}
\end{table*}

\subsection{Composed Video Retrieval}
Composed Video Retrieval (CoVR) enables the retrieval of specific videos from large databases by integrating visual queries with textual modification instructions, allowing for more precise and context-aware searches. This approach overcomes the limitations of traditional content-based video retrieval systems, which rely solely on visual features and often fail to capture user intent or nuanced context. The primary objective of CoVR is to improve search accuracy by leveraging multi-modal inputs. CoVR holds strong application potential across multiple domains, including online video platforms, live event discovery, and sports video retrieval.  On video platforms, it supports advanced content recommendation and management systems by identifying and suggesting videos that better align with user preferences and interests.

\begin{table}[]
\caption{Performance Comparison on  WebVid-CoVR-Test Dataset.}
\setlength{\tabcolsep}{5pt}
\label{tab:webvid}
\begin{tabular}{lccccc}
\hline
\multicolumn{1}{c}{\multirow{2}{*}{Method}} & \multirow{2}{*}{Backbone} & \multicolumn{4}{c}{WebVid-CoVR-Test} \\ \cline{3-6} 
\multicolumn{1}{c}{}                        &                           & R@1     & R@5     & R@10    & R@50   \\ \hline
CoVR(Avg)\pubinfo{24} \cite{ventura2024covr}                                   & CLIP                      & 44.37   & 69.13   & 77.62   & 93.00  \\
CoVR(Avg)\pubinfo{24} \cite{ventura2024covr}                                   & BLIP                      & 45.46   & 70.46   & 79.54   & 93.27  \\
CoVR-2(Avg)\pubinfo{24} \cite{ventura2024covr2}                                 & BLIP-2                    & 45.66   & 71.71   & 81.30   & 94.80  \\
CoVR(ft-CA)\pubinfo{24} \cite{ventura2024covr}                                 & BLIP                      & 53.13   & 79.93   & 86.85   & 97.69  \\
CoVR-2(ft-CA)\pubinfo{24} \cite{ventura2024covr2}                               & BLIP                      & 55.95   & 81.22   & 89.05   & 98.08  \\
CoVR-2(ft-CA)\pubinfo{24} \cite{ventura2024covr2}                               & BLIP-2                    & 59.82   & 83.84   & 91.28   & 98.24  \\
ECDE(ft-CA)\pubinfo{24} \cite{ventura2024covr2}                                 & BLIP                      & 60.12   & 84.32   & 91.27   & 98.72  \\ \hline
\end{tabular}
\end{table}

\begin{table}[]
\caption{Performance Comparison on EgoCVR Dataset.}
\setlength{\tabcolsep}{2pt}
\label{tab:egocvr}
\begin{tabular}{lccccccc}
\hline
\multicolumn{1}{c}{\multirow{3}{*}{Method}} & \multirow{3}{*}{Backbone} & \multicolumn{6}{c}{EgoCVR}                                          \\ \cline{3-8} 
\multicolumn{1}{c}{}                        &                           & \multicolumn{3}{c|}{Global}             & \multicolumn{3}{c}{Local} \\
\multicolumn{1}{c}{}                        &                           & R@1  & R@5  & \multicolumn{1}{c|}{R@10} & R@1     & R@5    & R@10   \\ \hline
BLIP-CoVR\pubinfo{24} \cite{ventura2024covr}                                   & BLIP                      & 5.4  & 15.2 & 24.3                      & 33.1    & 49.5   & 62.9   \\
BLIP-CoVR-ECDE\pubinfo{24} \cite{thawakar2024composed}                              & BLIP                      & 6.0  & 16.3 & 24.8                      & 33.4    & 49.3   & 63.0   \\
CIReVL\pubinfo{24} \cite{karthik2023vision}                                      & CLIP                      & 2.0  & 6.8  & 10.6                      & 33.6    & 49.7   & 61.4   \\
TFR-CVR\pubinfo{24} \cite{hummel2024egocvr}                                     & BLIP                      & 14.1 & 39.5 & 54.4                      & 44.2    & 61.0   & 73.2   \\ \hline
\end{tabular}
\end{table}

\begin{table}[]
\caption{Performance Comparison on Airplane, Tennis, and WHIRT.}
\setlength{\tabcolsep}{1pt}
\label{tab:crsir_three}
\begin{tabular}{lccccccccc}
\hline
\multicolumn{1}{c}{\multirow{3}{*}{Method}} & \multicolumn{9}{c}{Datasets}                                                                                      \\ \cline{2-10} 
\multicolumn{1}{c}{}                        & \multicolumn{3}{c|}{Airplane}             & \multicolumn{3}{c|}{Tennis}               & \multicolumn{3}{c}{WHIRT} \\
\multicolumn{1}{c}{}                        & R@1   & R@5   & \multicolumn{1}{c|}{R@10} & R@1   & R@5   & \multicolumn{1}{c|}{R@10} & R@1    & R@5     & R@10   \\ \hline
TIRG\pubinfo{19} \cite{vo2019composing}                                        & 15.05 & 37.28 & 52.20                     & 5.68  & 22.16 & 35.04                     & 0.83   & 4.18    & 4.56   \\
ComposeAE\pubinfo{21} \cite{anwaar2021compositional}                                   & 16.18 & 45.34 & 61.40                     & 5.90  & 23.58 & 37.55                     & 1.62   & 5.17    & 9.58   \\
Cosmo\pubinfo{21} \cite{lee2021cosmo}                                       & 20.91 & 49.31 & 62.66                     & 11.68 & 39.63 & 55.02                     & 1.71   & 7.23    & 13.15  \\
CLIP4Cir\pubinfo{22} \cite{baldrati2022effective}                                    & 20.53 & 51.19 & 63.85                     & 17.58 & 46.51 & 61.68                     & 2.10   & 9.86    & 17.62  \\
AACL\pubinfo{23} \cite{tian2023fashion}                                        & 14.29 & 45.53 & 61.08                     & 7.42  & 28.71 & 45.41                     & 2.28   & 7.93    & 14.64  \\
UncerRe\pubinfo{22} \cite{chen2022composed}                                     & 19.14 & 51.13 & 66.25                     & 13.97 & 43.78 & 60.59                     & 1.84   & 9.16    & 17.66  \\
SHF\pubinfo{24} \cite{wang2024scene}                                         & 62.15 & 96.22 & 98.43                     & 43.23 & 84.93 & 93.89                     & 5.08   & 21.60   & 36.33  \\ \hline
\end{tabular}
\end{table}

\begin{table}[]
\caption{Performance Comparison on PATTERNCOM Dataset.}
\setlength{\tabcolsep}{1pt}
\label{tab:patterncom}
\begin{tabular}{cccccccc}
\hline
\multicolumn{1}{c}{\multirow{2}{*}{Method}} & \multicolumn{7}{c}{PATTERNCOM}                                                        \\ \cline{2-8} 
\multicolumn{1}{c}{}                        & Color & Context & Density & Existence & Quantity & \multicolumn{1}{c}{Shape} & Avg   \\ \hline
WEICOM\pubinfo{24} \cite{psomas2024composed} \\ (CLIP)                                & 46.74 & 20.97   & 22.07   & 12.07     & 20.96    & 26.22                      & 24.83 \\
WEICOM\pubinfo{24} \cite{psomas2024composed} \\(RemoteCLIP)                          & 41.04 & 31.59   & 41.56   & 14.79     & 20.79    & 31.24                      & 30.19 \\ \hline
\end{tabular}
\end{table}

\begin{table}[]
\centering
\caption{Performance Comparison on ITCPR Dataset.}
\setlength{\tabcolsep}{6pt}
\label{tab:itcpr}
\begin{tabular}{lcccc}
\hline
\multicolumn{1}{c}{\multirow{2}{*}{\begin{tabular}[c]{@{}c@{}}Method\\ (Pre-trained on SynCPR)\end{tabular}}} & \multicolumn{4}{c}{ITCPR} \\ \cline{2-5} 
\multicolumn{1}{c}{}                                                                                            & R@1    & R@5    & R@10   & mAP    \\ \hline
CaLa\pubinfo{24} \cite{jiang2024cala}                                                                                                            & 39.33  & 60.85  & 68.66  & 49.29  \\
SPRC\pubinfo{24} \cite{bai2023sentence}                                                                                                            & 42.27  & 61.81  & 69.35  & 51.62  \\
FAFA\pubinfo{25} \cite{liu2025automaticsyntheticdatafinegrained}                                                                                                            & 46.54  & 66.21  & 73.12  & 55.60  \\ \hline
\end{tabular}
\end{table}

\begin{table*}[]
\centering
\caption{Performance Comparison on Sketch-based Image Dataset.}
\setlength{\tabcolsep}{10pt}
\label{tab:sketch_retrieval_results}
\begin{tabular}{lcccccccccc}
\hline
\multicolumn{1}{c}{\multirow{2}{*}{Method}} & \multicolumn{2}{c|}{ShoeV2}      & \multicolumn{2}{c|}{ChairV2}     & \multicolumn{2}{c|}{Sketchy}     & \multicolumn{2}{c|}{FS-COCO}     & \multicolumn{2}{c}{SketchyCOCO} \\
\multicolumn{1}{c}{}                        & R@5  & \multicolumn{1}{c|}{R@10} & R@5  & \multicolumn{1}{c|}{R@10} & R@5  & \multicolumn{1}{c|}{R@10} & R@5  & \multicolumn{1}{c|}{R@10} & R@5            & R@10           \\ \hline
Combiner\pubinfo{22} \cite{baldrati2022effective}                                    & 24.7 & 40.2                      & 35.7 & 39.9                      & 15.7 & 33.7                      & 11.6 & 22.1                      & 15.9           & 32.2           \\
TASK-former\pubinfo{22} \cite{sangkloy2022sketch}                                 & 27.7 & 44.1                      & 40.7 & 45.2                      & 17.8 & 35.2                      & 12.7 & 24.2                      & 19.4           & 34.7           \\
SceneTrilogy\pubinfo{23} \cite{chowdhury2023scenetrilogy}                                & 29.1 & 46.2                      & 43.4 & 46.8                      & 19.7 & 37.2                      & 14.5 & 28.3                      & 20.4           & 40.2           \\
Pic2Word\pubinfo{23} \cite{saito2023pic2word}                                    & 34.7 & 58.4                      & 55.7 & 62.1                      & 22.5 & 48.7                      & 16.7 & 32.6                      & 24.4           & 46.0           \\
SEARLE\pubinfo{23} \cite{baldrati2023zero}                                      & 38.4 & 64.8                      & 60.8 & 66.4                      & 25.3 & 54.2                      & 17.7 & 35.9                      & 26.0           & 50.4           \\
SCIR\pubinfo{24} \cite{koley2024you}                                        & 47.3 & 79.1                      & 73.5 & 81.4                      & 30.6 & 64.2                      & 22.7 & 43.5                      & 33.4           & 61.1           \\ \hline
\end{tabular}
\end{table*}

\begin{table}[]
\caption{Performance Comparison on Multiturn-Fashion-IQ test set.}
\setlength{\tabcolsep}{1pt}
\label{tab:multiturn}
\begin{tabular}{lccccccccc}
\hline
\multicolumn{1}{c}{\multirow{3}{*}{Method}} & \multicolumn{9}{c}{Multiturn-Fashion-IQ test set}                                                            \\ \cline{2-10} 
\multicolumn{1}{c}{}                        & \multicolumn{3}{c|}{Dress}             & \multicolumn{3}{c|}{Shirt}             & \multicolumn{3}{c}{Toptee} \\
\multicolumn{1}{c}{}                        & R@5  & R@8  & \multicolumn{1}{c|}{MRR} & R@5  & R@8  & \multicolumn{1}{c|}{MRR} & R@5     & R@8     & MRR    \\ \hline
TIRG\pubinfo{19} \cite{vo2019composing}                                        & 12.5 & 14.2 & 11.6                     & 13.8 & 16.7 & 12.9                     & 12.0    & 15.6    & 10.9   \\
RTIC\pubinfo{20} \cite{shin2020fashion}                                        & 11.8 & 18.8 & 10.2                     & 14.0 & 20.6 & 12.2                     & 13.2    & 18.9    & 11.7   \\
ComposeAE\pubinfo{21} \cite{anwaar2021compositional}                                   & 18.5 & 26.4 & 14.4                     & 19.8 & 25.2 & 14.5                     & 19.2    & 26.6    & 14.8   \\
CCNet\pubinfo{20} \cite{yu2020curlingnet}                                       & 12.7 & 17.2 & 10.5                     & 15.2 & 18.5 & 13.3                     & 13.6    & 16.2    & 12.1   \\
AUS\pubinfo{21} \cite{wu2021fashion}                                         & 13.4 & 15.3 & 10.5                     & 14.7 & 16.6 & 11.3                     & 12.4    & 13.3    & 10.6   \\
Dialog Manager\pubinfo{18} \cite{guo2018dialog}                              & 12.7 & 16.7 & 10.8                     & 13.9 & 17.7 & 11.6                     & 11.6    & 15.8    & 10.3   \\
IRR\pubinfo{23} \cite{wei2023conversational}                                         & 26.8 & 31.2 & 20.6                     & 25.8 & 30.4 & 19.8                     & 27.1    & 31.7    & 20.9   \\
CAFA\pubinfo{21} \cite{yuan2021conversational}                                        & 30.3 & 33.4 & 26.5                     & 29.8 & 33.5 & 25.6                     & 30.5    & 34.1    & 27.4   \\
FashionNTM\pubinfo{23} \cite{pal2023fashionntm}                                  & 48.3 & 52.8 & {--}                        & 45.1 & 49.8 & {--}                        & 43.8    & 48.8    & {--}      \\ \hline
\end{tabular}
\end{table}



\subsubsection{Benchmark Datasets}
\textbf{WebVid-CoVR} \cite{ventura2024covr} is an extensive dataset created automatically from Web-scraped video-caption pairs, resulting in 1.6 million triplets. Typically, videos have a duration of 16.8 seconds, modification texts consist of 4.8 words, and each target video is linked to 12.7 triplets. WebVid-CoVR also features validation and test sets sourced from the WebVid10M corpus. The validation set contains 7,000 triplets, while the test set comprises 3,200 triplets that have been carefully curated to guarantee high quality. The dataset stands out for its scale and natural domain coverage, making it suitable for training robust models capable of handling real-world applications.

\textbf{CIRR} \cite{liu2021image} and \textbf{FashionIQ} \cite{wu2021fashion} are primarily focused on composed image retrieval but also serve as benchmarks for evaluating the zero-shot performance of CoVR models. These datasets provide insights into how well CoVR methods can generalize to different types of data and tasks.

\textbf{EgoCVR} \cite{hummel2024egocvr} is a meticulously curated benchmark designed for the task of Fine-Grained Composed Video Retrieval, utilizing an extensive egocentric video dataset. This dataset consists of 2,295 queries, specifically crafted to emphasize high-quality temporal video understanding. Each query and target clip is derived from the same long-form video, with textual modifications requiring subtle changes in the depicted actions, thus necessitating robust video comprehension capabilities for effective performance. EgoCVR's construction involves collecting videos and corresponding narrations from the Ego4D dataset, focusing on diverse human-object interactions. To ensure the specificity and subtlety of action modifications, a rigorous manual annotation process was employed, contrasting with automated processes used in previous works. Consequently, this dataset highlights temporal modifications, with 1,811 samples (78.9\%) centered on temporal events, as opposed to 484 samples (21.1\%) involving object-centered changes. This focus significantly differs from the WebVid-CoVR-Test set, where 85\% of samples concentrate on object modifications rather than temporal ones.

\textbf{ICQ} \cite{zhang2024localizing} is a pioneering benchmark specifically designed for the task of localizing events in videos using multimodal queries (MQs). ICQ includes an evaluation dataset called ICQ-Highlight, featuring synthetic reference images and human-curated queries, offering a robust testbed for this novel task. The dataset evaluates model performance across four distinct reference image styles to ensure comprehensive coverage of diverse scenarios.

\subsubsection{Results and Analysis}

We report experimental results for composed video retrieval on two datasets: WebVid-CoVR-Test (Table~\ref{tab:webvid}) and EgoCVR (Table~\ref{tab:egocvr}). On the WebVid-CoVR-Test dataset, models fine-tuned with context-aware (CA) strategies show strong performance, particularly ECDE and CoVR-2(ft-CA), with Recall@10 exceeding 90\% and Recall@1 reaching up to 60\%. These results demonstrate the effectiveness of context modeling and pretraining strategies in open-domain video retrieval.

In contrast, EgoCVR presents a more challenging setting that emphasizes understanding fine-grained temporal variations in egocentric videos. The task includes two subtasks: \textit{Global} (retrieving the correct video from a large gallery) and \textit{Local} (identifying the relevant segment within a known video). Across all methods, performance on the Local task is consistently higher than on the Global task, highlighting the inherent difficulty of coarse-level retrieval across semantically similar long videos. In this setting, TFR-CVR achieves a notable gain in the Global subtask, improving Recall@1 from 5.4\% to 14.1\%, while the improvement in the Local task is relatively modest (from 33.1\% to 44.2\%). This asymmetric gain indicates that the model excels in video-level semantic discrimination. Nonetheless, even the best-performing models still fall short in precise temporal localization, suggesting that current approaches are limited in capturing fine-grained temporal dependencies.

In summary, although models in the CoVR line of work demonstrate strong performance on object-centric, open-domain benchmarks such as WebVid, substantial improvements are still needed for temporally grounded and semantically subtle scenarios like EgoCVR, particularly in fine-grained time reasoning and cross-frame semantic alignment.

\subsection{Composed Remote Sensing Image Retrieval}
Composed Remote Sensing Image Retrieval (CRSIR) enables users to perform more precise and expressive searches by combining both visual and textual inputs. Instead of relying on a single modality, users can submit a reference image together with a textual description that specifies desired geographic features, environmental conditions, or temporal information. This multi-modal approach enhances the system’s ability to interpret complex queries, leading to more accurate and context-aware retrieval results.

CRSIR is particularly valuable in applications that require high levels of specificity and customization. In environmental monitoring, for instance, it allows users to search for images of a particular landscape with added conditions such as vegetation type, season, or land use changes. In urban planning, the method supports detailed analysis of infrastructure development across time and space. In conclusion, CRSIR technology represents a significant advancement in remote sensing data applications, with both important academic implications and vast market potential.

\subsubsection{Benchmark Datasets}
\textbf{PATTERNCOM} \cite{psomas2024composed} is a large-scale, high-resolution remote sensing image retrieval dataset consisting of 38 classes, with each class containing 800 images of 256×256 pixels. In PATTERNCOM, specific categories are selected to be described in the query image, and a corresponding query text defines attributes related to each class. For instance, the query image for the "swimming pool" category is paired with text queries that describe attributes such as shape, with options such as "rectangular", "oval", and "kidney-shaped." The dataset includes six attributes, each comprising up to four different classes. Each attribute can be associated with two to five values per class. The dataset contains a total of over 21,000 queries, with the number of positive queries per class ranging from 2 to 1345.

\textbf{Airplane, Tennis, and WHIRT} \cite{wang2024scene} are organized in terms of quintets, consisting of a reference RS image and its scene graph, a target RS image and its scene graph, and a pair of modifier sentences. Scene graphs capture object attributes and spatial relationships between pairs. Modifier sentences detail differences between reference and target images. \textbf{Airplane} contains 1600 images of airplanes and 3461 pairs of modifier sentences, and \textbf{Tennis} includes 1200 images of tennis courts and 1924 modifier sentence pairs, where both datasets are sourced from UCM \cite{yang2010bag}, PatternNet \cite{zhou2018patternnet}, and NWPU-RESISC45 \cite{cheng2017remote}. \textbf{WHIRT} comprises 4940 images from WHDLD \cite{shao2018performance} and 3344 manually generated ⟨reference images, and  target image⟩ pairs.

\subsubsection{Results and Analysis}

We report composed remote sensing image retrieval results on two task subsets: the Airplane, Tennis, and WHIRT datasets (Table~\ref{tab:crsir_three}), and the PATTERNCOM dataset (Table~\ref{tab:patterncom}). The results in Table~\ref{tab:crsir_three} clearly demonstrate the substantial impact of scene complexity on retrieval performance. On the relatively simple and well-defined Airplane and Tennis datasets, early generic methods such as TIRG and CLIP4Cir achieve only around 15\%--20\% in Recall@1. In contrast, SHF, a method specifically designed for remote sensing imagery, shows a significant performance gain, achieving Recall@1 scores of 62.15\% and 43.23\% on Airplane and Tennis, respectively. This improvement suggests that conventional composition strategies are inadequate for remote sensing tasks, where architectures like SHF, capable of explicitly modeling spatial structure and hierarchical semantics, are essential. Conversely, all methods exhibit a marked performance drop on the more diverse and complex WHIRT dataset. Even SHF achieves only 5.08\% in Recall@1, underscoring the considerable challenge of generalizing to more heterogeneous remote sensing scenes.

Further experimental results indicate that generalization in remote sensing depends not only on model architecture but also on the incorporation of domain-specific knowledge during pretraining. On the PATTERNCOM dataset (Table~\ref{tab:patterncom}), the WEICOM model improves its average recall from 24.83\% to 30.19\% when the general-purpose CLIP encoder is replaced with RemoteCLIP, a backbone pretrained on remote sensing data. These findings highlight the importance of domain-adapted pretraining for capturing unique visual characteristics and enhancing vision-language alignment.


\subsection{Composed Person Retrieval}

Composed Person Retrieval (CPR) represents an innovative approach to identifying specific individuals by leveraging both visual and textual information. Traditional methods, such as Image-based Person Retrieval (IPR) \cite{luo2019bag} and Text-based Person Retrieval (TPR) \cite{li2017person,yang2023towards}, often fall short in effectively utilizing both types of data, leading to a loss in accuracy. CPR aims to address this limitation by simultaneously employing image and text queries to enhance the retrieval process. This dual-modality approach not only increases the descriptive power of the query but also refines the relevance of search results, providing more accurate identification of target individuals. CPR is particularly useful in social services and public security, where precise person identification is crucial.


\subsubsection{Benchmark Datasets}

\textbf{SynCPR} \cite{liu2025automaticsyntheticdatafinegrained} contains 1.15 million high-quality triplets, making it one of the largest synthetic datasets for CPR tasks. It includes diverse scenarios, broad age coverage, and comprehensive ethnic representation, ensuring high quality, realism, and diversity of person images.

\textbf{ITCPR} \cite{liu2025automaticsyntheticdatafinegrained} contains 2,225 annotated triplets, comprising 2,202 unique query combinations, from 1,199 identities. The gallery consists of 20,510 person images, among which 2,225 correspond directly to queries. The dataset is carefully reviewed to eliminate potential false-negative cases, ensuring reliable evaluation metrics. The textual annotations have an average sentence length of 9.54 words, with the longest sentence containing 32 words and the shortest containing 3 words. This dataset is exclusively designated for testing in the ZS-CPR task.



\subsubsection{Results and Analysis}

We report the evaluation results of Composed Person Retrieval on the ITCPR dataset, as shown in Table~\ref{tab:itcpr}. All models are pre-trained on the large-scale SynCPR dataset and tested under the zero-shot setting. Among the compared methods, FAFA~\cite{liu2025automaticsyntheticdatafinegrained} achieves the best performance, reaching 46.54\% in Recall@1 and 55.60\% in mAP, surpassing both CaLa~\cite{jiang2024cala} and SPRC~\cite{bai2023sentence}. These results demonstrate the effectiveness of incorporating fine-grained fusion strategies and synthetic data in improving retrieval accuracy.

\subsection{Composed Sketch-based Image Retrieval}
Composed Sketch-Text Image Retrieval \cite{koley2024you} aims to improve the accuracy and relevance of image retrieval by integrating sketch-based and textual inputs. This approach leverages sketches to capture object shapes and structures, while textual descriptions provide complementary details such as color, material, and texture. By combining coarse structural information with fine-grained attributes, it enables more expressive and flexible querying, especially useful when users lack a specific reference image.

This method offers broad application potential across various domains. In design and creative industries, designers can use sketches paired with text to search for inspiration or references in large databases, supporting fields such as fashion design and interior decoration. In e-commerce, consumers can input hand-drawn sketches or textual descriptions to find visually similar products, enhancing the online shopping experience. In law enforcement, investigators can use textual descriptions and suspect sketches to improve the efficiency of suspect identification and criminal investigations.

\subsubsection{Benchmark Datasets}


\textbf{ShoeV2} and \textbf{ChairV2} \cite{yu2016sketch}: Both datasets emphasize associations between sketch queries and corresponding images, with ShoeV2 comprising 2000 sketches and 6730 photos, while ChairV2 includes 400 sketches and 1800 photos. 

\textbf{Sketchy} \cite{sangkloy2016sketchy}: This dataset expands the scope by covering 125 categories, totaling 12,500 photos, each accompanied by at least five sketches. 

\textbf{FS-COCO} \cite{chowdhury2022fs} and \textbf{SketchyCOCO} \cite{gao2020sketchycoco}: FS-COCO contains 10,000 paired sketch-text-photo triplets, whereas SketchyCOCO offers 14,081 such triplets. The images and textual captions in these datasets originate from MS-COCO \cite{lin2014microsoft}, providing a rich ground for studying multi-modal retrieval scenarios.

\textbf{ImageNet-R(endition)} \cite{hendrycks2021many}: Comprising 30,000 images across 200 ImageNet classes \cite{deng2009imagenet} and spanning 16 domains, this dataset is particularly useful for examining domain-specific attributes and their transferability in image retrieval contexts.

\subsubsection{Results and Analysis}
We report retrieval results on five widely used benchmark datasets (Table~\ref{tab:sketch_retrieval_results}), including ShoeV2, ChairV2, Sketchy, FS-COCO, and SketchyCOCO. Across all datasets, the recently proposed SCIR method \cite{koley2024you} achieves the best performance, with Recall@10 reaching 81.4\% on ChairV2 and 79.1\% on ShoeV2. Notably, SCIR also leads on more complex datasets such as FS-COCO and SketchyCOCO, which require the model to jointly reason over sketch-based structural information and textual semantics. While earlier methods like Combiner and SEARLE incorporate both modalities, their fusion strategies are relatively shallow and offer limited cross-modal interaction, particularly in capturing fine-grained attributes. In contrast, SCIR introduces deeper alignment and richer interactions between sketch and text representations, resulting in substantial performance improvements, exceeding 10\% gain in Recall@10 on several benchmarks. These results underscore the importance of well-designed multimodal fusion mechanisms for retrieving images that require both global contour understanding and fine-grained attribute reasoning.

\subsection{Interactive/Conversational Retrieval}
Interactive/Conversational Retrieval (ICR) represents an advanced approach to image retrieval that leverages natural language interactions between users and systems to refine search outcomes progressively \cite{guo2018dialog, tan2019drill, wu2021deconfounded, yuan2021conversational, levy2024chatting, pal2023fashionntm, wei2023conversational, chen2023fashion, feng2023vqa4cir, barbany2024leveraging}. Unlike traditional methods relying solely on images or predefined textual attributes, ICR integrates user feedback through conversational interfaces, enhancing the accuracy and relevance of search results. This method enables users to provide iterative feedback in natural language, refining queries dynamically until they locate the desired image or item. The primary objective of ICR is to facilitate more intuitive, precise, and personalized searches by incorporating both visual and semantic information effectively. 
ICR has significant applications across various domains, including e-commerce, fashion, and social media.


\subsubsection{Benchmark Datasets}
Several benchmark datasets have been developed to evaluate and advance the capabilities of Interactive Conversational Retrieval systems. These include:

\textbf{Multi-turn FashionIQ} \cite{yuan2021conversational}, which extends the original FashionIQ \cite{wu2021fashion} dataset into a multi-turn setting. It includes 11,505 sessions across three clothing types, structured into transactions of 2-turns, 3-turns, and 4-turns. Each session's data is represented as a pair (In, Un), where In and Un denote the query image and the feedback text for each turn, respectively. This dataset is particularly useful for evaluating systems that require iterative refinement of search queries over multiple rounds of interaction.

\textbf{Multi-turn Shoes}~\cite{pal2023fashionntm}: The initial Shoes dataset \cite{berg2010automatic} encompasses a collection of images featuring women's footwear, spanning 10 distinct categories, which were sourced online and automatically annotated with various attributes. To enhance the applicability of these images for single-turn, feedback-oriented image retrieval tasks, work \cite{berg2010automatic} introduced supplementary natural language descriptions, creating about 10k training pairs and 4.6k test queries. Within the scope of this work, Pal et al. \cite{pal2023fashionntm} extended this dataset to accommodate multi-turn interactions, linking multiple single-turn interactions by matching the target image from one session to the query image of another. This extension facilitates research into multi-turn retrieval scenarios by concatenating several single-turn transactions and maintaining consistency with Multi-turn FashionIQ, thus offering valuable insights into memory retention and feedback handling across multiple turns. 

\textbf{Interactive Retrieval} \cite{wu2021deconfounded} is meticulously designed to support multi-turn interactive retrieval tasks involving complex scenes. Originating from the Visual Genome \cite{krishna2017visual}, this dataset includes detailed captions for various regions within each image, making it ideal for scenarios where users refine queries based on feedback. Comprising 8,960 training samples, approximately 10\% of the entire dataset, it evaluates model performance under limited labeled data conditions. For rigorous assessment, a subset featuring over 18 nouns per caption, totaling 7,000 samples, emphasizes challenging cases with potential confounders and spurious correlations. A user simulator mimics real human interactions by providing image-click and text-click feedback on candidate images. Validated through 2,500 user sessions conducted by six annotators, the simulator accurately predicts actual user clicks in 85\% and 71.4\% of sessions, respectively.

\subsubsection{Results and Analysis}

In existing multi-turn conversational retrieval studies, many methods provide only qualitative analyses, such as visualizations or case-based discussions, without reporting quantitative performance metrics. Therefore, we summarize a few representative approaches with available results, as shown in Table~\ref{tab:multiturn}. We report quantitative retrieval performance on the Multi-turn FashionIQ test set (Table~\ref{tab:multiturn}). In this task, \textbf{FashionNTM}~\cite{pal2023fashionntm} achieves the highest Recall@5 and Recall@8 across all subcategories (Dress, Shirt, Toptee), significantly outperforming earlier methods. For example, it reaches a Recall@5 of 48.3\% in the Dress category, indicating a clear overall performance advantage. This remarkable performance is likely due to its novel architecture based on a \textit{Cascaded Memory Neural Turing Machine (CM-NTM)}, which effectively manages the conversational state and integrates feedback from all previous turns to model the evolving user intent. In terms of evaluation metrics, in addition to the widely used Recall@K, we adopt MRR (Mean Reciprocal Rank) to better capture retrieval efficiency in interactive settings. MRR calculates the reciprocal of the rank at which the target image first appears and then averages this value over all queries. A higher MRR indicates that the system tends to retrieve relevant results earlier in the ranked list, which is particularly important in iterative feedback scenarios. Additionally, \textbf{IRR}~\cite{wei2023conversational} and \textbf{CAFA}~\cite{yuan2021conversational} also demonstrate competitive performance, with Recall@5 exceeding 25\% and MRR reaching 20.6\% and 26.5\%, respectively. In contrast, earlier methods such as \textbf{TIRG} and \textbf{ComposeAE} show limited capacity in handling multi-turn natural language feedback.


\section{FUTURE DIRECTIONS} \label{future}

Composed multi-modal retrieval has emerged as a rapidly evolving research field that, while achieving significant progress, continues to face numerous challenges and opportunities. Current CMR techniques, although demonstrating excellent performance under supervised learning paradigms, still exhibit notable limitations in zero-shot generalization, complex semantic understanding, and real-time performance. Concurrently, the rapid advancement of generative artificial intelligence, LLMs, and edge computing technologies provides novel technical approaches and solutions for CMR research. Furthermore, practical application scenarios demand higher requirements for personalization, explainability, and privacy protection, driving the field toward more practical and industrialized development. Based on current technological trends and practical application needs, we believe future research should focus on the following directions:

\textbf{Generative Data Construction and Quality Control}. 
(1) Controllable Synthetic Data Generation: The primary challenge facing current CMR methods is the insufficient quality and diversity of generated data. Future research needs to develop more refined generation control mechanisms, utilizing diffusion models and generative adversarial networks to achieve precise control over modification types, semantic intensity, and visual attributes. (2) {Multi-level Data Quality Assessment}: Constructing automated quality assessment frameworks that encompass multiple dimensions, including semantic consistency, visual rationality, and textual accuracy. By combining pre-trained multi-modal discriminators with human feedback reinforcement learning techniques, real-time quality monitoring and filtering of synthetic data can be achieved, thereby substantially improving the overall quality of training data. (3) {Hard Negative Data Augmentation}: Designing targeted hard sample generation strategies by constructing negative samples that are semantically similar but exhibit significant detail differences, thereby enhancing model discriminative capabilities. This approach can effectively alleviate the insufficient distinction between positive and negative samples in existing datasets, strengthening model robustness in complex retrieval scenarios.


\textbf{Complex Query Understanding Reasoning}. 
(1) {Temporal and Causal Reasoning}: Extending CMR capabilities to handle temporal relationships and causal logic, which is particularly important for video retrieval and dynamic scene understanding. 
(2) {Common-sense Knowledge Integration}: Integrating large-scale knowledge graphs and common-sense reasoning capabilities into CMR systems, enabling models to understand implicit semantic associations.
(3) {Multi-turn Interaction Optimization}: Developing conversational retrieval systems with memory capabilities capable of understanding cross-turn semantic dependencies and user intent evolution. Through maintaining dialogue states, understanding elliptical expressions, and handling coreference resolution, more natural and fluent interactive experiences can be provided.

\textbf{Explainability and System Trustworthiness}. 
(1) {Multi-granularity Explanation Generation}: Developing CMR systems capable of providing explanations at different abstraction levels. From fine-grained feature activation visualization to medium-grained semantic region annotation and high-level decision-logic explanation, corresponding explanatory information can be provided for users with different needs. This is significant for enhancing user trust and system transparency.
(2) {Uncertainty Quantification}: Introducing Bayesian deep learning and other techniques to quantitatively assess the reliability of retrieval results. When models have an ambiguous understanding of certain queries, they should actively seek clarification from users or provide multiple candidate explanations for user selection.
(3) {Adversarial Robustness}: Enhancing CMR system robustness against input perturbations and adversarial attacks. In particular, in security-sensitive application scenarios, the focus is on ensuring that systems maintain stable performance when faced with malicious input.

\textbf{Efficiency Optimization and Real-time Deployment}. (1) {Model Compression}: Developing lightweight, specialized CMR models. Through knowledge distillation, neural network pruning, and quantization techniques, model complexity can be significantly reduced while maintaining retrieval accuracy. Particularly for mobile and edge devices, adaptive model architectures are needed that can dynamically adjust computational complexity based on the device. (2) {Hierarchical Retrieval Architecture}: Designing multi-level retrieval systems that combine coarse-grained pre-filtering with fine-grained precise matching to substantially improve retrieval efficiency for large-scale databases. While ensuring retrieval quality, millisecond-level response speeds can be achieved to meet real-time application requirements.


\textbf{Cross-modal Extension and Emerging Applications}.
(1) {Immersive Experience Applications}: Expanding from current image-text combinations to richer modal combinations including audio, video, 3D models, and sensor data. For example, users can quickly retrieve and customize 3D objects, scene layouts, and other content in virtual environments through natural language descriptions and gesture interactions.
(2) {Scientific Research Support}: Applying CMR technology to scientific data analysis, such as ``finding images similar to reference cases but with clearer lesions'' in medical image diagnosis, or ``searching for materials with similar crystal structures but different doping elements'' in materials science. These applications require highly specialized domain knowledge integration.


\textbf{Standardized Evaluation and Fairness}.
(1) {Multi-dimensional Evaluation Framework}: Establishing more comprehensive evaluation systems that consider not only retrieval accuracy but also response time, user satisfaction, diversity, and other dimensions. Designing specialized evaluation protocols for different application scenarios to ensure consistency between evaluation results and actual application effectiveness.
(2) {Bias Detection and Mitigation}: Establishing systematic bias detection mechanisms to identify and mitigate algorithmic biases across dimensions such as gender, race, and age. Through fairness constraints and adversarial debiasing techniques, ensure that retrieval systems provide fair services for all user groups.






\section{CONCLUSION}   \label{CONC}
In this paper, we provided a comprehensive review of composed multi-modal retrieval (CMR), an emerging research field that integrates visual and textual information to achieve more flexible and precise content-based retrieval. We discussed the evolution from unimodal to cross-modal and then to composed multi-modal retrieval, highlighting its significance and wide-ranging applications in e-commerce, social media, and beyond. Furthermore, we explored key research directions, including supervised, zero-shot, and semi-supervised learning-based CMR, analyzing their methodologies, challenges, and advancements.  

Despite the remarkable progress, CMR still faces several open challenges, such as improving retrieval accuracy under zero-shot settings, handling large-scale datasets efficiently, and ensuring better generalization across different domains. Moreover, future research should focus on leveraging advancements in foundation models, generative AI, and self-supervised learning to enhance retrieval performance. We believe that continued exploration of CMR will play a pivotal role in shaping next-generation retrieval systems, enabling more intuitive and effective interactions between humans and vast multimodal information spaces.



%





\ifCLASSOPTIONcaptionsoff
  \newpage
\fi



%




{\small

\bibliographystyle{IEEEtran}
\bibliography{egbib}
}

%








\end{document}